\documentclass[sn-mathphys,icol]{sn-jnl}

\jyear{2021}%

\theoremstyle{thmstyleone}%
\theoremstyle{thmstyletwo}%

\theoremstyle{thmstylethree}%

\begin{document}

\title[Towards detecting the temporal fluctuations from GWs in asynchronous gauges]{\vspace*{-.1cm}\hspace*{-.5cm}\parbox[h][2\baselineskip][t]{12cm}{\centering Towards detecting the temporal fluctuations from gravitational waves in asynchronous gauges}}

\author*[1,3]{\fnm{Stefano} \sur{Bondani}}\email{stefano.bondani@uninsubria.it}
\author*[1,2,3]{\fnm{Sergio Luigi} \sur{Cacciatori}}\email{sergio.cacciatori@uninsubria.it}

\affil*[1]{\orgdiv{DiSAT}, \orgname{Universit\`a degli Studi dell'Insubria}, \orgaddress{\street{via Valleggio 11}, \city{Como}, \postcode{22100}, \state{Italy}\country{}}}

\affil*[2]{\orgname{INFN}, \orgaddress{\street{via Celoria 16}, \city{Milano}, \postcode{20133}, \state{Italy}\country{}}}

\affil*[3]{\orgname{Como Lake centre for AstroPhysics (CLAP)}, \orgaddress{\street{via Valleggio 11}, \city{Como}, \postcode{22100}, \state{Italy}\country{}}}

\abstract{The experimental possibility of detecting gravitational waves via their induced time perturbations is explored here, expanding from previous work. The oscillations of the time-time component in the metric are made explicit when working in asynchronous gauges: the desynchronization between a perturbed clock and a reference unperturbed clock constitutes the corresponding observable core target of a detector. To this end we explore the experimental techniques currently available for a preliminary assessment towards a feasibility study. We survey the state of the art in the fields of high precision timing and information preservation, necessary for achieving geodesic non-locality. A synthesis for a feasible prototype detector with the desired characteristics is presented. The optimum point between existing technologies is found around the 1 Hz frequency band, opening the window for the observation of classes of speculated sources of gravitational radiation such as intermediate mass black hole binaries.}




\maketitle

\section{Introduction}

The first detection of gravitational waves (GWs) celebrates its tenth anniversary \cite{LIGO} in 2025. In view of such recurrence, and in the midst of the largely successful onset of GW astronomy \cite{ligovirgo_O3, ligovirgo_O12, ligoNS, ligo_4}, the time is ripe to look ahead towards new GW science. Historically, the approach with regard to GW detection has trailed closely the widespread and commonly adopted description of GWs as the emerging spatial perturbations determined by the choice of the transverse-traceless (TT) gauge \cite{Pirani57, Peters63, Thorne72, Maggiore2007}. This path has had the merit of making GWs stand out clearly and rather efficiently from Einstein's equations, as well as leading somewhat smoothly to GWs eventually becoming detectable with the adoption of interferometers. By contrast, the TT-gauge-rooted description may inadvertently end up obscuring --~by virtue of its ever-recurring early adoption in most GW treatments~--~ the physical nature of what a GW ultimately is, projecting its core trait as a manifold's propagating curvature excitation, onto the particular results offered by the TT gauge itself, i.e.~the $+$ and $\times$ polarizations.
While fully justified in their own merit, the corresponding spatial perturbations oriented along the perpendicular plane with respect to the propagation direction of the GW, constitute an adequate description of the effects of gravitational radiation only as long as the TT gauge applies. Because spacetime, as the physical entity affected by the GW, is a four-dimensional manifold, and because the only invariant is the first (i.e.~linear) order Riemann tensor~--~or more generally, the GW itself~--~a change of gauge yields, in a perturbative approach, different emerging metric perturbations and corresponding observables. This was already correctly pointed out in \cite{light-time}: GW metric perturbations are gauge-dependent quantities, and a fully physical and gauge-invariant description could in principle only exist in terms of the Riemann curvature. Under such premises the wave contribution to the response of a detector would be wholly embodied in an integration of a projection of the Riemann curvature tensor perturbation along specific null geodesics of the unperturbed spacetime. In other words, metric perturbations depend directly on coordinate gauge choices. On the other hand, and again adhering to \cite{light-time}, the freedom to choose a gauge can be exploited to simplify computations, as it has attestedly been the case with the TT gauge: while ascribing physical significance to gauge-dependent quantities remains a perilous exercise, on a formal level any gauge is potentially as good as the TT, provided the solution is a tensor satisfying a wave equation\footnote{Strictly speaking, it must also satisfy the correct boundary and initial conditions: not all solutions in arbitrary gauges will correspond to physically meaningful GWs without proper projection.}, and when assuming  linearized diffeomorphisms, such tensor proves to be physical in the sense of a vanishing of its Lie derivative with respect to $\xi$ ($\xi^{\alpha}$ being the vector field generating a family of coordinate gauge transformations), for all $\xi$.\\
\subsection{The Asynchronous Traceless gauge}
In view of such considerations, and with the declared intention to expose different emerging degrees of freedom associated with a GW than those commonly arising under the TT gauge, it has been recently pointed out that with an appropriate gauge choice, the temporal share of the spatiotemporal perturbation carried by the GW can be made to surface \cite{bondani}.\\
In contrast to the manifestly synchronous TT gauge, the asynchronous-traceless (AT) gauge, first introduced in \cite{bondani}, displays a different set of emerging polarizations, some of which are related to the de-phasing --~under the perturbation carried by a GW~-- of coordinate time with respect to an observer non-local with the system (the observer at $\mathscr{I}^{\pm}$ is sufficient but not strictly necessary, as will be detailed in Section \ref{sec_detectors}).\\
As it is well known, a plane GW moving along the $z$ direction can be written in the TT gauge as:
\begin{align}
    h_{\mu\nu}=\begin{pmatrix}
        0 & 0 & 0 & 0 \\
        0 & h_{+}(w) & h_\times(w) & 0\\
        0 & h_\times(w) & -h_+(w) & 0\\
        0 & 0 & 0 & 0
    \end{pmatrix},
\end{align}
where all functions depend only on the variable $w=z-ct$. Let us consider the change of coordinates
\begin{align}
    x^\mu \longmapsto x^\mu+\xi^\mu_{(\alpha,\beta)}
\end{align}
with the class of diffeomorphisms $\xi^{\mu}(t,x,y,z)$ explicitly written as:
\begin{align}
    \xi^0_{(\alpha,\beta)}(t,x,y,z)=& \frac \alpha4 (x^2-y^2) h'_+(w)+\frac \beta2 xy h'_\times(w), \\
    \xi^1_{(\alpha,\beta)}(t,x,y,z)=& -\alpha\frac x2 h_+(w)-\beta\frac y2 h_+(w), \\
    \xi^2_{(\alpha,\beta)}(t,x,y,z)=& \alpha \frac y2 h_+(w)-\beta\frac x2 h_+(w), \\
    \xi^3_{(\alpha,\beta)}(t,x,y,z)=& \frac 14 (x^2-y^2) h'_+(w)+\frac 12 xy h'_\times(w),
\end{align}
satisfying $g_{\mu\nu}\mapsto g_{\mu\nu}+\partial_\mu \xi_\nu+\partial_\nu \xi_\mu$.
This leads to:
\begin{align}
    h_{\mu\nu}=\begin{pmatrix}
        h_o(w,x,y) & 0 & 0 & -h_o(w,x,y) \\
        0 & (1-\alpha) h_{+}(w) & (1-\beta) h_\times(w) & 0\\
        0 & (1-\beta) h_\times(w) & -(1-\alpha) h_+(w) & 0\\
        -h_o(w,x,y) & 0 & 0 & h_o(w,x,y)
    \end{pmatrix},
\end{align}
in which the term
\begin{align}
    h_o(w,x,y)=\frac \alpha2 (x^2-y^2) h''_+(w)+\beta xy h''_\times(w)
\end{align}
physically describes the desynchronization
as a function of all spatial coordinates.
Choosing $(\alpha,\beta)=(1,0)$, we get:
\begin{align}
    h_{\mu\nu}=\begin{pmatrix}
        h_o(w,x,y) & 0 & 0 & -h_o(w,x,y) \\
        0 & 0 & h_\times(w) & 0\\
        0 & h_\times(w) & 0 & 0\\
        -h_o(w,x,y) & 0 & 0 & h_o(w,x,y)
    \end{pmatrix}, \label{ATgauge}
\end{align}
with
\begin{align}
    h_o(w,x,y)=\frac 12 (x^2-y^2) h''_+(w).
\end{align}
One can also choose $(\alpha,\beta)=(0,1)$, resulting in the $h_+$ components surviving instead of the $h_{\times}$, and occupying  with opposite signs\footnote{Note that the signature is still Lorentzian.} the (1,1) and (2,2) slots. It's even possible to impose $(\alpha,\beta)=(1,1)$, so obtaining:
\begin{align}
    h_{\mu\nu}=\begin{pmatrix}
        h_o(w,x,y) & 0 & {\hspace{0.4cm} } & 0 & -h_o(w,x,y) \\
        0 & 0 & { } & 0 & 0\\
        0 & 0 & { } & 0 & 0\\
        -h_o(w,x,y) & 0 & { } & 0 & h_o(w,x,y)
    \end{pmatrix},
\end{align}
in which
\begin{align}
    h_o(w,x,y)=\frac 12 (x^2-y^2) h''_+(w)+xy h''_\times(w).
\end{align}
In this case, the metric is particularly simple and both the independent modes are encoded in $h_o$. We call it the $(1,1)$-AT gauge. If one moves into a Fermi-Walker free-falling gauge, then, one must compute the form: 
\begin{align}
 ds^2=&-c^2 dt^2\Big(1+R_{0i0j}x^i x^j\Big)- 2c dt dx^i\  \Big( \frac 23 R_{0jik}x^jx^k\Big)\cr&+dx^idx^j \Big( \delta_{ij}-\frac 13 R_{imjn}x^m x^n\Big). \label{FF}
\end{align}
The result is independent of the gauge and one can use any. However, the final expression is conveniently written in terms of the function $h_o(w,x,y)$ of the $(1,1)$-AT gauge.\\ Then, Eq.\,\eqref{FF} takes the form:
\begin{align}
  ds^2=&-c^2 dt^2\Big(1-h_o\Big)- \frac 43c dt   \Big( h_o dz-\frac z2 \partial_x h_o dx -\frac z2 \partial_y h_o dy \Big)\cr
  &+dx^idx^j \delta_{ij}  +\frac {z^2}6 \partial_a \partial_b h_o dx^a dx^b -\frac z3 \partial_a h_o dx^a dz +\frac 13 h_o dz^2,  
\end{align}
where $a,b$ take the values 1 and 2.\\
Thus, a free falling detector sees all components of the metric perturbed in a more or less complicate way, but the $tt$ component is exactly the same as in the $(1,1)$-AT gauge.\\ 
While in the TT-gauge, clocks (a clock being operationally defined as that which measures the local rate of physical time) are fixedly synchronized, resulting in spatial-only perturbations induced by a GW, in the new AT gauge, some of the metric perturbations along certain spatial axes are prevented, resulting in clocks desynchronizing over those same axes per effect of the passing GW. In other words, and echoing \cite{bondani}, some of the spatial perturbations seen by an interferometer get reabsorbed in the desynchronization of clocks in the AT gauge, where spatial contractions in the corresponding directions are impeded. The two are not separate or independent perturbations, but rather, emergent --~yet partial~-- manifestations of the same underlying 4D spacetime curvature perturbation.\smallskip\\
Interest for the detectability of GWs with clocks is not unprecedented: in 2016, the concept was proposed for a GW detector based on lattice atomic clocks to be employed as frequency probes of a shared laser light \cite{Kolkowitz}, effectively measuring a GW-induced Doppler shift. But while the authors in Ref.~\cite{Kolkowitz} introduced the possibility of using clocks as an alternative technique to measure the integral discrepancy produced by a GW over the spatial distance between them --~in full TT-gauge fashion~-- the proposal in \cite{bondani}, which forms the basis for this paper, is to use clocks to measure the differential shifts of time \emph{itself}, in agreement with the AT gauge recipe; this is, in other words, observing clocks losing phase alignment with themselves, and periodically regaining it, as a result of the oscillatory nature of gravitational radiation. This last point is certainly true for perturbations in $h_{00}$ but not necessarily so for mixed perturbations, like $h_{03}$ and $h_{30}$, for which the occurrence of a memory effect in the form of a cumulative desynchronization is to be expected\footnote{In a forthcoming follow-up work we will explore the possibility of probing observables related to $h_{03}$ perturbations. A closed circuital (i.e.~recursive) interferometric-based detector could in principle be sensitive to perturbations in $g_{tz}$: by means of a prolonged path length covered by the perturbed signal inside an appropriately sized triangular detector, the differential mixed perturbation $h_{03}=h_{30}$ gets cumulatively integrated, so that over sufficiently extended measurement times, which translates into a higher number of iterations w.r.t.~a classic interferometer cycle, the full perturbation piles up to produce a detectable signal in the form of a desynchronization between the clocks at the ends of the interferometric closed pattern. Usually, this perturbation is lost over the $\approx10^3$ iterations of a conventional (Earth-based, few-km arm length) interferometer cycle, in the sense that the perturbation is not able to build up over time, and cannot manifest over the $+$ and $\times$ components, which much more readily emerge from the background.}.

\noindent Occasionally, the GW-induced non-uniformity of time flow occurring at any given place has been mentioned, or vaguely treated, often times without derivation, for instance as an additional contribution when calculating the round-trip time of an observable photon in the local Lorentz gauge (see e.g.~Ref.\cite{rakhmanov}, in which it is referred to as the ‘localized gravitational redshift’).\\
Unbeknownst to the authors of \cite{bondani} at the time of publication, a much closer similarity to this approach can be found in \cite{loeb}, where what was indeed introduced is the concept of periodic modulations of the time-time component in the metric tensor. Although technically differing in the operational details about the detection process, as well as the resulting constraints regarding the frequency bands available for probing, the relevant core concept of the physical existence and potential detectability of a GW-induced oscillation \emph{of} time, is shared between these two works.\\
\noindent The phase shifts of the time-time component in the metric are physically associated to non-zero perturbation terms in $h_{00}$; in this paper we are mostly interested in the detectability of observables exclusively related to this term, which is the most straightforward way to exclude spatial components mixing into the observables. Ultimately this translates in the capability of observing a clock desynchronizing (and re-synchronizing) with its own rate, in phase with the GW oscillations.

\smallskip
\noindent In \textsection\,\ref{sec_frequency} we qualitatively evaluate how the different GW frequency bands are likely to become available for detection with this approach, and we anticipate the different experimental intrinsic challenges presented by low or high frequency GW events. In the remainder of Section \ref{sec_detectors}, we survey the state of the art in the fields of high precision timing (\textsection\,\ref{subsec_timing}), information preservation (\textsection\,\ref{subsec_infopres}), and additional experimental techniques such as those necessary to achieve slow light effect (\textsection\,\ref{subsec_slow}) and single photon counting (\textsection\,\ref{subsec_count}), needed for a complete detector. This is not intended to be a thorough review on every existing method or experiment in each subfield, but rather, a sample among the most promising techniques showing a few alternative routes to achieve a common result, a time-oriented GW detector, avoiding the need to conceive and develop new experimental systems or practices from scratch. 
In Section \ref{results} we anticipate how to implement a realization of a real-world detector (\textsection\,\ref{subsec_pipeline}), and the method to infer the reconstructed waveform from the available data points (\textsection\,\ref{subsec_waveform}). In Section \ref{sec_concl} we draw the conclusions at this stage of the work, and we outline the possible unfoldings of our research in the forthcoming future, highlighting which techniques will best fit a roadmap towards a full feasibility study.

\section{Detectors}\label{sec_detectors}
The physical quantity related to the oscillations in $h_{00}$, to be translated into an observable, is the temporal strain $h=\frac{\Delta t}{t}$: for any duration of time $t$, the influence of a fully positive half-period of a GW will, for instance, determine a dilation in the duration of $t$ the order of $\Delta t$, seen from an external unperturbed observer. Likewise, a fully negative half-period will produce a contraction in $t$ of the same order. Following the lead of \cite{bondani}, we adopt as reference a target strain sensitivity of $10^{-21}$. The detectability of specific GW frequency bands translates in the capability of measuring a duration of time $t$ --~which directly infers the intended observable frequency, as detailed in \textsection~\ref{sec_frequency}~-- with precision at least as good as $\Delta t$. The first experimental challenge towards a GW detector is therefore realized in determining how good the current standard in the field of high precision timing is, i.e.~at measuring specific intervals of time with sufficient resolution and precision\interfootnotelinepenalty=0\footnote{Although subtle, the distinction between resolution and precision is somewhat fuzzy in most contexts, but strictly, it should be intended as the former referring to the smallest time interval distinguishable by the measurement system, and the latter indicating consistency in the measurements. Because, for our purposes, the two terms turn out to indicate closely related features of a clock's performance, we will refer to these as mostly interchangeable terms.}.
\interfootnotelinepenalty=100
\smallskip

\noindent Furthermore, as extensively detailed in \cite{bondani}, the observer's clock needs to possess some degree of geodesic non-locality when compared to the observed clock. This, much intuitively, is because a perturbation in the flowing of time can only be observed when two clocks are under different gravitational fields, or in other words, are not showing the same perturbation at the same time. To this end, we carry over on the line of arranging two clocks in conditions of temporal non-locality: rather than having them sufficiently far away in space as to prevent both being gravitationally perturbed at the same time by the same GW (spatial non-locality), we opt for them to be local in space --~as spatially local as possible, in fact~-- but to be operated in a way that ensures the observation happens with a sufficient delay (hence, temporal non-locality) with respect to the perturbation acting on the system. This choice is motivated by the requirement to minimize any spatial propagation of signals, especially so during ongoing GW events, which would inevitably introduce degeneracies in the determination of the purely temporal nature of the perturbation in $h_{00}$. Such requirement translates in the need for an observable acting as proxy for the flowing of time, to be figuratively locked under perturbed conditions, only to be observed under unperturbed conditions --~hence, later on~-- ideally when the GW has left the system entirely, leaving it under flat Minkowski spacetime, or more generally under inertial conditions, for which the metric takes the form of Eq.\,\eqref{FF},
i.e.~the Fermi normal expansion around timelike geodesics. In absence of such a delay the phenomenon, although present, would not be observable. The physical support for the validity of this whole scheme is granted by the non-inertiality of the system under the GW perturbation: with the speed of light $c$ only locally constant, the switching between perturbed non-inertial and unperturbed inertial conditions ensures the kind of non-stationarity needed between the two regimes to guarantee the rate of time varying accordingly, and potentially being measurable.\\
The second main challenge is therefore involved with the locking and preservation, over a sufficiently long period of time, of the information contained in the physical observable --~say, a periodic electromagnetic signal forming the output of our clock~-- into an inert state, without appreciable loss of information. The duration of such a delay determines \emph{a priori} the class of GW sources aimed for detection: the final phase of the coalescence of a stellar mass black hole binary, for instance, takes only a handful of seconds before the binary merges; by contrast, a supermassive black hole binary can endure for years, while in band, before eventually merging; as a result, the duration of the whole GW event determines the extent needed by the system to preserve the information therein contained, which is itself indirectly informative of the corresponding classes of GW sources the detector is aiming at. With this approach we intend to find ways to experimentally probe our observable \textit{after} the GW event has taken place entirely, i.e.~post-merger/ringdown, which is, in principle, the most elementary working configuration, but not necessarily the easiest to achieve: it is conceivable that an advanced and more refined version of our proposal will enable on-the-fly observations, recurring to multiple and relative --~in lieu of single and absolute~-- measurements happening over a scattered modular multi-node detector. In other words, with the support of a refined reconstruction algorithm, it will be possible to infer the gravitational perturbation on the flowing of time even during the GW event itself, based on the inference from a network of detectors.\\
\subsection{Frequency bands}\label{sec_frequency}
The perturbation in the flowing of time carried by a GW begins acting coherently as soon as the temporal strain starts differing from zero in either direction. Let's assume the case of a fully positive strain half-period: after crossing the zero in the curve, any duration of time shorter than the half-period itself, will be dilated by a factor equal to the integrated strain up to that point; clearly, the longer the time spent under same-sign strain, the higher the cumulated perturbation will be. It is clear then, that the best case scenario in terms of maximizing detectability chances is ideally realized for a measurement of a time interval starting and ending at consecutive zeros of the waveform. This in turn is directly informative of the GW frequency under observation, and correspondingly of the different classes of GW-emitting sources. Events in the Ligo-Virgo-KAGRA Collaboration (LVK) \cite{avirgo,lvk2} frequency band, typically the final merger phases of stellar mass black hole binaries at few Mpc luminosity distance, will exhibit waveforms with half-period durations of approximately $10^{-2}$ s at $\approx 10^2$ Hz. The whole event (which we intend as limited from the moment the amplitude enters a minimum frequency threshold\footnote{Usually assumed between 20 and 30 Hz for LIGO.}, or less rigorously, to the last $n$ cycles until merger, with the optimal $n$ to be determined empirically) generally is over in a duration of a handful of seconds.\\
On the other hand, events in the future space-based observatory LISA \cite{lisa, lisa96, lisascird} frequency band will encompass a wider group of sources, mostly centered in frequency around its peak design sensitivity between $\approx 10^{-2}$ and $10^{-3}$ Hz. The half-period duration for a GW in the centi- to millihertz band lays approximately in the order of $10^2$ and up to $10^3$ s. Most astrophysical sources in the LISA band --~massive black hole binaries at redshift $\approx 3$, eccentric extreme mass ratio inspirals at redshift $<2$~-- will spend the better part of their inspiralling time at a quasi-stationary regime, taking arbitrarily long amounts of time before merging; only for some classes of sources --~e.g.~loud galactic binaries~-- this may not necessarily be the case. In general, though, it's fair to expect the signal from such sources to be mostly persistent, while still in band, from hours up to several years.
\begin{figure}[t]
\centering
{\includegraphics[width=.8\columnwidth]{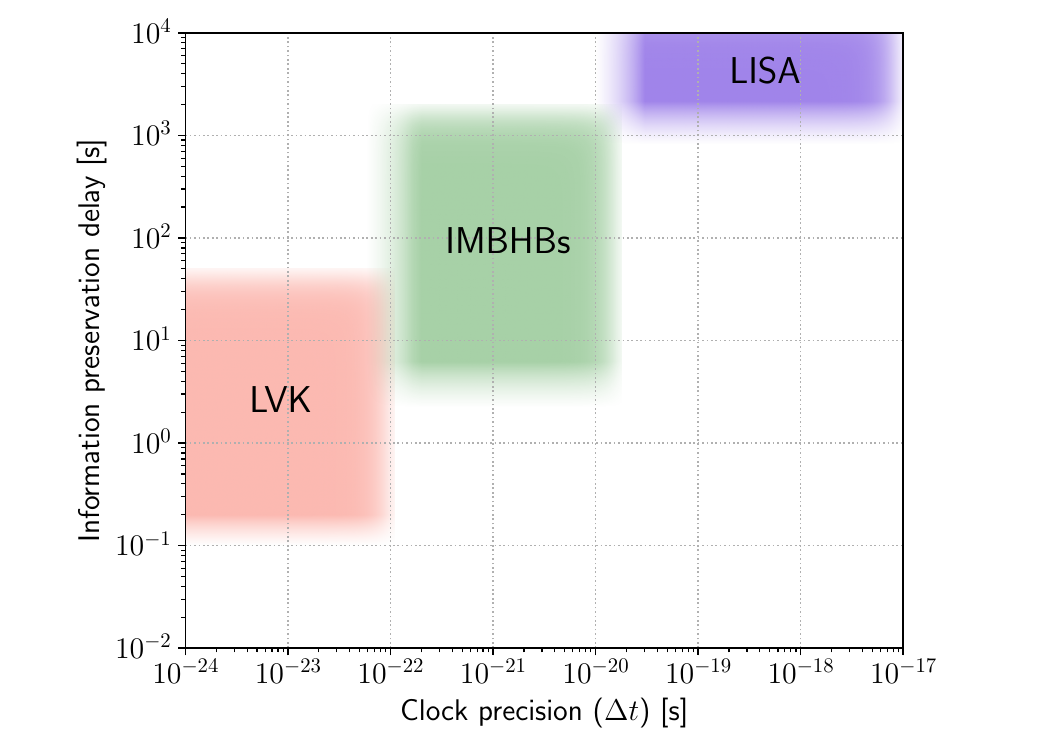}} 
\caption{Constraints in clock precision and duration of the information preservation delay, calculated for a minimum detectability threshold of $10^{-21}$ assuming circular binaries. The shaded areas are constrained on $x$ by a rough estimate of one half-period duration in the LVK, IMBHBs and LISA frequency bands. The constraints on $y$ are dictated by the typical range of durations before merger of the events falling into that band. The criteria for when an event enters a specific band and those constraining the frequency domains are fuzzy, hence the blurred perimeters of each area. The duration of GW events in the LISA band extend well beyond $10^4$\,\,s. The inclusion of this class of sources is only relevant when assuming to catch them in the very final phase of the coalescence. For less stringent constraints in threshold sensitivity (i.e.~$h>10^{-21})$, each area would move rightward, easing the demand on minimum clock precision. Similarly, restricting a typical event to a shorter pre-merger interval, or assuming eccentric orbits even at merger time, would shift the areas downward, relaxing the requirement on the  delay.}
\label{fig:bande}
\end{figure}

\noindent In the middle ground between the kHz and the mHz, at around 1 Hz, an hypothetical class of GW-emitting sources would radiate a signal with half-period duration in the order of the second. Speculating further on the nature of the sources, a class of objects like intermediate mass black hole binaries (IMBHBs), i.e.~binaries of black holes with masses around $10^{3\pm1}$ M$_{\odot}$ --~provided they existed in the first place~-- would emit gravitational radiation roughly centered around the 1-10 Hz frequency band, spending between several seconds and a few hours in the final inspiralling phase before merging.\\
Assuming the aforementioned half-period durations, in order to detect a temporal strain of order $h=\frac{\Delta t}{t}=10^{-21}$, the required timing resolutions needed for these three subgroups of frequency bands, lie over a range extending around $10^{-23}$ s, $10^{-19}$ s and $10^{-21}$ s, respectively\footnote{Because during low frequency events the system spends more time under same-sign GW amplitudes, the integrated time dilation or contraction at the end of a mHz GW half-period will be a few orders of magnitude higher than its equal-amplitude equivalent in the kHz, hence the difference in the minimum required clock precision for equal-magnitude detectability. It goes without saying that higher precisions are, \emph{ceteris paribus}, always desirable, especially when aiming at reconstructing the waveform (see \textsection\,\ref{subsec_waveform}).}.\\
Conversely, the amount of time needed to store information without loss before measurement (the delay between signal generation and observation) ranges from the few seconds needed for the highest frequency events, and up to arbitrarily long durations, increasing with wavelength. The plot in Fig.\,\ref{fig:bande} displays the constraints on clock precision and information preservation times, for typical classes of GW events and sources, assuming circular binaries and a target sensitivity of $10^{-21}$.\smallskip\\
The case could be made for eccentric binaries easing substantially the constraints on the required delay times inside the detector. Indeed, for a binary characterized by initial\footnote{An eccentric binary will eventually circularize by GW emission, reducing $e$ to 0 and shrinking $a$ to a radius.} values of semi-major axis and eccentricity $a_0$ and $e_0$, the timescale for GW infall is given by \cite{Maggiore2007}:
\begin{equation}
    \tau_{\lw{GW}}(e_0,a_0)=\tau_0(a_0)F(e_0),
\end{equation}
where
\begin{equation}
    \tau_0(a_0)=\frac{5}{256}\frac{c^5a_0^4}{G^3m^2\mu}
\end{equation}
is the timescale for GW coalescence for circular orbits for a binary of total mass $m$ and reduced mass $\mu$, and the term containing the eccentricity dependency is given by:
\begin{equation}
    \label{eq:F0}
    F(e_0)=\frac{48}{19}\frac{1}{g^4(e_0)}\int_0^{e_0}de\frac{g^4(e)(1-e^2)^{5/2}}{e(1+\frac{121}{304}e^2)}, 
\end{equation}
in which
\begin{equation}
    \label{eq:ge}
    g(e)=\frac{e^{12/19}}{1-e^2}\Bigl(1+\frac{121}{304}e^2\Bigr)^{870/2299}.
\end{equation}\smallskip
\begin{figure}[ht]
\centering
{\includegraphics[width=.8\columnwidth]{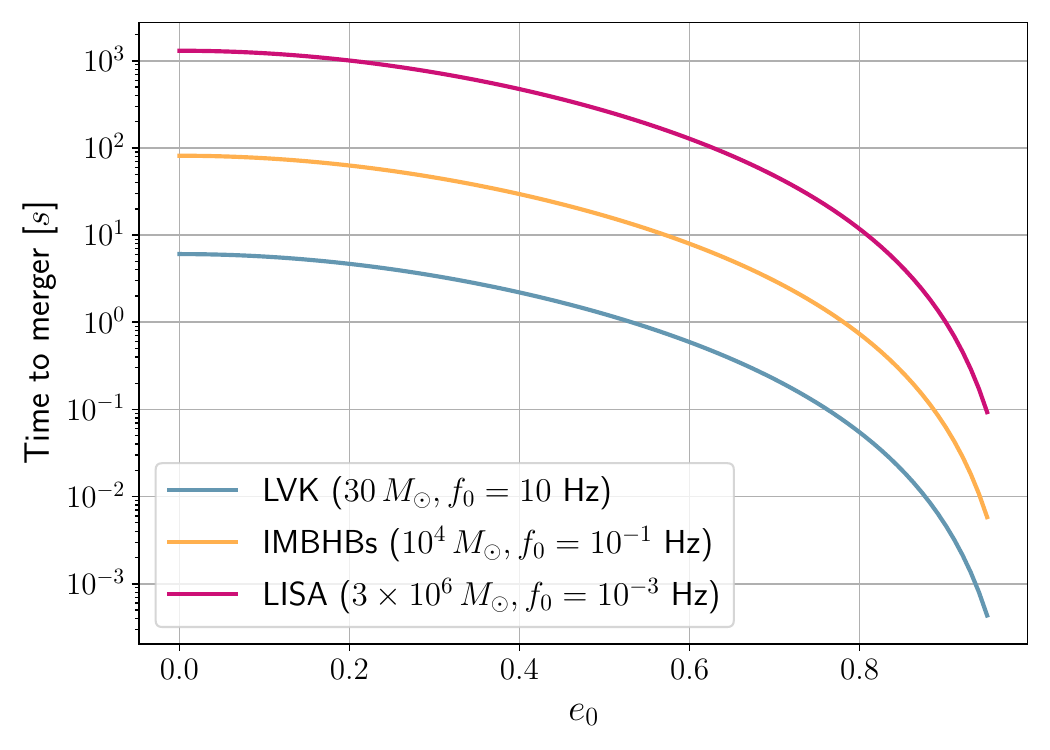}} 
\caption{Time before merger as a function of eccentricity $e_0$ when entering specific bands (at frequency $f_0$), for three equal-mass binaries representative of the typical classes of GW sources emitting inside the frequency bands best covered by LVK, LISA and in the IMBHBs band defined in the text. While eccentricity can substantially decrease the time to merger, it is also very uncommon for isolated binaries this close to merger having not almost completely circularized.}
\label{fig:tempi}
\end{figure}

\noindent The time before merger for a generic binary therefore will decrease with higher values of orbital eccentricity: for instance, a circular equal-mass binary of stellar mass black holes with masses $m_1=m_2=30\,M_{\odot}$, entering the LVK band at $\approx10$ Hz will take approximately $6$ s to merge\footnote{Assuming the actual merger to be realized when the two ISCOs touch.}, while the same binary will merge in 1 s when $e_0=0.5$ , and in a few ms when $e_0=0.9$. Similar considerations apply to classes of GW sources falling in the LISA frequency band, and the IMBHBs frequency band. In Fig.\,\ref{fig:tempi} the time before merger for different values of initial eccentricity when entering a specific band is plotted for three representative classes  of GW sources, labelled by the frequency band they would typically fall into. In addition to the aforementioned example (LVK band, blue curve in the Figure), the plot assumes $10^4\, M_{\odot}$ for the IMBHs entering the band at 0.1 Hz (orange curve) and $3\times10^6 \,M_{\odot}$ BHs entering at 1 mHz  (LISA band, magenta curve). While a low frequency circular binary surely maximizes the requirement for the delay times, exceeding $10^3$\,\,s with these parameters, the same binary will merge in under 10 s if the residual eccentricity when entering the band is above 0.8. On the other hand, most isolated binaries generally circularize enough during their secular evolution, that when reaching the merger phase their residual eccentricity is largely negligible \cite{Blanchet2024}. Only inside dense globular clusters, or alike environments, can non-isolated binaries form and maintain high values of eccentricity, e.g.~by the the Kozai mechanism \cite{kozai1, kozai2}, until late in their dynamical evolution phase \cite{miller}. Moreover, the waveform of an even moderately eccentric binary will show rather intricate features, departing from the more tractable sinusoid of a circular binary, hindering  the waveform reconstruction process as it will be presented in \textsection~\ref{subsec_waveform}. For these reasons we deem safe to assume in the rest of this work to consider circular binaries only, which in turn will provide upper limits on the experimental constraints related to the duration of information preservation requirements in \textsection~\ref{subsec_infopres}.

\noindent We will now turn our attention to currently existing experimental techniques, expressing the state of the art in their own subfield, and we will try to isolate those that would best fit the design and scope of each segment of a feasible GW detector.

\subsection{High precision timing}\label{subsec_timing}
The field of high precision timing has undergone a state of steady and fast-paced progress in the last few decades, thanks in particular to the advancements in the science of optical lattice atomic clocks (OLCs) \cite{atomic_rev, takamoto2005, ludlow2006, riehle2015}: faithfully complying with 1981 Nobel Prize Arthur Schawlow's advice to “never measure anything but frequency”, precise measurements of time have indeed been mostly measurements of frequency for roughly the last sixty years. While microwave based techniques held the role of best available technology until the mid-2000s, optical atomic clocks have since then progressively enhanced the achievable fractional frequency uncertainty\footnote{To avoid conflating domains, we treat
$\delta\nu / \nu_{0}$ as a quantity directly mappable onto timing stability via  $\delta t / t \simeq \delta\nu / \nu_{0}$. This way, a clock’s performance in frequency  space sets effective bounds on resolvable time-related observables.} $\delta\nu/\nu_0$ ($\nu_0$ being the unperturbed atomic transition frequency) from $10^{-15}$ down to $10^{-18}$ and less \cite{history}.\\
In terms of experiments involving physical observables with controllable duration, processes happening in the attosecond ($10^{-18}$\,\,s) timescale have now become an established benchmark \cite{ludlow2013,  attoclock, 10-18s, attosecond, astrain}.
Attosecond streaking, for instance, is naturally well-suited to measure the ultra-short time scales of attosecond-long X-ray pulses \cite{43as}.\\
Furthermore, while absolute or fractional uncertainties in the $10^{-18}$ range are nowadays rooted in experimental certitude \cite{2_18, nicholson-18, huntemann2016, McGrew2018, boulder2021}, consistent progress in the late 2010s has pushed the limits in absolute or fractional frequency uncertainties down to the $10^{-19}$ level \cite{qlogic, optical, shift19}.
Soon after, in the early 2020s, Ref.~\cite{10-21b} reported a relative statistical uncertainty of $8.9\times10^{-20}$. Lastly, in 2022, Ref.~\cite{redshift-mm} reached a fractional frequency measurement uncertainty of $7.6\times10^{-21}$.\smallskip\\
While the use of atomic clocks embodies the current standard in high precision timing, the introduction of the research on nuclear clocks promises to exceed the current levels of timing stability. This is thanks to the higher quality factors and lower sensitivity to perturbations of nuclear oscillators, which can count on small magnitudes of nuclear electromagnetic moments as their main vantage points over atomic oscillators. 
The workings of nuclear clocks usually require EM sources able to excite ultra-narrow nuclear resonance transitions.\\
While the science of actual nuclear clocks is at an ongoing development phase, mostly centered around thorium clocks (more in \textsection~\ref{sec_nuclear}), temporally controlled frequency changes of nuclear transitions can serve to monitor the shortest time intervals to date\footnote{Incidentally, this result shifted back to time, from frequency, the domain in which record-low ultra-precise timing is performed.}: in 2021 the authors in Ref.~\cite{heeg} performed coherent control of nuclear excitons reporting a 40 mrad, equivalent to 10 zeptoseconds (1 zs = $10^{-21}$ s), temporal stability of the phase control in their system. Almost simultaneously, Ref.~\cite{zepto} reported the smallest detectable phase shift\interfootnotelinepenalty=0\footnote{The concepts of phase shift and time delay are in general not interchangeable at the experimental level \cite{phasedelay}: in wave mechanics the relationship $ {\delta t} = \delta\phi(2\pi\nu)^{-1} $ to infer a time delay from a phase shift when frequency $\nu$ is known, strictly holds true only under certain conditions, such as for linear media and with $\nu=\text{const}$. Interpreting a phase shift as a measure of time calls for an experimental setup capable of producing a well-defined carrier frequency with linear behavior and accurate pulse characterization, as well as good control over quantum interference effects. We will assume these requirements are met in producing the results mentioned here.
} to date at 1.43 zs, with a remarkable calculated timing stability of the temporal shift under 50 yoctoseconds (1 ys = $10^{-24}$ s, see also the ensuing considerations in Refs.~\cite{yocto_re} and \cite{yocto_re2}).
\interfootnotelinepenalty=100

\smallskip
\noindent On top of atomic and nuclear clocks, different techniques can achieve comparable performances in high precision timing: in 2011 the authors in \cite{zepto_pulse} were able to control the precision between two delayed pulses down to 300 zs at 2$\sigma$ statistical significance; in 2020 molecular electron interferometry was performed to measure the phase shift of electron waves in the far field with precision better than 247 zs \cite{zepto_photo}.

\smallskip\noindent Regardless of the specific adopted technique, or the physical nature of the corresponding resulting phenomenon, the primary requirement for a clock-acting device inside a GW detector, will be for it to be sensitive to the rate of time above a certain precision threshold when generating its output. Additionally, the measurement of time intervals ought to be obtained or converted in the form of gapped EM signals with intra-separation of well known and sufficient precision. Since the explicit purpose of the detector is to isolate by design only the time perturbations induced by GWs, any spatial perturbation, and especially those acting over the path covered by a propagating EM signal, needs to undergo a minimization process; henceforth, compactness of the experimental setup is the next major prerequisite. In the following subsections we will survey in more detail the aforementioned experimental achievements and assess their versatility towards our scope and requirements. 

\subsubsection{Optical lattice atomic clocks}
Adopting OLCs might appear to be the most obvious way in the immediate future for the purpose of time-oriented GW detection, thanks to well known technology and associated systematics. Controlling and minimizing the clock's intrinsic instability is crucial to produce reliable measurements over time\footnote{This is commonly quantified by indicators such as the Allan deviation \cite{allan1,allan2,allan3}, which measures the oscillator's frequency stability over different time intervals and can serve to compare, once the noise sources are known, the stability of different OLCs.}. The shared requirement among many existing such setups is to possess high atom numbers ($\gt10^3$) to reduce quantum projection noise, and large trap depths to suppress tunnelling. 
To reach a fractional frequency measurement uncertainty of $7.6\times10^{-21}$, clock interrogation in Ref.~\cite{redshift-mm} proceeds, after trapping a large number ($\simeq10^5$) of ultra-cold $^{87}$Sr atoms inside an in-vacuum optical cavity, by probing the ultra-narrow $^1S_0(g)\to{}^3P_0(e)$ transition at 100\,nK, with the resulting excitation fraction observed as the decay of fringe contrast measured by fluorescence Ramsey spectroscopy\footnote{ Ramsey spectroscopy consists in repeated observation of the phase difference between atomic states resulting from the interaction with a stable laser.}. Extended coherence times ($>{}$30 s) are essential for this technique to achieve the reported results (see also \cite{half-minute} on atomic coherence times) both in terms of accuracy and precision. Clocks are arranged in a 1D lattice, with single clock stability of 3.1$\times10^{-18}$ at 1\,s. The record-low uncertainty of 7.6$\times10^{-21}$ between two uncorrelated regions is achieved after averaging times in the few hours timescale.
The size extension of the experimental setup is below half a meter.\\
The OLC setup adopted in Ref.~\cite{10-21b} to reach an uncertainty of $8.9\times10^{-20}$ uses a ‘multiplexed’ system to minimize instabilities from the oscillator; the resulting uncertainty is obtained after $\simeq$ 3 hours of averaging. As with most OLCs, the spatial volume occupied by this clock is again below the cubic meter.\\
Currently, the most compact system to host an OLC has been developed with size $465\times588\times415$ mm$^3$ \cite{compactOLC}.\\
Because the statistical averaging required to reduce uncertainties inherently spans extended periods of time, up into hours timescales, this technique fits the purpose for a GW detector only when the GW half-period duration can match the averaging time, making it a suitable candidate predominantly for long-wavelength events. However, in general the uncertainty reached after few second averaging times can be adequate, at $\simeq10^{-18}$, for the loudest GW events, especially so for OLCs relying on quantum entanglement, which are presented in the next paragraph.
\paragraph{\emph{Entangled systems}}
Regardless of the specifics of single clocks, the performance of OLCs can be enhanced when multiple such systems are arranged in an entangled network configuration \cite{quantum_improvement, network, network-entangled}. Although the notion of entangled clock networks originated with the intention to create an Earth-sized `world clock' \cite{network}, the same concept can be rescaled and adapted to fit a compact GW detector, reducing the distance between the nodes, but mostly preserving the functioning between them. Compared with conventional correlation spectroscopy techniques, whose uncertainty is generally limited by dephasing of the probe laser, quantum-enhanced measurements on entangled systems of phase-locked OLCs can reduce by a factor $\sqrt{2}$ and up to 4 the measurement uncertainty \cite{network-entangled}, possibly surpassing the standard quantum limit, to (asymptotically) reach the Heisenberg limit. In general the main source of decoherence in entangled systems is due to magnetic field noise perturbing the qubit transition frequency, limiting to short probe durations only. To mitigate this issue, superconducting solenoids could be employed, drastically reducing magnetic field fluctuations. To demonstrate the enhancement given by entanglement, the specific setup adopted in Ref.~\cite{network-entangled} employs two remote $^{88}$Sr$^+$ ions entangled using the 422\,nm $S_{1/2}\to P_{1/2}$ transition; a common 674\,nm laser then drives the $^5S_{1/2}\longleftrightarrow {}^4D_{5/2}$ optical transition. The authors found that entangled states yield the lowest experimental uncertainty, with a minimum at Ramsey time\footnote{Ramsey time refers to the duration between two coherent pulses applied to the quantum system in superposition.} $T_R=10\,$ms; the uncertainty is quantified by the entanglement enhancement, defined as the ratio between the number of measurements required to reach a given precision without entanglement and the number of measurements required using the entangled state.\\
To reach a fractional frequency uncertainty of approximately $4\times 10^{-17}$ after 1\,\,s of averaging time, a network of 10 entangled clocks, each containing 10 Al$^+$ ions was employed in Ref.~\cite{network}. The authors add that the quantum gain gets more marked when employing neutral atom clocks, reaching a stability of\,~$\approx2\times10^{-18}$ after 1\,\,s of averaging time in a system of 10 clocks containing $10^3$~neutral atoms each. A projected fractional frequency uncertainty beyond $10^{-20}$ after $10^2$\,\,s averaging time is foreseen with the future employment of erbium or other species with $\mu$Hz linewidth transitions.

\noindent The main drawback of an extended system primarily consists in introducing potential degeneracies in identifying the GW signal as purely time-perturbing. Even so, while single OLCs can vary in size from $10^{-1}$ to 1 m, the two-clocks entangled setup adopted in Ref.~\cite{network} has a total extension of 2 m, which is a factor $\simeq10^3$ shorter than the arms in LVK's interferometers, and in a first approximation is safely negligible for all physically relevant GW frequency bands.\\

\subsubsection{Nuclear and X-ray clocks}\label{sec_nuclear}
Although a full realization of a nuclear clock is yet to come, at the current proof-of-concept stage the main advantage of nuclear, against atomic clocks, consists in enhanced accuracy and stability, mainly against stray EM fields, granted by minimal magnitudes of their electromagnetic moments due to a factor $10^4$ difference in diameter between atoms and nuclei, combined with the role played by the strong force coupling between nucleons; moreover, nuclear clocks offer the possibility to be interrogated without recurring to suspension in vacuum chambers, which is generally required for atomic clocks.\\ 
The research on nuclear clocks has historically mostly focused on $^{229}$Th \cite{nuclear0,nuclear2,nuclear3,nuclear4,nuclear5,nuclear6, nuclear7, nuclear8}, because of its closely spaced energy levels at $\approx8.4$ eV ($\lambda=148.3821(4)$ nm), which entails the possibility of producing the resonant excitation with coherent EM radiation, i.e.~lasers \cite{nuclear7}. The absolute transition frequency uncertainty at $\approx 2\times10^6$ GHz between the ground and the metastable isomeric state $^{229\textrm{m}}$Th has been recently resolved down to the kHz level \cite{th_freq}. As of today, talking in terms of fully realized nuclear clocks may be somewhat improper or at least premature, ‘frequency standard’ being a more fitting definition of what constitutes the current state of the progress in this branch\footnote{More so, the recent upgrade from hypothetical to a realized system \cite{th_freq} only set a reference starting point at $10^{-12}$ in fractional frequency uncertainty.}. Nevertheless, with a reasonably long half-life of the isomer at $\approx10^3$ s \cite{nuclear6} and an expected fractional frequency uncertainty as low as $1\times10^{-20}$ \cite{nuclear8} (i.e.~comparable with the best atomic OLCs, but with enhanced stability), this is the subject of a rapidly evolving research field, which promises to deliver a proper timekeeping clock with the precision necessary for GW detection in the foreseeable future.\smallskip\\
A different technique, adopted by the authors in Ref.~\cite{heeg} consists in employing\,\,$^{57}$Fe M{\"o}ssbauer nuclei and X-ray sources, using mechanical motions of the resonant absorber to shape the X-ray pulses in the temporal domain, and obtaining detectable systematic phase drifts in the few-zs timescale. In particular, the first in a couple of pulses induces a nuclear exciton in the target; a transition dipole moment is therefore induced in the target, of which magnitude and phase are known. The phase control, the authors assert, reaches a stability of 40 mrad, corresponding to systematic phase drifts detectable down to 10 zs.
Furthermore, the authors note, by means of magnetic switching, a `split and control unit' could store the X-ray pulse for a variable time, delaying the control pulse at will \cite{storage}, a feature which, if properly adapted, possibly fits the purpose of producing a controllable interval between subsequent photons, granting the possibility to aim at different specific GW frequencies.\\
On a concomitant variation of this experiment, the authors in Ref.~\cite{zepto} adopted an ensemble of  M{\"o}ssbauer nuclei embedded in a ferromagnetic film, to perform zs single-photon interferometry using electromagnetically induced excitations of magnons, the quasi-particles of the magnetic degrees of freedom in the solid embedding a quantum system. Their result consists in the smallest experimentally controlled time delay in a photon field to date at 1.3 zs, calculated as the ratio between phase shifts and carrier frequency. Moreover, the emission of a single $\gamma$-ray during relaxation of the collectively excited nuclei within the lifetime of 141 ns, naturally presents the supplementary characteristics required for an individual and detectable EM signal in our detector. Performed at the Deutsches Elektronen-Synchrotron (DESY) in Hamburg, one of only four existing endstations\footnote{The other three being the European Synchrotron Radiation Facility (ESRF) in Grenoble, the Advanced Photon Source (APS) in Chicago, and the Super Photon Ring – 8 GeV (SPring-8) in the Hyogo Prefecture, Japan.} for nuclear resonant scattering capable of producing similar experiments, this is in theory a viable 4-point-array clock-concept, or at least a very precise tuner for nuclear clock transitions, suitable, performance-wise, for the requirements of a GW detector. In practice, though, the limitations in terms of repeatability of the observations ($\simeq0.1-1$ mHz rate), size of the synchrotron facilities ($10^2 - 10^3$ m) and the considerable costs associated to single-shot measurements may, at present time, place this technique beyond reach for adoption.\\
Lastly, a promising prospect comes from the development of a $^{45}$Sc nuclear clock, which is currently in progress \cite{45Sc}. With a transition energy of 12.4 keV and a moderately long lifetime of 0.47 s, $^{45}$Sc offers an ultra-narrow nuclear resonance that significantly surpasses current M{\"o}ssbauer resonances in terms of stability and precision thanks to a quality factor $\textrm{Q}=10^{19}$, six orders of magnitude larger than the 14.4 keV resonance of $^{57}$Fe. Its realization can be foreseen in the near to medium-term future once \emph{i)} further advancements in the resonant spectral flux of the modern narrow-band X-ray free-electron lasers (XFELs) will reach the required necessary flux to optimally excite the isomer; \emph{ii)} frequency combs capable of operating effectively at energy levels up to 12.4 keV are further developed\footnote{Current methods, like using XFEL oscillators \cite{xfelo} or intra-cavity high harmonic generation (HHG) \cite{hhg}, are still under development \cite{progX, progX2} and their full capabilities have not yet been realized. Note, however, that, thanks to the HHG process occurring in less than half of a laser oscillation period, HHG is behind the results in \cite{attosecond}, which observed, using XUV interferometers  and high-harmonic spectroscopy, delays in the attosecond timescale (200\,mrad phase difference) from the emission of H isotopes.}; \emph{iii)} full target optimization for maximum nuclear forward scattering (NFS) signal strength is achieved, minimizing the linewidth uncertainty of the actual resonance.\\

\subsubsection{Polarization and photoionization clocks}\label{sec_photo}
Motivated by research on rather disparate applications including, among \mbox{others}, coherent quantum control of molecular dynamics on femtosecond (1 fs = $10^{-15}$~s) timescales \cite{molecular}, electron dynamics control \cite{electrondynamics, electrondynamics2}, or interference of ultra-short free-electron wave packets \cite{packets}, the production of pairs of ultra-short laser pulses with controllable temporal delay has been the subject of active science since the 1990s. Interpulse delays were initially enacted in Michelson or Mach-Zehnder interferometer systems by means of mechanical adjustments of the path lengths. Subsequently, aiming to increase precision and stability in pulse-to-pulse delay, an all-optical technique based on spectral pulse shaping was developed and its results presented in 2011: the authors in Ref.~\cite{zepto_pulse} combined a fs polarization pulse shaper in conjunction with a polarizer and two linear spectral phase masks to study the interference signal resulting from two temporally delayed pulses. The design is based on the proposal contained in Ref.~\cite{compact}, which consisted in a compact ($130\times200$ mm$^2$) pulse shaper, and it mimics the behavior of a common path interferometer without requiring moving parts. The separation between the two pulses in each pair is controllable in the time domain with $2\sigma$-precision down to 280 zs and standard deviation of 11 zs. More in detail, the spectral phase of two  orthogonally polarized electric field components of an incident Ti:sapphire laser is simultaneously and independently modulated, in a spectral transmission window centered around $\lambda=790$~nm: the input pulse is split --~applying appropriate linear spectral phase functions~-- into two identical replicas with crossed linear polarizations, delayed in the time domain. The polarizer then projects the two polarization directions onto a common plane. The duration of each pulse is less than $12$ fs, and interpulse delays are varied between few to several fs.
While the requirement for a system producing pairs of photons delayed in the time domain with sufficiently controllable precision meets the requirement for our detector, the pulse-to-pulse delay time falls short for useful GW detection purposes. The possibility to navigate this limitation by slowing the group velocity of the EM signal to obtain appropriate time gaps is explored in \textsection~\ref{subsec_slow}.\\
A modified version of this experiment from 2019 is described in \cite{zepto680}: employing higher-order harmonic generation (HHG), two converging replicas of a driving fs pulse result in an attosecond pulse train, producing an interference pattern in the extreme UV. The design effectively creates a common path interferometer and the separation between the two laser foci is controlled with a spatial light modulator. With pulses $\approx$ 30 fs long centered around $\lambda=785$ nm, the ability to control the relative optical phase between the two driving laser pulses allows a fine-tuning of the delay between the two pulses with resolution of 12.8 attoseconds and precision as low as 680 zs at the 13th harmonic (lower and higher harmonics yielding worse precision). This experimental setup can fit a compact (few cm in size) volume to operate. With the previous experiment, it shares a fully electro-optical design, with no need for mechanical elements to operate; this entails rather restricted delay ranges available, which will be again tentatively addressed in \textsection~\ref{subsec_slow}. Using mechanical elements, while lessening stability and precision, would on the other hand allow longer delays, more appropriate for GW observations.\\
Another experiment performing ultra-precise timing came in 2020, when the authors in \cite{zepto_photo} performed photoionization of H$_2$ molecules, addressing it in the time domain: since the birth of a photoelectron wave from a molecular orbital does not occur simultaneously across the molecule, a birth time delay --~manifesting as a phase shift of the electron wave in the far field~-- was observed using electron interferometry, and measured up to the expected value of 247 zs for the average bond length of H$_2$. This result, though, is not conclusive, since a recent similar experiment conducted to accurately determine photoionization delays, showed unexpectedly larger times, reaching up to 700 attoseconds \cite{atto_photo}.\\
Photoelectron spectroscopy is also at the core of the RABBITT approach \cite{astrain, rabbitt2}, short for `reconstruction of attosecond beating by interference of two-photon transitions': initially intended for ultrashort laser pulses characterization, the experiment in Ref.~\cite{atto_beat} focuses on two-photon transition interference patterns, offering the capability to measure the time delay between the ionization of different electrons in helium, neon and argon sample atoms. Attosecond precision is achieved by reconstructing phase shifts between different harmonic orders. The dynamics for the elements beyond helium, the authors argue, are characterized by multiple residual ion states, making the corresponding theoretical predictions in their paper subject to debate.\\
Similarly, a method dating back to 2005 to achieve attosecond-scale temporal resolution under the name of FROG-CRAB, for `frequency-resolved optical gating for complete reconstruction of attosecond bursts', converts attosecond pulses into electron wave packets, and the resulting associated streaking spectrograms is then analyzed \cite{frog}. Ionization of atoms is conducted by an attosecond XUV pulse combined with a synchronized IR laser field. The temporal profile of the XUV pulse is reconstructed from the resulting energy shifts in the photoelectrons across a range of delay times and applying the FROG algorithm, achieving a precision of $\approx 70$ attoseconds. \\
Finally, under the same class of photoionization clocks, a \lq Zeptosecond Angular Streak Camera\rq~was presented in 2022 \cite{zepto_camera}: the time delay in photoionization as well as the duration of a pump pulse are measured by employing an XUV pump pulse and a polarized IR streaking pulse. The time of ionization is mapped to the angle of the emitted electron through the spin or the effective angular streaking caused by the IR pulse; the momentum distribution of the photoelectrons is then analyzed, determining a time delay. A precision in the zs is achieved thanks to high information redundancy in the full vectorial momentum distribution of the emitted photoelectrons. While not directly acting as a conventional clock, this system can complement a pulse generator, tuning the interval between coupled pulses.\\
As previously suggested, many of the presented techniques would greatly benefit, for our purposes, from the possibility of a lossless reduction in the propagation velocity of the EM signals they produce, which will be the subject of next subsection.

\subsection{Slow light}\label{subsec_slow}
Individually controlling the group velocity of pairs of closely gapped photons whose separation in the time domain is known down to the precision needed for GW detection, would --~in absence of information loss~-- allow the tuning of the interval between them to exactly match the GW frequency range of interest, as detailed in \textsection~\ref{sec_frequency}. Moreover, by diversifying the intra-signal interval between distinct pairs of photons, different GW frequency bands automatically become accessible within the same detector.\\
Leveraging the concept of electromagnetically induced transparency (EIT)\footnote{EIT happens when a narrow window of transparency within an otherwise absorptive medium arises from the destructive interference between two laser beams \cite{eit_first, eit}.}, the group velocity of light was recently slowed down by a factor exceeding $\approx10^4$ \cite{transparency}. In particular, the coupling of two collective resonances -- the in-plane electric dipole and the out-of-plane quadrupole surface lattice resonances -- within an all-dielectric metasurface made up of periodic silicon nm-thick nanodisks, produced collective EIT-like bands with Q factors $\simeq 10^3$ under normal incidence. While the behavior of the resonance resembles EIT, the transparency in this specific experiment is in fact induced by Fano interference\footnote{The interference between a continuum and a quasi-bound state, resulting in an asymmetric spectral line shape.} between the two resonances. The authors in \cite{transparency} obtained a group delay, i.e.~the ratio between transmission phase shift and angular frequency, \mbox{$\tau_g > 4$ ps} and group index $n_g$, i.e.~the ratio between speed of light in vacuum and light group velocity $v_g$ in the medium, as high as 12540 under normal incidence. Under oblique incidence, the dependence of Q factor, $\tau_g$ and $n_g$ on the incidence angle $\theta$ shows inverse quadratic behavior as in $(Q, n_g, \tau_g) \propto 1/\sin^2\theta$. Diverging Q factors and group delays for $\theta < 0.5^{\circ}$ appear to convey ultra-strong slow light effect. For instance, at $\theta=0.1^{\circ}$ the authors report reaching a group delay $\tau_g=55.5$~ps, and a group index $n_g>1.6\times10^5$. While the absolute delay reached with this experiment (note again, using nm-thick disks) is far from a range useful for GW detectability, the relevance of this technology is reasonably conceivable when envisioning its potential if properly scalable, or adaptable; ultra-cold atoms and low decoherence materials become necessary tools to counteract both the transparency window and coherence time quickly degrading under ordinary operating conditions. In general, minimizing distortion and pulse broadening commands solid control over absorption and decoherence losses: phase matching (PM) strategies can help maintain temporal coherence in nonlinear optical processes, ensuring efficient interaction between photons without disrupting their relative separation; by engineering refractive index profiles or angular alignment, to prevent temporal walk-off, spectral broadening, and coherence degradation, phase matching techniques (whether implemented via birefringent PM, nonlinear waveguide PM, chirped or ordinary quasi- PM, group velocity PM, etc\dots) can directly support the fidelity of inter-photon separation.\\
Whether the slow light effect would persist sufficiently homogeneous over successive iterations will arguably be the subject of forthcoming experimental testing.  Suppression of absorption and losses within the medium depends on the robustness against the emergence of imaginary components inside the temporal dispersion relation $\omega^2 \propto k$
for an optical field with angular frequency \(\omega\) and wave number \(k\); the same considerations apply for the spatial relation $k(\omega)$. With the group velocity expressed as $v_g = \frac{\partial \omega}{\partial k}$,
the ability to control and maintain a steep slope will ensure the group index remaining sufficiently uniform across the relevant spectral window over successive iterations.\\
Unsurprisingly, the interest for achieving strong slowing of light is at least a few decades old. In 1999, proper EIT in gaseous Na at 50 nK temperature (i.e.~below Bose–Einstein condensation) was used to slow $v_g$ down to 17 m\,s$^{-1}$ \cite{gas17}. The light speed in the medium is inversely proportional to the atom density $N$ as in:
\begin{equation}
    v_g=\frac{c}{n(\omega_p)+\omega_p\frac{\mathrm{d}n}{\mathrm{d}\omega_p}}\sim\propto\frac{1}{\omega_p}\frac{\lvert\Omega_c
    \rvert^2}{N}
\end{equation}
where $n(\omega_p)$ is the refractive index at probe frequency $\omega_p$ (rad\,s$^{-1}$) and $\Omega_c$ is the Rabi frequency\footnote{Rabi frequency defines the oscillation rate of atomic energy levels under oscillating EM fields. It relates directly with the field's amplitude and the transition dipole moment of the two affected atomic levels.} for the coupling laser. The resulting pulse delays for atom clouds a few $10^2\,\mu$m in length and total number of atoms $\simeq10^6$ are greater than $7\,\mu$s.\\
In the same year, resonant light propagation through a vapor of\,\,\,$^{85}$Rb contained in a cell with antirelaxation wall coating was investigated \cite{vapor}, obtaining light pulse delays as large as 13 ms, corresponding to a remarkably low $v_g\approx 8$~m\,s$^{-1}$. This result is conveyed by the ultra-narrow resonance obtained in an atomic vapor with slow ground state alignment relaxation, where the relaxation rate ($\gamma_{\textrm{rel}}\simeq2\pi$ Hz) inversely relates to the pulse time delay.\\
A similar EIT-based experiment, performed in 2001 using solid crystals\footnote{Specifically, praseodymium-doped yttrium orthosilicates (Pr$^{3+}$:Y$_2$SiO$_5$).}, rather than gas vapor, obtained $v_g$ as low as 45 m s$^{-1}$ and a corresponding time delay of $\,\simeq 67\,\mu$s \cite{solid}. The inhomogeneous broadening of the ground state transition is generally the limiting factor in minimizing $v_g$.\\
A theoretical, slightly different concept to achieve ultra-slow or even stopped light is articulated in the form of `optomechanically induced transparency' (OMIT) \cite{opto}: by manipulating the dissipation rates of two coupled active-passive cavities, a steep dispersion can generate at the transparency window. Crucially, when the decay rates align, stopped light can ideally be achieved. This result might offer profound implications for a GW detector, by potentially allowing full control over the intra-signal separation, opening the window to multi-frequency detectability. A hybrid optomechanical system featuring a 1D Bose–Einstein condensate trapped in a shallow optical lattice \cite{mikaeili22} presents a method to transition the system from normal mode splitting to the OMIT regime by manipulating the $s$-wave scattering frequency of atomic collisions while maintaining a constant pumping rate for the cavity: by controlling the depth of the optical lattice, the strength of atom-atom interactions, and the total number of atoms involved, a record-low $v_g = 1$ mm s$^{-1}$ and $\tau_g\gt150$ ms were achieved. This ultra-slow light occurs in a frequency region where the reflection coefficient of the cavity reaches its maximum, enabling its detection in the output observable. The experimental realization hinges on the control of the $s$-wave scattering frequencies, which can be correlated to the transverse trapping frequency of the condensate; the ability to control both the $s$-wave scattering frequency and the pumping rate of the coupling laser determines the amplification of peak time delay, which promises to surpass the performance of EIT. 

\smallskip
\noindent For this approach to become suitable as a segment in a GW detector, homogeneity of the slowing of light over multiple successive iterations is essential, as well as the minimization \emph{and} homogeneity of information loss --~specifically, about the interval separation between EM pulse couples~-- both over single and successive iterations.  

\subsection{Information preservation}\label{subsec_infopres}
Once a periodic EM signal has been produced and controlled with sufficient precision down to the required time interval appropriate for the specific GW frequency of interest --~with the desirable support of slow light implementation~--  it is necessary to preserve the information encoded in the inter-signal separation, delaying its observation until the supposed GW perturbation has left the system entirely. Because, as explained in  detail in \textsection\,\ref{sec_frequency}, different classes of GW sources will merge, once they've entered the final inspiralling phase, after widely varying extents of time, the duration over which to losslessly preserve the information encoded in the separation between each couple of EM pulses will necessarily dictate the aimed-at frequency band and indirectly, the corresponding classes of GW-emitting sources. Aiming at lower GW frequencies will demand longer such durations, and vice versa.\\
The challenge to store and preserve EM radiation for an adequate length of time can be undertaken in a variety of different ways, but the requirement for compactness is only the first in a series of limiting factors. For this reason, km-scale fiber loops or large optical cavities are initially disfavored, pushing the design towards miniaturized, highly stable photonic architectures.\\
To begin with, storage and recall of optical pulses has been demonstrated with the use of EIT itself \cite{100mus, halted}: the main obstacles in resorting to EIT for extended storage of information are dephasing of the signal, thermal perturbations and general coherence loss; even residual interactions within the medium may impact negatively the precision with which the separation between couples of photons is preserved.
Notably, EIT can be combined with quantum memory techniques by converting the optical excitation into a spin coherence in an ultra-cold atomic ensemble, i.e.~storing the light instead of merely slowing it: this has allowed to enhance the performance offered by plain EIT, which has progressively allowed exceeding the 1 second mark \cite{1sec, 1sec_b, 2sec}, to then reach durations of several seconds \cite{1min} and eventually up to one minute \cite{1min_b}.\\
Employing compact optical fiber delay lines might offer some advantages, particularly in terms of minimizing propagation losses when utilizing hollow-core photonic crystal fibers \cite{hollow, hollow2} (at the expense of dispersion losses) or ultra-low-loss silica fibers \cite{silica, silica2}, by looping photons through a confined path. However, because every pass introduces slight losses, obtaining direct optical storage over several seconds appears to be impractical without amplification or quantum memory techniques.\\
Among other potential candidates, whispering gallery mode (WGM) resonators \cite{wmg, wmg2} offer minimal dispersion and losses: by confining light through continuous total internal reflection along a curved dielectric boundary (be it microspheres, microtoroids, or crystalline resonators), WGMs can achieve high-Q factors and compact confinement; on the downside, photon storage times can only reach ms-long durations \cite{wmg3}. This limitation can be in part addressed by cascading multiple rings into coupled ring resonators \cite{crr, crr2}, but extending into the multi-second range needed for useful GW detection still appears implausible, even for the higher frequency events, which require the lowest total such durations.\\
Yet another similar situation is encountered with high-finesse Fabry-Pérot cavities \cite{fabry1, fabry2, fabry3}: while these mirror-based resonant structures can be miniaturized down to chip size \cite{fabry_micro} to fit the compactness requirement, even the longest achieved decay times (besides being obtained with extended structures, which would violate the compactness requirement) cannot exceed the ms-range mark \cite{fabry_long}. Currently, limited research has been performed to extend photon storage times with the help of cryogenic, superconducting or cascading configurations, as well as emerging nonlinear effects in coupled cavities \cite{fabry_nl1, fabry_nl2} but even this technology can hardly be predicted to enhance storage times beyond the sub-second regime, at least in the near future.\\
Notably, outright trapping of single photons can be achieved in a coupled cavity-atom system through a two-photon excitation process to populate the embedded eigenstates \cite{ee, ee2, ee3}: the validity of this mechanism, including its reverse release process, was demonstrated in Ref.~\cite{infopres}, in which the authors claimed for an ideal system arbitrarily high efficiencies and infinite lifetime, regardless of the presence of leakage channels. In realistic scenarios the mechanism promises to store single photons for times much longer than what single-cavity or single-atom configurations can achieve.\\
Another among the most promising alternatives for long term storage of information involves employing spin-wave-based memories in atomic ensembles or rare-earth-doped crystals: in 2021, by using a zero-first-order-Zeeman magnetic field and dynamical decoupling to protect spin coherence, a coherent optical storage of light in an atomic frequency comb (AFC) memory was achieved and maintained for over one hour \cite{1hour}. This, to our knowledge, is currently the longest observed storage time of optical memories. Because the information encoded in the photons is stored in the collective atomic excitations, the process could potentially lead to a loss of temporal information due to the fixed nature of the comb's structure. For this reason the difference in arrival times likely requires for it to be finely tuned to align precisely with the comb's structure. On the other hand, different AFCs, if available, could offer a range of distinct periodic absorption spectra to potentially cover several GW frequencies at once. It's because of the hard limitation imposed by current information preservation capabilities, that a detector devoted to time-oriented GW observations with the design presented here can only push its scope up to the higher frequency tail in the LISA band, as discussed in \textsection\,\ref{sec_frequency}; without a working solution to perform observations \emph{during} the GW event, all the frequency bands extending beyond the mHz, e.g.~the $\mu$Ares \cite{muares} and the PTA band \cite{pta0,pta1,pta2,pta3}, are at this stage excluded.\smallskip\\
Lastly, regardless of the specific technique (or combination thereof) adopted to store and preserve our coupled photons, it would greatly benefit to exploit properties similar to those offered by Airy photons \cite{airy0, airy2} or Bessel beams \cite{bessel_1, bessel_2}: the quantum analogs of Airy beams\footnote{Airy beams are self-accelerating, non-diffracting, self-healing optical wave packets \cite{airy_beams}, which emerge in 1D as the only exactly shape-preserving solution of the potential-free Schrödinger equation \cite{airy_pre}; in 2D and in gaussian 3D approximation, the solutions are a two-dimensional Airy beam, and a parabolic accelerating trajectory \cite{parabolic}. Conversely, Bessel beams are inherently two-dimensional due to their radial symmetry. Airy behavior is not limited to EM radiation, but can also describe the propagation of particles \cite{airy_e, airy_n}.}, Airy photons are characterized by non-diffracting propagation through curved trajectories, preserving their spatial profile over relatively extended distances. Conversely, Bessel beams naturally maintain a straight-line propagation path. Arbitrary trajectories can be engineered for Airy beams by employing suitable periodic media \cite{airy_period1, airy_period2, airy_propag2} and the ongoing research on attenuation correction \cite{airy_comp1, airy_comp} offers the prospect of adjustable exponential intensity increase/decrease over finite distances. A true non-diffracting beam would require an infinite radius of the beam's amplitude field which would in turn demand infinite power. Real-world (approximated) Bessel beams are typically generated using axicons (conical lenses) or annular apertures, creating a beam with a central core and concentric rings; Airy beams are generally produced by applying a cubic phase modulation to a Gaussian beam, resulting in their characteristic parabolic trajectory. Complete spatiotemporal control of propagation-invariant (2+1)D Airy single-photon optical \emph{bullets} was recently demonstrated \cite{airy1}. The propagation distance generally is limited by a beam's finite energy, and the non-diffractiveness is enhanced by a low density, low temperature medium \cite{airy_propag}. While, under controlled environment conditions, the non-diffractive range of Bessel and Airy beams can only be observed to reach the few tens of meters limit offered by laboratory constraints \cite{airy_propag2, bessel_propag3}, the research on the atmospheric propagation of Bessel beams \cite{bessel_propag1, bessel_propag2} suggests that km-scale ranges are already available even under a perturbed environment. The prospect of in-vacuum or highly controlled environment realization of a Bessel/Airy beam --~or a tweaked adaptation thereof~-- offers the tempting prospect of extended lossless propagation of the EM signal inside a GW detector.\\

\subsection{Signal counting}\label{subsec_count}
The final step in carrying out a single and self-contained measurement in the detector consists in correctly observing and registering the time of arrival (ToA) of the clock-acting EM signal we have first produced, then slowed down, and subsequently preserved for a length of time coherent with the duration of a GW event at the frequency of interest. By registering the photons' ToA with comparable precision to that with which their separation was known at the moment of their emission, a deviation from the expected interval between them would flag for a possible GW event, retrodicting the induced perturbation in the flowing of time back when the photon pair itself was produced.\\
While sampling rates have their own relevance in granting an accurate waveform reconstruction, the focus in this final step of the system's architecture is almost exclusively directed towards the timing precision the counter can achieve, which is commonly referred to as the instrumental timing jitter.\\
The adopted single photon counting technique \cite{counting} will require achieving precision, i.e.~timing jitter, sufficient for GW detectability. The counting method adopted in \cite{zepto} (the experiment mentioned in \textsection~\ref{sec_nuclear} inferring a 1 zs precision in the tuning of the quantum phase of a collectively excited $^{57}$Fe nuclear state via transient magnons) is based on time-gated avalanche photodiodes (APDs) counting the delay between single photons \cite{diode}: by selectively activating a gate voltage pulse, this special kind of APD can enter Geiger-mode temporarily, ensuring that the APD is active only during specific time intervals, which reduces noise and improves SNRs. Still, while convenient in granting single photon operability, APDs --~even the high-performance germanium–silicon  ones \cite{germanium-diode}~-- can generally only reach $\sim$ a few ps in resolution \cite{spapd}. While still useful for claiming the inferred results in \cite{zepto}, this technique still falls short for the ultra-high precision needed  in a direct measurement.\\
Similarly, superconducting nanowire single-photon detectors can offer high detection efficiency, minimal dark count rate and dead time with observed time jitters as low as 3 ps \cite{spnano_rev0, spnano_rev1, spnano_rev2, spnano, spd}. 
The precision achievable with this technique is material- and temperature- dependent: theoretically, optimal conditions in terms of geometry, absorption efficiency, nanowire fabrication and composition, could push the time jitter performance to sub-ps levels. \\
A slightly better resolution is obtained applying a bandpass sampling method to time measurement electronics, which in 2023 allowed to reach a precision better than 0.4 ps \cite{passabanda}.\\
The use of advanced cross-correlation algorithms, digital signal processing  including filtering and phase measurement are in general a good way to enhance the resolution of time measurements. Advanced spectroscopy  techniques, such as the time-stretch dispersive Fourier transformation, can map the spectrum of a broadband optical pulse into a time-stretched waveform, offering the possibility to produce a real-time analysis of spectral dynamics up to attosecond precision \cite{fft, fft2}.\\
An all-fibre photonic generation of optical pulse trains, employing a fibre-delay line-based repetition-rate stabilization method, was realized in Ref.\,\cite{Jung2015}, achieving a 980 attosecond absolute rms timing jitter accumulated over 0.01 s.\\
The relative timing jitter between closely separated solitons, i.e.~self-localized wave packets exhibiting particle-like behavior in nonlinear media, was measured in Ref.~\cite{att_jitter} using a balanced optical cross-correlation (BOC) method with a temporal resolution of 5 zs/\raisebox{0.4ex}{$\surd${}}Hz, finding a rms upper estimate of 60 attoseconds for the intramolecular\footnote{The term `soliton molecules' is used to describe a bound state of optical solitons, arranged in stable configurations akin to those typical of conventional molecular structures.} timing jitter. In this experiment, a Ti:sapphire fs laser generates temporal soliton molecules; a Michelson interferometer (with arms' length 83 mm) splits and recombines the soliton molecules; then the BOC device is used to detect the cross-correlation signals between the two copies of the soliton molecules. Moreover, optical heterodyne methods based on comb-line interference offer even better performance over the BOC method, demonstrating a  $\simeq530$ ys$/\surd{}$Hz timing jitter-detection noise floor for a single mode-locked laser with an integrated timing jitter of 16.3 attoseconds \cite{zs_jitter}. 
A $\simeq160$ ys$/\surd{}$Hz timing jitter noise floor between Ti:sapphire mode-locked lasers was demonstrated in Ref.~\cite{ys_jitter}, to reach an integrated timing jitter resolution of one attosecond. For a review on timing jitter measurement techniques with resolution around the attosecond see also Ref.~\cite{Kim:16}.\\ 
Finally, a further technique compatible with the requirement to detect phase-coherent temporal deviations caused by gravitational perturbations, Hong-Ou-Mandel (HOM) interferometry measures how indistinguishable the arrival times of single photons are, to sub-fs precision: based on the HOM quantum effect, by which two indistinguishable photons incident simultaneously on a beam splitter interfere such that they exit together through the same output port, eliminating the probability of coincident detection at separate outputs, HOM interferometry can measure the change in relative arrival time between two photons with ultra-high accuracy. Unlike classical interferometry, which requires stabilization to isolate from phase differences, HOM interferometry can detect interference effects without needing to account for changes in the relative phase of the photons. When the source has a joint spectral amplitude with a \(\operatorname{sinc}\)-like structure (arising from phase-matching in spontaneous parametric down-conversion with a finite crystal length), its Fourier transform into the time domain produces a temporal Airy pattern, which represents the probability distribution of the photon-pair emission interval; the HOM dip then, reflects the destructive interference at zero delay within such distribution. A typical HOM interferometer is generally compact, with arms' length not exceeding 1–2 meters. Rather inconveniently, the HOM effect is realized only as long as the temporal separation between the two photons is within --~or comparable to~-- their coherence time, which for traveling photons can hardly exceed a few ms. Recently, the measurement of the change in relative arrival time between two photons reached a best precision and accuracy of 4.7 and 0.5 attoseconds, respectively \cite{hom}. While employing shorter down-conversion crystals and higher-efficiency photodetectors, the authors argue, would yield an approximate 50-fold improvement in precision, lowering the limit down to the range of 100 zs, coherence time of the photons would still be the limiting factor in adopting this technique --~since the two-photon interval needs to be coherent with a GW half-period~-- and the attosecond precision only applies locally around zero delay, within the coherence envelope. Regardless, this technique carries the potential to become relevant for accurate waveform reconstruction should it be adaptable into a cascading or recursive system, with multiple in-series instances of the HOM process. This is especially true for LVK events, for which the typical half-period around the peak sensitivity falls around the $\approx 10^{-2}-10^{-3}$ s mark which, at coherence times in the ms range, would require at most a 10-fold increase either in coherence times or in iterability of the process to obtain a train of signals employable for high frequency GWs observation; the highest frequency GW sources would be already available for detection within a single instance of HOM interferometry, especially during their final merger phase, where half periods can get shorter than 1 ms.\\


\section{Implementation and results extraction}\label{results}

The full form of a detector devoted to the observation of purely temporal GW-induced perturbations, i.e.~oscillations in $h_{00}$, can now be sketched by assembling together the most fitting techniques from each of the previous Sections, each fulfilling a specific task. 
\subsection{Detector pipeline}\label{subsec_pipeline}
\begin{figure}[t]
\centering
{\includegraphics[width=.8\columnwidth]{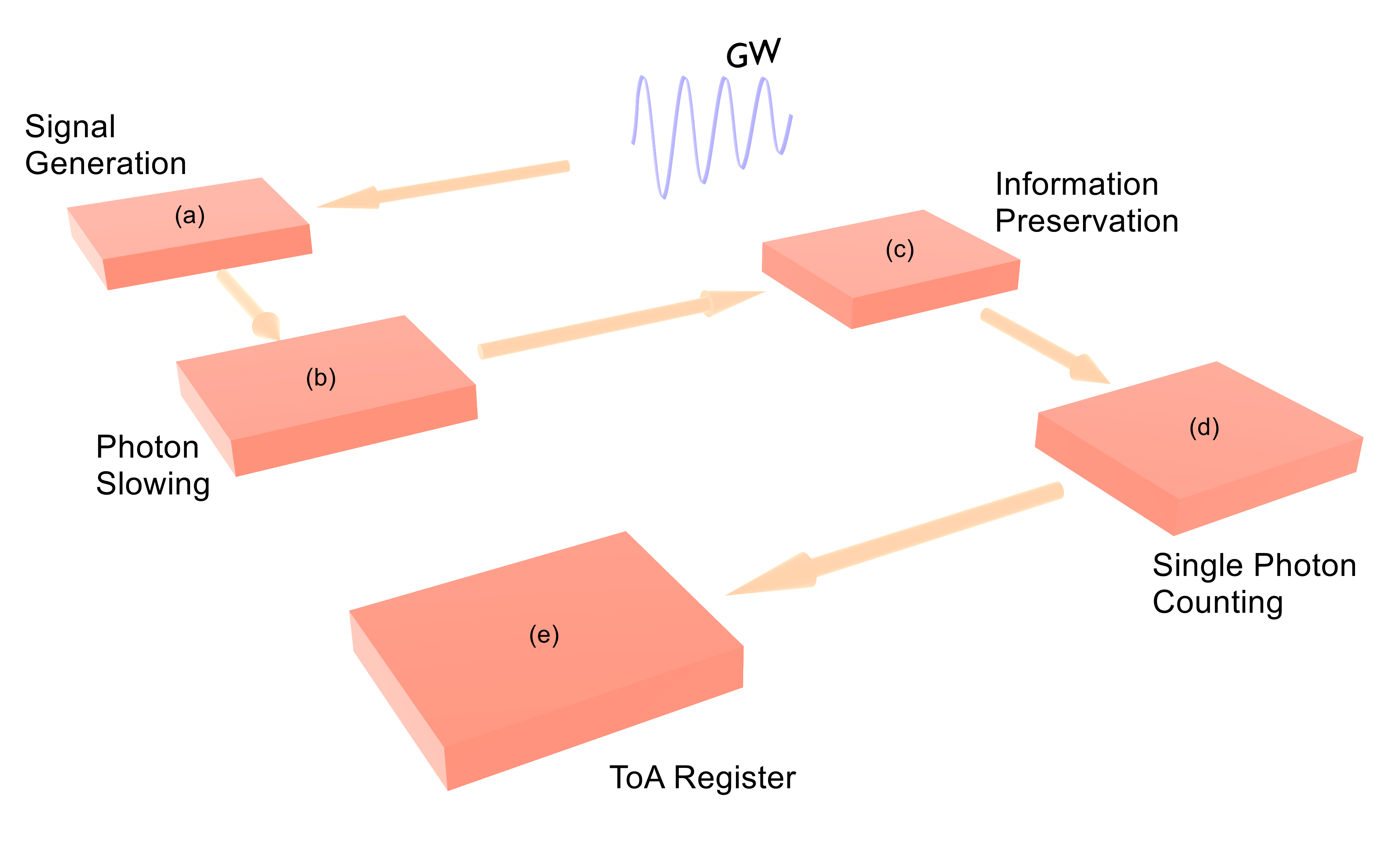}} 
\caption{Complete pipeline for a full standalone detector. (a) Signal generation: the first step acts as the ticking of a clock, adopting any suitable technique from \textsection\,\ref{subsec_timing}; (b) Photon slowing (\textsection\,\ref{subsec_slow}): this step is optional but convenient to improve the adaptability of the generated signal to fit adequate spacing for GW observation; (c) Information preservation (\textsection\,\ref{subsec_infopres}): this step ensures the ($\approx$ lossless) delay which is crucial to obtain geodesic non-locality between signal generation and observation, assuming the former happens under GW perturbation and the latter under flat Minkowski or unperturbed conditions; (d) Photon counting (\textsection\,\ref{subsec_count}): the final step measures the ToA of each photon, with the necessary precision; (e) ToA register: the interval between each couple of photons is stored in a database, and a deviation from the expected unperturbed value flags for a GW detection candidate.}
\label{fig:pipeline}
\end{figure}

In the most general description, let's recapitulate the requirement as it follows: a device acting as a clock producing a highly predictable periodic signal in the form of gapped photons, the internal gap being coherent with the GW frequency of interest and known down to precision compatible with GW observations (taking into consideration only temporal strains $h<10^{-18}$); a means of reducing the photons' group velocity, to match their separation with the GW frequency of interest, granting a high degree of spatial and temporal homogeneity, and minimal loss induction; a mechanism ensuring the actual measurement is sufficiently delayed, wrt to the moment of signal generation, by an amount of time coherent with the kind of GW event of interest (source-dependent and generally increasing with wavelength), counting on the assumption that some of the signals will catch the window in time in which the source --~a binary in most cases~-- will have meanwhile merged, offering the unperturbed conditions in which to operate the meaningful comparison measurements; a device capable of counting single photons and measuring the temporal interval between them with sufficient precision; finally, a data analysis framework to flag for detection candidates whenever there is a discrepancy between the measured and the predicted unperturbed interval. The scheme of such a system is sketched in Fig.\,\ref{fig:pipeline}. To isolate GW signals from spurious noise, as well as being able to reconstruct the sky localization of the source, a minimum of three non-coplanar nodes of the same detector, or eventually three altogether separate and independent detectors, are necessary, in a likewise arrangement to the early LIGO-Virgo network, or the LISA configuration. 

\noindent The four key stages of ultra-precise signal generation, slow-light-induced temporal tuning, lossless information preservation, and ultra-high-resolution signal measurement, require to employ, adapt or combine the most suitable techniques which we tentatively reviewed in the previous Section. The sequence consisting of a selection from each subgroup into a single instrument will indicate the GW frequency range most readily available for probing with already existing experimental facilities. The main driver exerting a strong constraint in determining the available frequency range is the combination of clock's precision and duration of the information preservation, as anticipated in \textsection\,\ref{sec_frequency}. The performance of slow light implementation constitutes a softer constraint, in that its limitations can be more easily circumvented by scaling, adapting or combining the existing techniques to increase the final separation inside each pair of photons.\\ 
Regarding ultra-precise timing, the system presented in Ref.~\cite{zepto} offers the most desirable performance for the required purpose in a detector, but measurement cycle rates and operational costs render it a non-viable choice at this time. While awaiting for nuclear clocks to reach a mature development, OLCs offer a proven and established technology, and are especially indicated to provide a reference calibrating clock; this could serve to tune the precision of a polarization pulse shaper, which best fits the detector's purpose of creating gapped couples of photons whose separation is known with sub-attosecond precision.\\
To precisely control the internal interval between them, EIT and especially OMIT can alter the the group velocity of light down to the mm s$^{-1}$ range, which can stretch an intial fs-scale inter-photon interval to more than $10^2$ ms, constraining the most immediately accessible GWs frequency betweeen the LVK band and the high frequency tail of the IMBHBs band.\\
EIT-based memories pushing storage times to more than 30 s constrain the available frequency below the few Hz band, comprising the entire LVK band and a good part of the IMBHBs band. AFC memories can potentially push this limit towards the high frequency tail of the LISA band, with the strong caveat regarding the inter-photon interval necessitating a perfect match with the comb internal structure. For these reasons, we deem safe to assume that the most appealing intersection between the capabilities of all the available techniques, falls inside the IMBHBs frequency band, around and below the 1 Hz band, towards the low frequency tail of the LVK band. This is enough support to claim that by repurposing and combining currently existing experimental techniques, a readily available GW detector could be sensible to GW signals from binaries of IMBHs, objects whose existence, it's worth noting, has until now eluded detection.\smallskip\\
\noindent Once a technique devoted to each segment is identified and placed in the detector workflow, a few more considerations are in order. The first has to do with the dynamics of the gravitational potential $U$ shifting the fractional frequency measured by a clock as in 
\begin{equation}
    \frac{\Delta\nu}{\nu_0}=\frac{\Delta U}{c^2}
\end{equation}

\noindent in which the potential is affected by all known sources of noise, be it tidal, seismic, or the orbital motion around a central mass either of Earth, should the detector be Earth-based, or of the spacecraft, should it be space-based, not differently from what happens with interferometers. In both, it becomes crucial to properly address the problem of comparing the reading of clocks with different worldlines. An accurate treatment of the problem for static and stationary cases is found in \cite{GR-chronometry}, together with a discussion about emitter-observer maps for distant communicating clocks.\\
Moreover, the gravitational potential on Earth is subject to an altitude-based gradient which becomes relevant even at small vertical displacements for clocks with the inaccuracies needed for GW science. In particular, the fractional frequency difference between two clocks on Earth’s surface, where the mean gravitational acceleration is $g$, vertically separated by a distance $\Delta r$ is \cite{nuclear8}:
\begin{equation}
\frac{\Delta\nu_{\scalebox{.55}{G} }}{ \nu_{\textrm{clock}}}= \frac{g\Delta r}{c^2}
\end{equation}

\noindent which results in a $1 \times 10^{-19}$ fractional frequency shift for a 1 mm discrepancy in clock height. Similar considerations can be found in Ref.~\cite{Takamoto2020} and as part of Ref.~\cite{redshift-mm}, whose result of a fractional frequency uncertainty of $7.6\times10^{-21}$ was mentioned in \textsection\,\ref{subsec_timing}. For a more thorough treatment of gravitational frequency shifts in clocks, see also Ref.~\cite{Schiller2008}. Finally, an accurate treatment of clocks dynamics in the presence of gravity, as well as the gravitational effects in atomic clocks, which again, go beyond the scope of this paper, can be found in Ref.~\cite{deloc}.


\subsection{Waveform reconstruction}\label{subsec_waveform}
This kind of experimental structure will offer an increasingly higher degree of refinement in the determination of the reconstructed waveform, the more points are available for interpolation. While, in fact, an interferometer has the advantage of outputting somewhat continuous measurements, which fairly smoothly identify a waveform candidate via matched filtering, a GW detector of the kind presented here will, for a start, offer only an individual data point, expressing a single-value amplitude in temporal strain. This, alone, can only give information on the presence --~or lack thereof~-- of gravitational radiation, or more precisely, of a perturbation in the flowing of time at a specific moment in the past. To observe the maximum allowed temporal strain amplitude, the two photons must be ideally coincident with consecutive zeros in the waveform, thus avoiding any self-canceling sectors of opposite sign; the resulting point is obtained as the integral under the curve between the two. Increasing the density of data points, i.e.~producing more photon couples per GW period\footnote{Equal spacing among different couples is not strictly necessary, although offering a simpler workflow to operate.}, will gradually populate enough points between the maxima (and minima), so that a fitting algorithm can more or less confidently reconstruct the most likely waveform. Conversely, points of zero amplitude will result from those couples who happen to fall exactly across the zero with equal distance on each side (assuming a regular sinusoidal waveform here), implying a zero-sum integral in time flow deviation between them. If the spacing is correctly set half a period long for the GW being observed, this automatically means the two photons sit over the adjacent maximum and minimum in the waveform, either side of the zero. Predictably, all the points in the middle will result from the residual perturbation surviving over the self-canceling portions of opposite sign.
\begin{figure}[t]
\centering
{\includegraphics[width=.8\columnwidth]{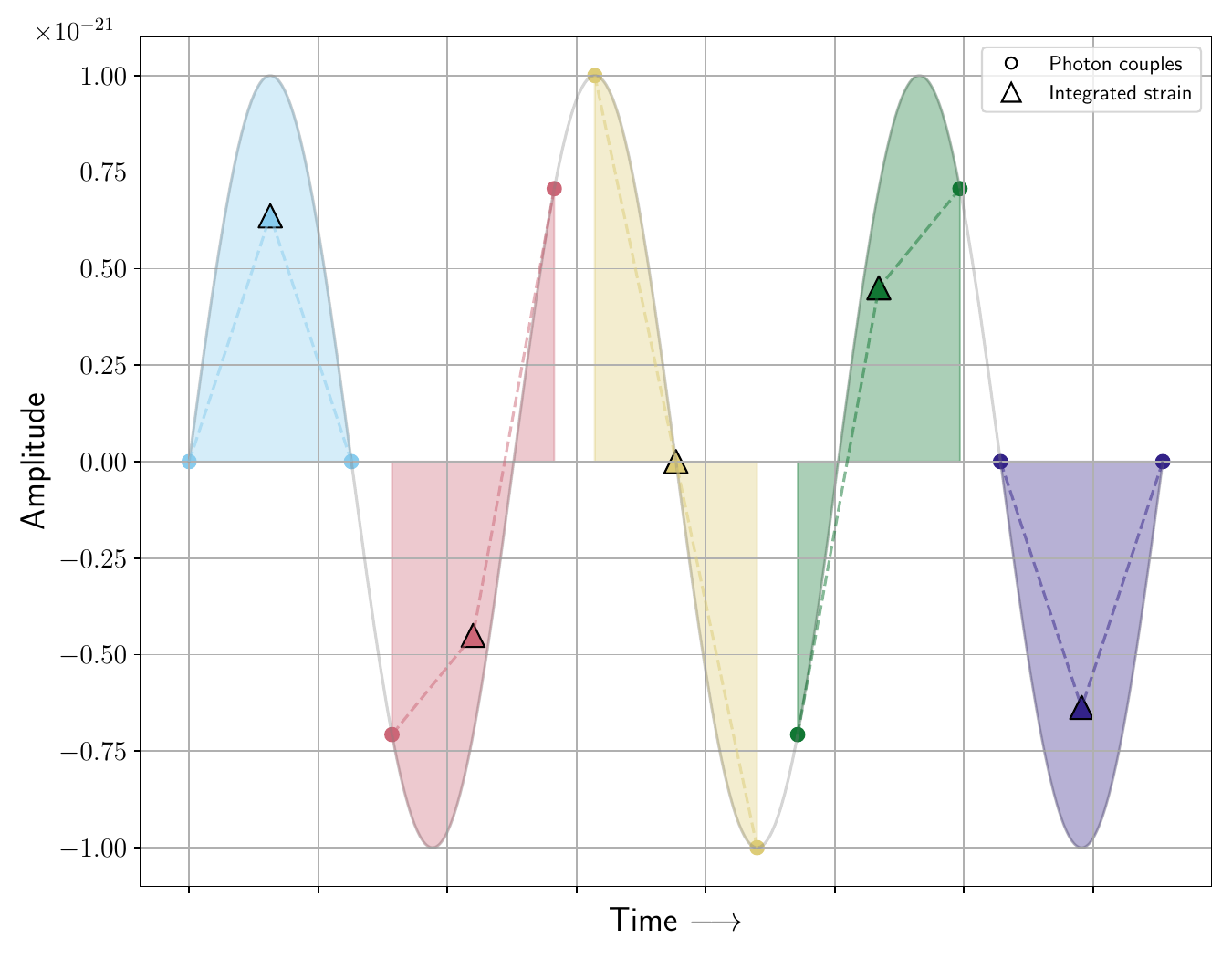}} 
\caption{Example of five different points (triangles) obtained from an equal number of photon couples (circles) and normalized to $\pi$. Each couple is equally spaced to match a half period for this specific waveform. The more time spent under same sign amplitude gives rise to larger values of the temporal strain. Notice that at the maxima and minima, the discrepancy --~by underestimation in this case~-- with the underlying waveform is maximized (it gets overestimated when not normalizing to $\pi$), and needs to be corrected. Conversely, towards the zeros (yellow triangle) the reconstruction gets increasingly more accurate by default.}
\label{fig:punti}
\end{figure}

\noindent An example of five different reconstructed points is shown in Fig.\,\ref{fig:punti}: each couple of photons is equally spaced, internally, to coincide with one half-period of the underlying wave. The integral between them (i.e.~the shaded areas) determine the total time deviation occurred in-between. The resulting values (the triangles in the plot) are normalized to $\pi$ and horizontally shifted halfway between their parent couple for image clarity. Notice how the longer time spent under same sign amplitude, implies larger absolute values in strain. More importantly, notice also how without a correction function, the reconstruction is decreasingly faithful, the larger the absolute value of the amplitude is. Therefore, before implementing a properly manufactured fitting function to interpolate between the reconstructed points to obtain the best fit for the GW waveform, a correction function, vertically shifting the resulting points to match the correct amplitude in the waveform, is advisable. For instance, a simple correction function of the form:
\begin{equation}\label{eq:corr}
y_i' = y_i (1 + \alpha)
\end{equation}
\noindent vertically shifts each point proportionally to its initial value $y_i$ by a single parameter ${0<}\,{\lvert\alpha\lvert}\,{<1}$ which can be determined numerically. Without the normalization to $\pi$ adopted to plot the points in Fig.\,\ref{fig:punti}, the points will generally overestimate the underlying amplitude and will dictate the condition on $\alpha<0$. Once a value for $\alpha$ has been determined, a fitting process can interpolate the available points based on previous assumptions or knowledge about the waveform. For a series of data points in which \( y_{\max} \) and \( y_{\min} \) are the maximum and minimum values of the amplitude in the series, \( P \) is the assumed or known period, a basic fitting function reads:
\begin{equation}\label{eq:fit}
    \hat{y}(x) = A  \sin\left( \frac{2\pi x}{P} + \phi \right) + C
\end{equation}
where the amplitude and the vertical offset \[A = \frac{1}{2}\bigl(y_{\max} - y_{\min}\bigr)\text{,}\quad C = \frac{1}{2}\bigl(y_{\max} + y_{\min}\bigr) \] 
are fixed by the two extrema, while the phase \[ \phi = \frac{\pi}{2} - \frac{2\pi}{P} x_{\max} \] is set so the maximum in the fit occurs at \( x_{\max} \), i.e.~the \( x \)-coordinate of \( y_{\max} \). The $\pi/2$ term is canceled if Eq.\,(\ref{eq:fit}) is redefined around the $\cos$ instead of the $\sin$ of the argument.
\begin{figure}[t]
\centering
{\includegraphics[width=.8\columnwidth]{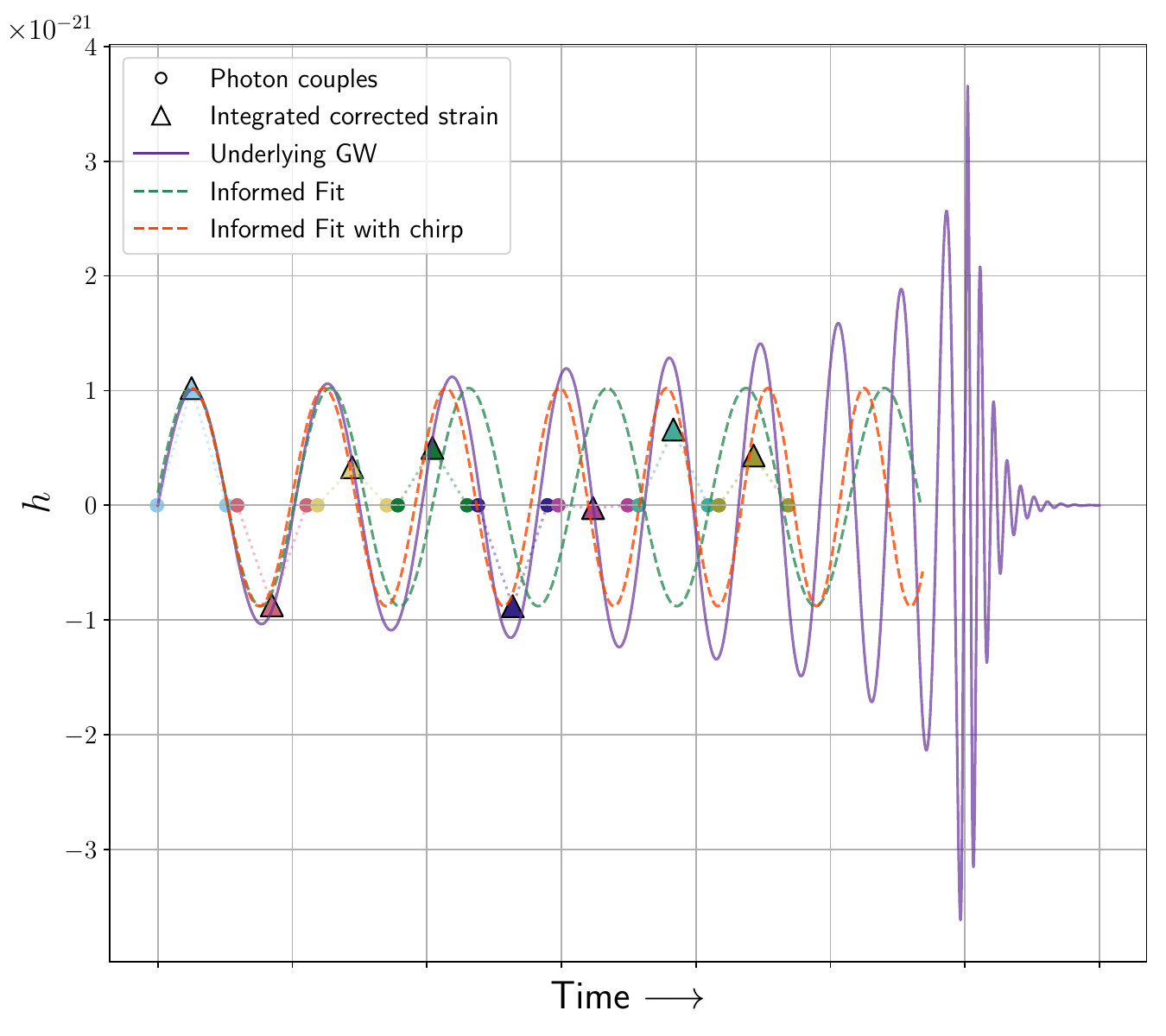}} 
\caption{Example of two different fitting functions for the inspiralling phase of the same underlying GW (purple line), based on eight reconstructed points (triangles), obtained from likewise number of photon couples (circles, shared colors). The green dashed line shows the basic period-informed fit of Eq.\,(\ref{eq:fit}). The orange dashed line shows a correction thereof, which aims to reproduce the onset of chirping, as per Eq.\,(\ref{eq:fit_chirp}). While this ensures a few more periods are confidently reconstructed, compared with the basic fit, both lose validity as late as a couple of periods before merger, where the increase in amplitude is also not addressed.}
\label{fig:fit}
\end{figure}
\noindent Clearly, in a real world scenario, several periods and/or a thicker point series are needed to properly assess the correct values of \( y_{\max} \) and \( y_{\min} \), which makes this approach slightly weaker for long wavelength GWs. In addition, the more features are present in the waveform (eccentric orbits, three body interactions, etc\dots) the denser the collection of reconstructed points will need to be. Moreover, in proximity of and during a merger, a chirping correction to any fitting function becomes a necessity. A basic correction to account for the onset of chirping is obtained with:
\begin{equation}\label{eq:fit_chirp}
\hat{y} = A  \sin\left(2\pi \left[f_0 (x - x_{\min}) + \frac{1}{2} \beta (x - x_{\min})^2\right] + \phi \right) + C
\end{equation}
where $f_0$ is the initial GW frequency and $x_{\min}$ the moment in time the first photon taken into account is emitted. The chirp rate \( \beta \) is found numerically to minimize the rms fitting error. This defines a phase-locked, linearly chirped sine wave whose instantaneous frequency increases linearly as:
\begin{equation}
f(x) = f_0 + \beta (x - x_{\min}).
\end{equation} 
The resulting two fits to a set of eight reconstructed points over 11 half-periods, corrected with a scaling parameter $\alpha=-0.56$ in Eq.\,(\ref{eq:corr}), are plotted in Fig.\,\ref{fig:fit}. Both are informed fits, i.e.~assuming to know the period of the underlying GW. This is of course intended behavior, since the GW frequency in our detector is \textit{aimed-at}, meaning that the detector operates by searching for a specific frequency and, as such, determines the internal gap inside each photon couple. The fitting function from Eq.\,(\ref{eq:fit}) is acceptable as long as we are sufficiently far from the final merger phase. The corrected fit of Eq.\,(\ref{eq:fit_chirp}) does a better job at reconstructing the inception of chirping, but fails a couple of periods before merger. Finding a more refined, possibly nonlinear, fitting function, covering the entire chirp (including the dramatic increase in amplitude in the very final periods), merger and ringdown, while working with a handful of reconstructed points, goes well beyond the scope of this paper. 


\section{Conclusions}\label{sec_concl}
In this work we carried out a first assessment towards the feasibility of experimentally detecting the time oscillations induced by GWs on clocks. The theoretical support offered by the asynchronous traceless gauge grants a nonzero, oscillating, $h_{00}$ term which becomes detectable once geodesic non-locality is achieved, provided clock precision exceeds a certain minimum threshold. The temporal strain $\Delta t/t$ is then observed when a duration of time $t$ is measured with precision better than $\Delta t$. Because the effect of each equal-amplitude half period of a GW cancels the previous half period's action on a clock's reading, the oscillations in the flowing of time must be written on an observable, to be interrogated for deviations wrt unperturbed conditions. To realize this, a delay is introduced between signal generation and observation, during which the GW event supposedly ends in a merger, leaving the system in unperturbed conditions, which was labeled as flat Minkowski in the text as a placeholder for any subsequent conditions subtracted of the perturbation from that specific GW event. This program inherently places most of the interest around the final phase of the inspiral of compact binaries, right before the final merger.\smallskip\\
All in all, this is a tentative approach to the challenge of conceiving the design of a full-fledged, off-the-shelf, GW detector, resting entirely on existing experimental techniques. This choice aims primarily at cost reduction, whereas developing and engineering any portion of the detector for which no suitable technology is available or adaptable, remains an open eventuality.\\
A representative, while non-comprehensive, excursus of some of the best techniques fit for purpose in each segment of a GW detector of this kind has been presented. In terms of performance, high frequency GW events put more demands on a clock's precision, while low frequency ones shift the challenge towards achieving prolonged lossless preservation of information. Among the several possible solutions to achieve both goals, the most immediately fitting design is provided by pulse shapers producing photon pairs whose separation is controllable with precision better than 1 attosecond. Optomechanically induced transparency can help reducing the velocity group of light, until the interval inside each couple matches the half-period of the GW frequency the detector is aiming at. Airy photons, if employable, can ensure non-diffractiveness. A combination of electromagnetically induced transparency with quantum memory techniques, single atom cavity systems or atomic frequency comb memories, can all separately perform preservation of information for durations exceeding several tens of seconds. The possibility to close the cycle of a single measure rests on the ability to directly read this information into the ToA of photons with the necessary precision: the timing jitter noise floor of tightly synchronized, mode-locked lasers can reach $\simeq10^2$ ys $/\surd{}$Hz, and integrated timing jitters of attosecond precision. Hong-Ou-Mandel is another possible pathway to address this problem. An indirect reading, i.e.~inferring an higher precision from a less precise one, has not been investigated but remains an open and concrete prospect. Considering and combining the limitations offered by each specific technique, the most accessible frequency band today, i.e.~employing already existing technologies only, extends around 1 Hz frequency, enclosing the low frequency tail of the LVK band, and the high frequency tail of the LISA band. This constitutes the exciting possibility of having a ready-to-use detector able to probe the existence of IMBHs by receiving the GW signal from binaries of such elusive objects.\smallskip\\
Furthermore, the attitude towards minimizing any spatial propagation of signals, even over distances well below the km scale, while admittedly over-constraining, rests on the consideration that treating and controlling temporal non-locality is better achieved in total absence of spurious elements coming from mixed perturbation terms in the metric.\\
Nevertheless, we don't rule out a future detour from this standpoint, especially when addressing the possibility of detecting long wavelength GWs, which necessarily will demand producing measurements \textit{during} the event itself, rather than after it. This will most likely require a mixing of temporal and spatial non-locality, to perform relative, rather than absolute, measurements of time, under different phases of the same GW.\\
With minimal tweaking, i.e.~producing photon couples with variable internal separation, our system presents the standout benefit of allowing multi-frequency GW observations, which is inherently precluded to interferometers. Moreover, the prospect of low frequency GWs to be observed from Earth becomes accessible, which again is not a possibility accessible to interferometers because of size and noise constraints.

\bmhead{Acknowledgments}
The authors thank Lars Bocklage for sharing his knowledge regarding his experiment. 
S.B.~would like to dedicate this work to the memory of the late Rai Weiss, who kindly but firmly endorsed the early ideas leading to this paper.


\bibliography{gwbiblio}


\begin{thebibliography}{167}
\ifx \bisbn   \undefined \def \bisbn  #1{ISBN #1}\fi
\ifx \binits  \undefined \def \binits#1{#1}\fi
\ifx \bauthor  \undefined \def \bauthor#1{#1}\fi
\ifx \batitle  \undefined \def \batitle#1{#1}\fi
\ifx \bjtitle  \undefined \def \bjtitle#1{#1}\fi
\ifx \bvolume  \undefined \def \bvolume#1{\textbf{#1}}\fi
\ifx \byear  \undefined \def \byear#1{#1}\fi
\ifx \bissue  \undefined \def \bissue#1{#1}\fi
\ifx \bfpage  \undefined \def \bfpage#1{#1}\fi
\ifx \blpage  \undefined \def \blpage #1{#1}\fi
\ifx \burl  \undefined \def \burl#1{\textsf{#1}}\fi
\ifx \doiurl  \undefined \def \doiurl#1{\url{https://doi.org/#1}}\fi
\ifx \betal  \undefined \def \betal{\textit{et al.}}\fi
\ifx \binstitute  \undefined \def \binstitute#1{#1}\fi
\ifx \binstitutionaled  \undefined \def \binstitutionaled#1{#1}\fi
\ifx \bctitle  \undefined \def \bctitle#1{#1}\fi
\ifx \beditor  \undefined \def \beditor#1{#1}\fi
\ifx \bpublisher  \undefined \def \bpublisher#1{#1}\fi
\ifx \bbtitle  \undefined \def \bbtitle#1{#1}\fi
\ifx \bedition  \undefined \def \bedition#1{#1}\fi
\ifx \bseriesno  \undefined \def \bseriesno#1{#1}\fi
\ifx \blocation  \undefined \def \blocation#1{#1}\fi
\ifx \bsertitle  \undefined \def \bsertitle#1{#1}\fi
\ifx \bsnm \undefined \def \bsnm#1{#1}\fi
\ifx \bsuffix \undefined \def \bsuffix#1{#1}\fi
\ifx \bparticle \undefined \def \bparticle#1{#1}\fi
\ifx \barticle \undefined \def \barticle#1{#1}\fi
\bibcommenthead
\ifx \bconfdate \undefined \def \bconfdate #1{#1}\fi
\ifx \botherref \undefined \def \botherref #1{#1}\fi
\ifx \url \undefined \def \url#1{\textsf{#1}}\fi
\ifx \bchapter \undefined \def \bchapter#1{#1}\fi
\ifx \bbook \undefined \def \bbook#1{#1}\fi
\ifx \bcomment \undefined \def \bcomment#1{#1}\fi
\ifx \oauthor \undefined \def \oauthor#1{#1}\fi
\ifx \citeauthoryear \undefined \def \citeauthoryear#1{#1}\fi
\ifx \endbibitem  \undefined \def \endbibitem {}\fi
\ifx \bconflocation  \undefined \def \bconflocation#1{#1}\fi
\ifx \arxivurl  \undefined \def \arxivurl#1{\textsf{#1}}\fi
\csname PreBibitemsHook\endcsname

\bibitem{LIGO}
\begin{barticle}
\bauthor{\bsnm{{Abbott et al.}}, \binits{B.P.}}:
\batitle{{Observation of gravitational waves from a binary black hole merger}}.
\bjtitle{Phys. Rev. Lett.}
\bvolume{116},
\bfpage{061102}
(\byear{2016}).
\doiurl{10.1103/PhysRevLett.116.061102}
\end{barticle}
\endbibitem

\bibitem{ligovirgo_O3}
\begin{barticle}
\bauthor{\bsnm{{Abbott et al.}}, \binits{R.}}:
\batitle{{GWTC}-2: Compact binary coalescences observed by {LIGO} and {V}irgo during the first half of the third observing run}.
\bjtitle{Phys. Rev. X}
\bvolume{11},
\bfpage{021053}
(\byear{2021}).
\doiurl{10.1103/PhysRevX.11.021053}
\end{barticle}
\endbibitem

\bibitem{ligovirgo_O12}
\begin{barticle}
\bauthor{\bsnm{{Abbott et al.}}, \binits{B.P.}}:
\batitle{{GWTC}-1: A gravitational-wave transient catalog of compact binary mergers observed by {LIGO} and {V}irgo during the first and second observing runs}.
\bjtitle{Phys. Rev. X}
\bvolume{9},
\bfpage{031040}
(\byear{2019}).
\doiurl{10.1103/PhysRevX.9.031040}
\end{barticle}
\endbibitem

\bibitem{ligoNS}
\begin{botherref}
\oauthor{\bsnm{{Abbott et al.}}, \binits{R.}}:
Observation of gravitational waves from two neutron star-black hole coalescences.
Astrophysical Journal Letters
\textbf{915}(1)
(2021).
\doiurl{10.3847/2041-8213/ac082e}
\end{botherref}
\endbibitem

\bibitem{ligo_4}
\begin{barticle}
\bauthor{\bsnm{{Cahillane}}, \binits{C.}},
\bauthor{\bsnm{{Mansell}}, \binits{G.}}:
\batitle{{Review of the Advanced LIGO Gravitational Wave Observatories Leading to Observing Run Four}}.
\bjtitle{Galaxies}
\bvolume{10}(\bissue{1}),
\bfpage{36}
(\byear{2022})
{\href{https://arxiv.org/abs/2202.00847}{{arXiv:2202.00847}}}
{[gr-qc]}.
\doiurl{10.3390/galaxies10010036}
\end{barticle}
\endbibitem

\bibitem{Pirani57}
\begin{barticle}
\bauthor{\bsnm{Pirani}, \binits{F.A.E.}}:
\batitle{Invariant formulation of gravitational radiation theory}.
\bjtitle{Phys. Rev.}
\bvolume{105},
\bfpage{1089}--\blpage{1099}
(\byear{1957}).
\doiurl{10.1103/PhysRev.105.1089}
\end{barticle}
\endbibitem

\bibitem{Peters63}
\begin{barticle}
\bauthor{\bsnm{Peters}, \binits{P.C.}},
\bauthor{\bsnm{Mathews}, \binits{J.}}:
\batitle{Gravitational radiation from point masses in a keplerian orbit}.
\bjtitle{Phys. Rev.}
\bvolume{131},
\bfpage{435}--\blpage{440}
(\byear{1963}).
\doiurl{10.1103/PhysRev.131.435}
\end{barticle}
\endbibitem

\bibitem{Thorne72}
\begin{barticle}
\bauthor{\bsnm{Press}, \binits{W.H.}},
\bauthor{\bsnm{Thorne}, \binits{K.S.}}:
\batitle{Gravitational-wave astronomy}.
\bjtitle{Annual Review of Astronomy and Astrophysics}
\bvolume{10},
\bfpage{335}--\blpage{374}
(\byear{1972})
\end{barticle}
\endbibitem

\bibitem{Maggiore2007}
\begin{bbook}
\bauthor{\bsnm{Maggiore}, \binits{M.}}:
\bbtitle{Gravitational Waves: Volume 1: Theory and Experiments}.
\bpublisher{OUP Oxford},
\blocation{ISBN 0198570740}
(\byear{2007})
\end{bbook}
\endbibitem

\bibitem{light-time}
\begin{barticle}
\bauthor{\bsnm{Koop}, \binits{M.J.}},
\bauthor{\bsnm{Finn}, \binits{L.S.}}:
\batitle{Physical response of light-time gravitational wave detectors}.
\bjtitle{Phys. Rev. D}
\bvolume{90},
\bfpage{062002}
(\byear{2014}).
\doiurl{10.1103/PhysRevD.90.062002}
\end{barticle}
\endbibitem

\bibitem{bondani}
\begin{barticle}
\bauthor{\bsnm{{Bondani}}, \binits{S.}},
\bauthor{\bsnm{{Cacciatori}}, \binits{S.L.}}:
\batitle{{Non canonical polarizations of gravitational waves}}.
\bjtitle{European Physical Journal C}
\bvolume{83}(\bissue{4}),
\bfpage{310}
(\byear{2023})
{\href{https://arxiv.org/abs/2212.10580}{{arXiv:2212.10580}}}
{[gr-qc]}.
\doiurl{10.1140/epjc/s10052-023-11502-1}
\end{barticle}
\endbibitem

\bibitem{Kolkowitz}
\begin{barticle}
\bauthor{\bsnm{Kolkowitz}, \binits{S.}},
\bauthor{\bsnm{Pikovski}, \binits{I.}},
\bauthor{\bsnm{Langellier}, \binits{N.}},
\bauthor{\bsnm{Lukin}, \binits{M.D.}},
\bauthor{\bsnm{Walsworth}, \binits{R.L.}},
\bauthor{\bsnm{Ye}, \binits{J.}}:
\batitle{Gravitational wave detection with optical lattice atomic clocks}.
\bjtitle{Phys. Rev. D}
\bvolume{94},
\bfpage{124043}
(\byear{2016}).
\doiurl{10.1103/PhysRevD.94.124043}
\end{barticle}
\endbibitem

\bibitem{rakhmanov}
\begin{barticle}
\bauthor{\bsnm{Rakhmanov}, \binits{M.}}:
\batitle{Response of test masses to gravitational waves in the local lorentz gauge}.
\bjtitle{Phys. Rev. D}
\bvolume{71},
\bfpage{084003}
(\byear{2005}).
\doiurl{10.1103/PhysRevD.71.084003}
\end{barticle}
\endbibitem

\bibitem{loeb}
\begin{botherref}
\oauthor{\bsnm{{Loeb}}, \binits{A.}},
\oauthor{\bsnm{{Maoz}}, \binits{D.}}:
{Using Atomic Clocks to Detect Gravitational Waves}.
arXiv e-prints,
1501--00996
(2015)
{\href{https://arxiv.org/abs/1501.00996}{{arXiv:1501.00996}}}
{[astro-ph.IM]}.
\doiurl{10.48550/arXiv.1501.00996}
\end{botherref}
\endbibitem

\bibitem{avirgo}
\begin{barticle}
\bauthor{\bsnm{{Abbott}}, \binits{B.P.}},
\bauthor{\bsnm{{et al.}}},
\bauthor{\bsnm{{{KAGRA} Collaboration}}, \binits{L.S.C.}},
\bauthor{\bsnm{{Virgo Collaboration}}}:
\batitle{{Prospects for observing and localizing gravitational-wave transients with Advanced LIGO, Advanced Virgo and KAGRA}}.
\bjtitle{Living Reviews in Relativity}
\bvolume{21}(\bissue{1}),
\bfpage{3}
(\byear{2018})
{\href{https://arxiv.org/abs/1304.0670}{{arXiv:1304.0670}}}
{[gr-qc]}.
\doiurl{10.1007/s41114-018-0012-9}
\end{barticle}
\endbibitem

\bibitem{lvk2}
\begin{barticle}
\bauthor{\bsnm{{Akutsu}}, \binits{T.}},
\bauthor{\bparticle{et} \bsnm{al.}}:
\batitle{{Overview of {KAGRA}: Detector design and construction history}}.
\bjtitle{Progress of Theoretical and Experimental Physics}
\bvolume{2021}(\bissue{5}),
\bfpage{05}--\blpage{101}
(\byear{2021})
{\href{https://arxiv.org/abs/2005.05574}{{arXiv:2005.05574}}}
{[physics.ins-det]}.
\doiurl{10.1093/ptep/ptaa125}
\end{barticle}
\endbibitem

\bibitem{lisa}
\begin{botherref}
\oauthor{\bsnm{{Amaro-Seoane et al.}}, \binits{P.}}:
{Laser Interferometer Space Antenna}.
arXiv e-prints,
1702--00786
(2017)
{\href{https://arxiv.org/abs/1702.00786}{{arXiv:1702.00786}}}
{[astro-ph.IM]}
\end{botherref}
\endbibitem

\bibitem{lisa96}
\begin{bchapter}
\bauthor{\bsnm{Folkner}, \binits{W.}},
\bauthor{\bsnm{Hellings}, \binits{R.}},
\bauthor{\bsnm{Maleki}, \binits{L.}},
\bauthor{\bsnm{Bender}, \binits{P.}},
\bauthor{\bsnm{Fallen}, \binits{J.}},
\bauthor{\bsnm{Stebbins}, \binits{R.}},
\bauthor{\bsnm{Danzmann}, \binits{K.}},
\bauthor{\bsnm{Cornelisse}, \binits{J.}},
\bauthor{\bsnm{Jafry}, \binits{Y.}},
\bauthor{\bsnm{Reinhard}, \binits{R.}}:
\bctitle{{LISA} - {L}aser {I}nterferometer {S}pace {A}ntenna for gravitational wave measurements}.
In: \bbtitle{33rd Aerospace Sciences Meeting and Exhibit}
(\byear{1996}).
\doiurl{10.2514/6.1995-829}.
\burl{https://arc.aiaa.org/doi/abs/10.2514/6.1995-829}
\end{bchapter}
\endbibitem

\bibitem{lisascird}
\begin{botherref}
\oauthor{\bsnm{Research}, \binits{E.S.}},
\oauthor{\bsnm{Centre}, \binits{T.}}:
{LISA} Science Requirements Document.
(ESA-L3-EST-SCI-RS-001)
(2018).
\url{https://www.cosmos.esa.int/documents/678316/1700384/SciRD.pdf/25831f6b-3c01-e215-5916-4ac6e4b306fb?t=1526479841000}
\end{botherref}
\endbibitem

\bibitem{Blanchet2024}
\begin{botherref}
\oauthor{\bsnm{Blanchet}, \binits{L.}}:
Post-newtonian theory for gravitational waves.
Living Reviews in Relativity
\textbf{27}(1)
(2024).
\doiurl{10.1007/s41114-024-00050-z}
\end{botherref}
\endbibitem

\bibitem{kozai1}
\begin{barticle}
\bauthor{\bsnm{{Kozai}}, \binits{Y.}}:
\batitle{{Secular perturbations of asteroids with high inclination and eccentricity}}.
\bjtitle{Astronomical Journal}
\bvolume{67},
\bfpage{591}--\blpage{598}
(\byear{1962}).
\doiurl{10.1086/108790}
\end{barticle}
\endbibitem

\bibitem{kozai2}
\begin{barticle}
\bauthor{\bsnm{Lidov}, \binits{M.L.}}:
\batitle{The evolution of orbits of artificial satellites of planets under the action of gravitational perturbations of external bodies}.
\bjtitle{Planetary and Space Science}
\bvolume{9}(\bissue{10}),
\bfpage{719}--\blpage{759}
(\byear{1962}).
\doiurl{10.1016/0032-0633(62)90129-0}
\end{barticle}
\endbibitem

\bibitem{miller}
\begin{barticle}
\bauthor{\bsnm{Miller}, \binits{M.C.}},
\bauthor{\bsnm{Hamilton}, \binits{D.P.}}:
\batitle{Four-body effects in globular cluster black hole coalescence}.
\bjtitle{The Astrophysical Journal}
\bvolume{576}(\bissue{2}),
\bfpage{894}
(\byear{2002}).
\doiurl{10.1086/341788}
\end{barticle}
\endbibitem

\bibitem{atomic_rev}
\begin{barticle}
\bauthor{\bsnm{{Ludlow}}, \binits{A.D.}},
\bauthor{\bsnm{{Boyd}}, \binits{M.M.}},
\bauthor{\bsnm{{Ye}}, \binits{J.}},
\bauthor{\bsnm{{Peik}}, \binits{E.}},
\bauthor{\bsnm{{Schmidt}}, \binits{P.O.}}:
\batitle{{Optical atomic clocks}}.
\bjtitle{Reviews of Modern Physics}
\bvolume{87}(\bissue{2}),
\bfpage{637}--\blpage{701}
(\byear{2015})
{\href{https://arxiv.org/abs/1407.3493}{{arXiv:1407.3493}}}
{[physics.atom-ph]}.
\doiurl{10.1103/RevModPhys.87.637}
\end{barticle}
\endbibitem

\bibitem{takamoto2005}
\begin{barticle}
\bauthor{\bsnm{Takamoto}, \binits{M.}},
\bauthor{\bsnm{Hong}, \binits{F.-L.}},
\bauthor{\bsnm{Higashi}, \binits{R.}},
\bauthor{\bsnm{Katori}, \binits{H.}}:
\batitle{An optical lattice clock}.
\bjtitle{Nature}
\bvolume{435},
\bfpage{321}--\blpage{4}
(\byear{2005}).
\doiurl{10.1038/nature03541}
\end{barticle}
\endbibitem

\bibitem{ludlow2006}
\begin{barticle}
\bauthor{\bsnm{{Ludlow}}, \binits{A.D.}},
\bauthor{\bsnm{{Boyd}}, \binits{M.M.}},
\bauthor{\bsnm{{Zelevinsky}}, \binits{T.}},
\bauthor{\bsnm{{Foreman}}, \binits{S.M.}},
\bauthor{\bsnm{{Blatt}}, \binits{S.}},
\bauthor{\bsnm{{Notcutt}}, \binits{M.}},
\bauthor{\bsnm{{Ido}}, \binits{T.}},
\bauthor{\bsnm{{Ye}}, \binits{J.}}:
\batitle{{Systematic Study of the $^{87}$Sr Clock Transition in an Optical Lattice}}.
\bjtitle{Physical Review Letters}
\bvolume{96}(\bissue{3}),
\bfpage{033003}
(\byear{2006})
{\href{https://arxiv.org/abs/physics/0508041}{{arXiv:physics/0508041}}}
{[physics.atom-ph]}.
\doiurl{10.1103/PhysRevLett.96.033003}
\end{barticle}
\endbibitem

\bibitem{riehle2015}
\begin{barticle}
\bauthor{\bsnm{Riehle}, \binits{F.}}:
\batitle{Towards a redefinition of the second based on optical atomic clocks}.
\bjtitle{Comptes Rendus Physique}
\bvolume{16}(\bissue{5}),
\bfpage{506}--\blpage{515}
(\byear{2015}).
\doiurl{10.1016/j.crhy.2015.03.012}.
\bcomment{The measurement of time / La mesure du temps}
\end{barticle}
\endbibitem

\bibitem{history}
\begin{barticle}
\bauthor{\bsnm{{Derevianko}}, \binits{A.}},
\bauthor{\bsnm{{Gibble}}, \binits{K.}},
\bauthor{\bsnm{{Hollberg}}, \binits{L.}},
\bauthor{\bsnm{{Newbury}}, \binits{N.R.}},
\bauthor{\bsnm{{Oates}}, \binits{C.}},
\bauthor{\bsnm{{Safronova}}, \binits{M.S.}},
\bauthor{\bsnm{{Sinclair}}, \binits{L.C.}},
\bauthor{\bsnm{{Yu}}, \binits{N.}}:
\batitle{{Fundamental physics with a state-of-the-art optical clock in space}}.
\bjtitle{Quantum Science and Technology}
\bvolume{7}(\bissue{4}),
\bfpage{044002}
(\byear{2022})
{\href{https://arxiv.org/abs/2112.10817}{{arXiv:2112.10817}}}
{[gr-qc]}.
\doiurl{10.1088/2058-9565/ac7df9}
\end{barticle}
\endbibitem

\bibitem{ludlow2013}
\begin{barticle}
\bauthor{\bsnm{Hinkley}, \binits{N.}},
\bauthor{\bsnm{Sherman}, \binits{J.A.}},
\bauthor{\bsnm{Phillips}, \binits{N.B.}},
\bauthor{\bsnm{Schioppo}, \binits{M.}},
\bauthor{\bsnm{Lemke}, \binits{N.D.}},
\bauthor{\bsnm{Beloy}, \binits{K.}},
\bauthor{\bsnm{Pizzocaro}, \binits{M.}},
\bauthor{\bsnm{Oates}, \binits{C.W.}},
\bauthor{\bsnm{Ludlow}, \binits{A.D.}}:
\batitle{An atomic clock with $10^{-18}$ instability}.
\bjtitle{Science}
\bvolume{341}(\bissue{6151}),
\bfpage{1215}--\blpage{1218}
(\byear{2013})
{\href{https://arxiv.org/abs/https://www.science.org/doi/pdf/10.1126/science.1240420}{{https://www.science.org/doi/pdf/10.1126/science.1240420}}}.
\doiurl{10.1126/science.1240420}
\end{barticle}
\endbibitem

\bibitem{attoclock}
\begin{bbook}
\bauthor{\bsnm{Cirelli}, \binits{C.}},
\bauthor{\bsnm{Pfeiffer}, \binits{A.N.}},
\bauthor{\bsnm{Smolarski}, \binits{M.}},
\bauthor{\bsnm{Eckle}, \binits{P.}},
\bauthor{\bsnm{Keller}, \binits{U.}}:
In: \beditor{\bsnm{Plaja}, \binits{L.}},
\beditor{\bsnm{Torres}, \binits{R.}},
\beditor{\bsnm{Za{\"i}r}, \binits{A.}} (eds.)
\bbtitle{The Attoclock: A Novel Ultrafast Measurement Technique with Attosecond Time Resolution},
pp. \bfpage{135}--\blpage{158}.
\bpublisher{Springer},
\blocation{Berlin, Heidelberg}
(\byear{2013}).
\burl{https://doi.org/10.1007/978-3-642-37623-8_9}
\end{bbook}
\endbibitem

\bibitem{10-18s}
\begin{barticle}
\bauthor{\bsnm{{Bloom}}, \binits{B.J.}},
\bauthor{\bsnm{{Nicholson}}, \binits{T.L.}},
\bauthor{\bsnm{{Williams}}, \binits{J.R.}},
\bauthor{\bsnm{{Campbell}}, \binits{S.L.}},
\bauthor{\bsnm{{Bishof}}, \binits{M.}},
\bauthor{\bsnm{{Zhang}}, \binits{X.}},
\bauthor{\bsnm{{Zhang}}, \binits{W.}},
\bauthor{\bsnm{{Bromley}}, \binits{S.L.}},
\bauthor{\bsnm{{Ye}}, \binits{J.}}:
\batitle{{An optical lattice clock with accuracy and stability at the 10$^{-18}$ level}}.
\bjtitle{Nature}
\bvolume{506}(\bissue{7486}),
\bfpage{71}--\blpage{75}
(\byear{2014})
{\href{https://arxiv.org/abs/1309.1137}{{arXiv:1309.1137}}}
{[physics.atom-ph]}.
\doiurl{10.1038/nature12941}
\end{barticle}
\endbibitem

\bibitem{attosecond}
\begin{botherref}
\oauthor{\bsnm{Mustary}, \binits{M.H.}},
\oauthor{\bsnm{Xu}, \binits{L.}},
\oauthor{\bsnm{Wu}, \binits{W.}},
\oauthor{\bsnm{Haram}, \binits{N.}},
\oauthor{\bsnm{Laban}, \binits{D.E.}},
\oauthor{\bsnm{Xu}, \binits{H.}},
\oauthor{\bsnm{He}, \binits{F.}},
\oauthor{\bsnm{Sang}, \binits{R.T.}},
\oauthor{\bsnm{Litvinyuk}, \binits{I.V.}}:
Attosecond delays of high-harmonic emissions from hydrogen isotopes measured by {XUV} interferometer.
Ultrafast Science
\textbf{2022}
(2022)
{\href{https://arxiv.org/abs/https://spj.science.org/doi/pdf/10.34133/2022/9834102}{{https://spj.science.org/doi/pdf/10.34133/2022/9834102}}}.
\doiurl{10.34133/2022/9834102}
\end{botherref}
\endbibitem

\bibitem{astrain}
\begin{barticle}
\bauthor{\bsnm{Paul}, \binits{P.M.}},
\bauthor{\bsnm{Toma}, \binits{E.S.}},
\bauthor{\bsnm{Breger}, \binits{P.}},
\bauthor{\bsnm{Mullot}, \binits{G.}},
\bauthor{\bsnm{Augé}, \binits{F.}},
\bauthor{\bsnm{Balcou}, \binits{P.}},
\bauthor{\bsnm{Muller}, \binits{H.G.}},
\bauthor{\bsnm{Agostini}, \binits{P.}}:
\batitle{Observation of a train of attosecond pulses from high harmonic generation}.
\bjtitle{Science}
\bvolume{292}(\bissue{5522}),
\bfpage{1689}--\blpage{1692}
(\byear{2001})
{\href{https://arxiv.org/abs/https://www.science.org/doi/pdf/10.1126/science.1059413}{{https://www.science.org/doi/pdf/10.1126/science.1059413}}}.
\doiurl{10.1126/science.1059413}
\end{barticle}
\endbibitem

\bibitem{43as}
\begin{barticle}
\bauthor{\bsnm{Gaumnitz}, \binits{T.}},
\bauthor{\bsnm{Jain}, \binits{A.}},
\bauthor{\bsnm{Pertot}, \binits{Y.}},
\bauthor{\bsnm{Huppert}, \binits{M.}},
\bauthor{\bsnm{Jordan}, \binits{I.}},
\bauthor{\bsnm{Ardana-Lamas}, \binits{F.}},
\bauthor{\bsnm{W\"{o}rner}, \binits{H.J.}}:
\batitle{Streaking of 43-attosecond soft-{X}-ray pulses generated by a passively {CEP}-stable mid-infrared driver}.
\bjtitle{Opt. Express}
\bvolume{25}(\bissue{22}),
\bfpage{27506}--\blpage{27518}
(\byear{2017}).
\doiurl{10.1364/OE.25.027506}
\end{barticle}
\endbibitem

\bibitem{2_18}
\begin{barticle}
\bauthor{\bsnm{{Bothwell}}, \binits{T.}},
\bauthor{\bsnm{{Kedar}}, \binits{D.}},
\bauthor{\bsnm{{Oelker}}, \binits{E.}},
\bauthor{\bsnm{{Robinson}}, \binits{J.M.}},
\bauthor{\bsnm{{Bromley}}, \binits{S.L.}},
\bauthor{\bsnm{{Tew}}, \binits{W.L.}},
\bauthor{\bsnm{{Ye}}, \binits{J.}},
\bauthor{\bsnm{{Kennedy}}, \binits{C.J.}}:
\batitle{{JILA SrI optical lattice clock with uncertainty of $2.0\times10^{-18}$}}.
\bjtitle{Metrologia}
\bvolume{56}(\bissue{6}),
\bfpage{065004}
(\byear{2019})
{\href{https://arxiv.org/abs/1906.06004}{{arXiv:1906.06004}}}
{[physics.atom-ph]}.
\doiurl{10.1088/1681-7575/ab4089}
\end{barticle}
\endbibitem

\bibitem{nicholson-18}
\begin{barticle}
\bauthor{\bsnm{{Nicholson}}, \binits{T.L.}},
\bauthor{\bsnm{{Campbell}}, \binits{S.L.}},
\bauthor{\bsnm{{Hutson}}, \binits{R.B.}},
\bauthor{\bsnm{{Marti}}, \binits{G.E.}},
\bauthor{\bsnm{{Bloom}}, \binits{B.J.}},
\bauthor{\bsnm{{McNally}}, \binits{R.L.}},
\bauthor{\bsnm{{Zhang}}, \binits{W.}},
\bauthor{\bsnm{{Barrett}}, \binits{M.D.}},
\bauthor{\bsnm{{Safronova}}, \binits{M.S.}},
\bauthor{\bsnm{{Strouse}}, \binits{G.F.}},
\bauthor{\bsnm{{Tew}}, \binits{W.L.}},
\bauthor{\bsnm{{Ye}}, \binits{J.}}:
\batitle{{Systematic evaluation of an atomic clock at $2\times10^{-18}$ total uncertainty}}.
\bjtitle{Nature Communications}
\bvolume{6},
\bfpage{6896}
(\byear{2015})
{\href{https://arxiv.org/abs/1412.8261}{{arXiv:1412.8261}}}
{[physics.atom-ph]}.
\doiurl{10.1038/ncomms7896}
\end{barticle}
\endbibitem

\bibitem{huntemann2016}
\begin{barticle}
\bauthor{\bsnm{Huntemann}, \binits{N.}},
\bauthor{\bsnm{Sanner}, \binits{C.}},
\bauthor{\bsnm{Lipphardt}, \binits{B.}},
\bauthor{\bsnm{Tamm}, \binits{C.}},
\bauthor{\bsnm{Peik}, \binits{E.}}:
\batitle{Single-ion atomic clock with $3\ifmmode\times\else\texttimes\fi{}{10}^{\ensuremath{-}18}$ systematic uncertainty}.
\bjtitle{Phys. Rev. Lett.}
\bvolume{116},
\bfpage{063001}
(\byear{2016}).
\doiurl{10.1103/PhysRevLett.116.063001}
\end{barticle}
\endbibitem

\bibitem{McGrew2018}
\begin{barticle}
\bauthor{\bsnm{McGrew}, \binits{W.}},
\bauthor{\bsnm{Zhang}, \binits{X.}},
\bauthor{\bsnm{Fasano}, \binits{R.}},
\bauthor{\bsnm{Schäffer}, \binits{S.}},
\bauthor{\bsnm{Beloy}, \binits{K.}},
\bauthor{\bsnm{Nicolodi}, \binits{D.}},
\bauthor{\bsnm{Brown}, \binits{R.}},
\bauthor{\bsnm{Hinkley}, \binits{N.}},
\bauthor{\bsnm{Milani}, \binits{G.}},
\bauthor{\bsnm{Schioppo}, \binits{M.}},
\bauthor{\bsnm{Yoon}, \binits{T.}},
\bauthor{\bsnm{Ludlow}, \binits{A.}}:
\batitle{Atomic clock performance enabling geodesy below the centimetre level}.
\bjtitle{Nature}
\bvolume{564}(\bissue{7734}),
\bfpage{87}--\blpage{90}
(\byear{2018}).
\doiurl{10.1038/s41586-018-0738-2}
\end{barticle}
\endbibitem

\bibitem{boulder2021}
\begin{barticle}
\bauthor{\bsnm{{Boulder Atomic Clock Optical Network Bacon Collaboration}}},
\bauthor{\bsnm{{Beloy}}, \binits{K.}},
\bauthor{\bsnm{{Bodine}}, \binits{M.I.}},
\bauthor{\bsnm{{Bothwell}}, \binits{T.}},
\bauthor{\bsnm{{Brewer}}, \binits{S.M.}},
\bauthor{\bsnm{{Bromley}}, \binits{S.L.}},
\bauthor{\bsnm{{Chen}}, \binits{J.-S.}},
\bauthor{\bsnm{{Desch{\^e}nes}}, \binits{J.-D.}},
\bauthor{\bsnm{{Diddams}}, \binits{S.A.}},
\bauthor{\bsnm{{Fasano}}, \binits{R.J.}},
\bauthor{\bsnm{{Fortier}}, \binits{T.M.}},
\bauthor{\bsnm{{Hassan}}, \binits{Y.S.}},
\bauthor{\bsnm{{Hume}}, \binits{D.B.}},
\bauthor{\bsnm{{Kedar}}, \binits{D.}},
\bauthor{\bsnm{{Kennedy}}, \binits{C.J.}},
\bauthor{\bsnm{{Khader}}, \binits{I.}},
\bauthor{\bsnm{{Koepke}}, \binits{A.}},
\bauthor{\bsnm{{Leibrandt}}, \binits{D.R.}},
\bauthor{\bsnm{{Leopardi}}, \binits{H.}},
\bauthor{\bsnm{{Ludlow}}, \binits{A.D.}},
\bauthor{\bsnm{{McGrew}}, \binits{W.F.}},
\bauthor{\bsnm{{Milner}}, \binits{W.R.}},
\bauthor{\bsnm{{Newbury}}, \binits{N.R.}},
\bauthor{\bsnm{{Nicolodi}}, \binits{D.}},
\bauthor{\bsnm{{Oelker}}, \binits{E.}},
\bauthor{\bsnm{{Parker}}, \binits{T.E.}},
\bauthor{\bsnm{{Robinson}}, \binits{J.M.}},
\bauthor{\bsnm{{Romisch}}, \binits{S.}},
\bauthor{\bsnm{{Sch{\"a}ffer}}, \binits{S.A.}},
\bauthor{\bsnm{{Sherman}}, \binits{J.A.}},
\bauthor{\bsnm{{Sinclair}}, \binits{L.C.}},
\bauthor{\bsnm{{Sonderhouse}}, \binits{L.}},
\bauthor{\bsnm{{Swann}}, \binits{W.C.}},
\bauthor{\bsnm{{Yao}}, \binits{J.}},
\bauthor{\bsnm{{Ye}}, \binits{J.}},
\bauthor{\bsnm{{Zhang}}, \binits{X.}}:
\batitle{{Frequency ratio measurements at 18-digit accuracy using an optical clock network}}.
\bjtitle{Nature}
\bvolume{591}(\bissue{7851}),
\bfpage{564}--\blpage{569}
(\byear{2021}).
\doiurl{10.1038/s41586-021-03253-4}
\end{barticle}
\endbibitem

\bibitem{qlogic}
\begin{barticle}
\bauthor{\bsnm{Brewer}, \binits{S.M.}},
\bauthor{\bsnm{Chen}, \binits{J.-S.}},
\bauthor{\bsnm{Hankin}, \binits{A.M.}},
\bauthor{\bsnm{Clements}, \binits{E.R.}},
\bauthor{\bsnm{Chou}, \binits{C.W.}},
\bauthor{\bsnm{Wineland}, \binits{D.J.}},
\bauthor{\bsnm{Hume}, \binits{D.B.}},
\bauthor{\bsnm{Leibrandt}, \binits{D.R.}}:
\batitle{$^{27}${{A}l}$^{+}$ quantum-logic clock with a systematic uncertainty below ${10}^{\ensuremath{-}18}$}.
\bjtitle{Phys. Rev. Lett.}
\bvolume{123},
\bfpage{033201}
(\byear{2019}).
\doiurl{10.1103/PhysRevLett.123.033201}
\end{barticle}
\endbibitem

\bibitem{optical}
\begin{barticle}
\bauthor{\bsnm{Marti}, \binits{G.E.}},
\bauthor{\bsnm{Hutson}, \binits{R.B.}},
\bauthor{\bsnm{Goban}, \binits{A.}},
\bauthor{\bsnm{Campbell}, \binits{S.L.}},
\bauthor{\bsnm{Poli}, \binits{N.}},
\bauthor{\bsnm{Ye}, \binits{J.}}:
\batitle{Imaging optical frequencies with 100 $\ensuremath{\mu}${H}z precision and 1.1 $\ensuremath{\mu}$m resolution}.
\bjtitle{Phys. Rev. Lett.}
\bvolume{120},
\bfpage{103201}
(\byear{2018}).
\doiurl{10.1103/PhysRevLett.120.103201}
\end{barticle}
\endbibitem

\bibitem{shift19}
\begin{barticle}
\bauthor{\bsnm{Kim}, \binits{K.}},
\bauthor{\bsnm{Aeppli}, \binits{A.}},
\bauthor{\bsnm{Bothwell}, \binits{T.}},
\bauthor{\bsnm{Ye}, \binits{J.}}:
\batitle{Evaluation of lattice light shift at low ${10}^{\ensuremath{-}19}$ uncertainty for a shallow lattice {S}r optical clock}.
\bjtitle{Phys. Rev. Lett.}
\bvolume{130},
\bfpage{113203}
(\byear{2023}).
\doiurl{10.1103/PhysRevLett.130.113203}
\end{barticle}
\endbibitem

\bibitem{10-21b}
\begin{barticle}
\bauthor{\bsnm{{Zheng}}, \binits{X.}},
\bauthor{\bsnm{{Dolde}}, \binits{J.}},
\bauthor{\bsnm{{Lochab}}, \binits{V.}},
\bauthor{\bsnm{{Merriman}}, \binits{B.N.}},
\bauthor{\bsnm{{Li}}, \binits{H.}},
\bauthor{\bsnm{{Kolkowitz}}, \binits{S.}}:
\batitle{{Differential clock comparisons with a multiplexed optical lattice clock}}.
\bjtitle{Nature}
\bvolume{602}(\bissue{7897}),
\bfpage{425}--\blpage{430}
(\byear{2022})
{\href{https://arxiv.org/abs/2109.12237}{{arXiv:2109.12237}}}
{[physics.atom-ph]}.
\doiurl{10.1038/s41586-021-04344-y}
\end{barticle}
\endbibitem

\bibitem{redshift-mm}
\begin{barticle}
\bauthor{\bsnm{{Bothwell}}, \binits{T.}},
\bauthor{\bsnm{{Kennedy}}, \binits{C.J.}},
\bauthor{\bsnm{{Aeppli}}, \binits{A.}},
\bauthor{\bsnm{{Kedar}}, \binits{D.}},
\bauthor{\bsnm{{Robinson}}, \binits{J.M.}},
\bauthor{\bsnm{{Oelker}}, \binits{E.}},
\bauthor{\bsnm{{Staron}}, \binits{A.}},
\bauthor{\bsnm{{Ye}}, \binits{J.}}:
\batitle{{Resolving the gravitational redshift across a millimetre-scale atomic sample}}.
\bjtitle{Nature}
\bvolume{602}(\bissue{7897}),
\bfpage{420}--\blpage{424}
(\byear{2022})
{\href{https://arxiv.org/abs/2109.12238}{{arXiv:2109.12238}}}
{[physics.atom-ph]}.
\doiurl{10.1038/s41586-021-04349-7}
\end{barticle}
\endbibitem

\bibitem{heeg}
\begin{barticle}
\bauthor{\bsnm{Heeg}, \binits{K.P.}},
\bauthor{\bsnm{Kaldun}, \binits{A.}},
\bauthor{\bsnm{Strohm}, \binits{C.}},
\bauthor{\bsnm{Ott}, \binits{C.}},
\bauthor{\bsnm{Subramanian}, \binits{R.}},
\bauthor{\bsnm{Lentrodt}, \binits{D.}},
\bauthor{\bsnm{Haber}, \binits{J.}},
\bauthor{\bsnm{Wille}, \binits{H.-C.}},
\bauthor{\bsnm{Goerttler}, \binits{S.}},
\bauthor{\bsnm{Rüffer}, \binits{R.}},
\bauthor{\bsnm{H.}, \binits{C.}}:
\batitle{{Coherent {X}-ray optical control of nuclear excitons}}.
\bjtitle{Nature}
\bvolume{590}(\bissue{7846}),
\bfpage{401}--\blpage{404}
(\byear{2021}).
\doiurl{10.1038/s41586-021-03276-x}
\end{barticle}
\endbibitem

\bibitem{zepto}
\begin{barticle}
\bauthor{\bsnm{Bocklage}, \binits{L.}},
\bauthor{\bsnm{Gollwitzer}, \binits{J.}},
\bauthor{\bsnm{Strohm}, \binits{C.}},
\bauthor{\bsnm{Adolff}, \binits{C.F.}},
\bauthor{\bsnm{Schlage}, \binits{K.}},
\bauthor{\bsnm{Sergeev}, \binits{I.}},
\bauthor{\bsnm{Leupold}, \binits{O.}},
\bauthor{\bsnm{Wille}, \binits{H.-C.}},
\bauthor{\bsnm{Meier}, \binits{G.}},
\bauthor{\bsnm{Röhlsberger}, \binits{R.}}:
\batitle{Coherent control of collective nuclear quantum states via transient magnons}.
\bjtitle{Science Advances}
\bvolume{7}(\bissue{5}),
\bfpage{3991}
(\byear{2021})
{\href{https://arxiv.org/abs/science.org/doi/pdf/10.1126/sciadv.abc3991}{{science.org/doi/pdf/10.1126/sciadv.abc3991}}}.
\doiurl{10.1126/sciadv.abc3991}
\end{barticle}
\endbibitem

\bibitem{phasedelay}
\begin{barticle}
\bauthor{\bsnm{Albrecht}, \binits{A.W.}},
\bauthor{\bsnm{Hybl}, \binits{J.D.}},
\bauthor{\bsnm{Gallagher~Faeder}, \binits{S.M.}},
\bauthor{\bsnm{Jonas}, \binits{D.M.}}:
\batitle{{Experimental distinction between phase shifts and time delays: Implications for femtosecond spectroscopy and coherent control of chemical reactions}}.
\bjtitle{The Journal of Chemical Physics}
\bvolume{111}(\bissue{24}),
\bfpage{10934}--\blpage{10956}
(\byear{1999})
{\href{https://arxiv.org/abs/https://pubs.aip.org/aip/jcp/article-pdf/111/24/10934/19287691/10934\_1\_online.pdf}{{https://pubs.aip.org/aip/jcp/article-pdf/111/24/10934/19287691/10934\_1\_online.pdf}}}.
\doiurl{10.1063/1.480457}
\end{barticle}
\endbibitem

\bibitem{yocto_re}
\begin{barticle}
\bauthor{\bsnm{Shvyd’ko}, \binits{Y.}},
\bauthor{\bsnm{Schindelmann}, \binits{P.}}:
\batitle{On yoctosecond science}.
\bjtitle{Nature}
\bvolume{608}(\bissue{7922}),
\bfpage{16}--\blpage{17}
(\byear{2022}).
\doiurl{10.1038/s41586-022-04870-3}
\end{barticle}
\endbibitem

\bibitem{yocto_re2}
\begin{barticle}
\bauthor{\bsnm{Heeg}, \binits{K.P.}},
\bauthor{\bsnm{Bocklage}, \binits{L.}},
\bauthor{\bsnm{Strohm}, \binits{C.}},
\bauthor{\bsnm{Ott}, \binits{C.}},
\bauthor{\bsnm{Lentrodt}, \binits{D.}},
\bauthor{\bsnm{Haber}, \binits{J.}},
\bauthor{\bsnm{Wille}, \binits{H.-C.}},
\bauthor{\bsnm{R\"{u}ffer}, \binits{R.}},
\bauthor{\bsnm{Gollwitzer}, \binits{J.}},
\bauthor{\bsnm{Adolff}, \binits{C.F.}},
\bauthor{\bsnm{Schlage}, \binits{K.}},
\bauthor{\bsnm{Sergeev}, \binits{I.}},
\bauthor{\bsnm{Leupold}, \binits{O.}},
\bauthor{\bsnm{Meier}, \binits{G.}},
\bauthor{\bsnm{Keitel}, \binits{C.H.}},
\bauthor{\bsnm{R\"{o}hlsberger}, \binits{R.}},
\bauthor{\bsnm{Pfeifer}, \binits{T.}},
\bauthor{\bsnm{Evers}, \binits{J.}}:
\batitle{Reply to: On yoctosecond science}.
\bjtitle{Nature}
\bvolume{608}(\bissue{7922}),
\bfpage{18}--\blpage{19}
(\byear{2022}).
\doiurl{10.1038/s41586-022-04871-2}
\end{barticle}
\endbibitem

\bibitem{zepto_pulse}
\begin{barticle}
\bauthor{\bsnm{K\"{o}hler}, \binits{J.}},
\bauthor{\bsnm{Wollenhaupt}, \binits{M.}},
\bauthor{\bsnm{Bayer}, \binits{T.}},
\bauthor{\bsnm{Sarpe}, \binits{C.}},
\bauthor{\bsnm{Baumert}, \binits{T.}}:
\batitle{Zeptosecond precision pulse shaping}.
\bjtitle{Opt. Express}
\bvolume{19}(\bissue{12}),
\bfpage{11638}--\blpage{11653}
(\byear{2011}).
\doiurl{10.1364/OE.19.011638}
\end{barticle}
\endbibitem

\bibitem{zepto_photo}
\begin{barticle}
\bauthor{\bsnm{{Grundmann}}, \binits{S.}},
\bauthor{\bsnm{{Trabert}}, \binits{D.}},
\bauthor{\bsnm{{Fehre}}, \binits{K.}},
\bauthor{\bsnm{{Strenger}}, \binits{N.}},
\bauthor{\bsnm{{Pier}}, \binits{A.}},
\bauthor{\bsnm{{Kaiser}}, \binits{L.}},
\bauthor{\bsnm{{Kircher}}, \binits{M.}},
\bauthor{\bsnm{{Weller}}, \binits{M.}},
\bauthor{\bsnm{{Eckart}}, \binits{S.}},
\bauthor{\bsnm{{Schmidt}}, \binits{L.P.H.}},
\bauthor{\bsnm{{Trinter}}, \binits{F.}},
\bauthor{\bsnm{{Jahnke}}, \binits{T.}},
\bauthor{\bsnm{{Sch{\"o}ffler}}, \binits{M.S.}},
\bauthor{\bsnm{{D{\"o}rner}}, \binits{R.}}:
\batitle{{Zeptosecond birth time delay in molecular photoionization}}.
\bjtitle{Science}
\bvolume{370}(\bissue{6514}),
\bfpage{339}--\blpage{341}
(\byear{2020})
{\href{https://arxiv.org/abs/2010.08298}{{arXiv:2010.08298}}}
{[physics.atom-ph]}.
\doiurl{10.1126/science.abb9318}
\end{barticle}
\endbibitem

\bibitem{allan1}
\begin{barticle}
\bauthor{\bsnm{Allan}, \binits{D.W.}}:
\batitle{Statistics of atomic frequency standards}.
\bjtitle{Proceedings of the IEEE}
\bvolume{54}(\bissue{2}),
\bfpage{221}--\blpage{230}
(\byear{1966}).
\doiurl{10.1109/PROC.1966.4634}
\end{barticle}
\endbibitem

\bibitem{allan2}
\begin{bchapter}
\bauthor{\bsnm{{Allan}}, \binits{D.W.}}:
\bctitle{{Clock Characterization Tutorial}}.
In: \bbtitle{Proceedings of the Fifteenth Annual Precise Time and Material Interval (PTTI) Applications and Planning Meeting. Held at Naval Research Laboratory},
pp. \bfpage{459}--\blpage{475}
(\byear{1984})
\end{bchapter}
\endbibitem

\bibitem{allan3}
\begin{barticle}
\bauthor{\bsnm{Greenhall}, \binits{C.A.}},
\bauthor{\bsnm{Howe}, \binits{D.A.}},
\bauthor{\bsnm{Percival}, \binits{D.B.}}:
\batitle{Total variance, an estimator of long-term frequency stability [standards]}.
\bjtitle{IEEE Transactions on Ultrasonics, Ferroelectrics, and Frequency Control}
\bvolume{46}(\bissue{5}),
\bfpage{1183}--\blpage{1191}
(\byear{1999}).
\doiurl{10.1109/58.796124}
\end{barticle}
\endbibitem

\bibitem{half-minute}
\begin{bchapter}
\bauthor{\bsnm{Young}, \binits{A.W.}},
\bauthor{\bsnm{Eckner}, \binits{W.J.}},
\bauthor{\bsnm{Milner}, \binits{W.R.}},
\bauthor{\bsnm{Kedar}, \binits{D.}},
\bauthor{\bsnm{Norcia}, \binits{A.}},
\bauthor{\bsnm{Oelker}, \binits{E.}},
\bauthor{\bsnm{Schine}, \binits{N.}},
\bauthor{\bsnm{Ye}, \binits{J.}},
\bauthor{\bsnm{Kaufman}, \binits{A.M.}}:
\bctitle{Half-minute-scale atomic coherence and high relative stability in a tweezer clock}.
(\byear{2020}).
\burl{https://api.semanticscholar.org/CorpusID:263511276}
\end{bchapter}
\endbibitem

\bibitem{compactOLC}
\begin{botherref}
\oauthor{\bsnm{Chen}, \binits{Y.}},
\oauthor{\bsnm{Zhou}, \binits{C.}},
\oauthor{\bsnm{Tan}, \binits{W.}},
\oauthor{\bsnm{Guo}, \binits{F.}},
\oauthor{\bsnm{Zhao}, \binits{G.}},
\oauthor{\bsnm{Xia}, \binits{J.}},
\oauthor{\bsnm{Meng}, \binits{J.}},
\oauthor{\bsnm{Chang}, \binits{H.}}:
Development of compact and robust physical system for strontium optical lattice clock.
Applied Sciences
\textbf{14}(4)
(2024).
\doiurl{10.3390/app14041551}
\end{botherref}
\endbibitem

\bibitem{quantum_improvement}
\begin{barticle}
\bauthor{\bsnm{{Huelga}}, \binits{S.F.}},
\bauthor{\bsnm{{Macchiavello}}, \binits{C.}},
\bauthor{\bsnm{{Pellizzari}}, \binits{T.}},
\bauthor{\bsnm{{Ekert}}, \binits{A.K.}},
\bauthor{\bsnm{{Plenio}}, \binits{M.B.}},
\bauthor{\bsnm{{Cirac}}, \binits{J.I.}}:
\batitle{{Improvement of Frequency Standards with Quantum Entanglement}}.
\bjtitle{Physical Review Letters}
\bvolume{79}(\bissue{20}),
\bfpage{3865}--\blpage{3868}
(\byear{1997})
{\href{https://arxiv.org/abs/quant-ph/9707014}{{arXiv:quant-ph/9707014}}}
{[quant-ph]}.
\doiurl{10.1103/PhysRevLett.79.3865}
\end{barticle}
\endbibitem

\bibitem{network}
\begin{barticle}
\bauthor{\bsnm{{K{\'o}m{\'a}r}}, \binits{P.}},
\bauthor{\bsnm{{Kessler}}, \binits{E.M.}},
\bauthor{\bsnm{{Bishof}}, \binits{M.}},
\bauthor{\bsnm{{Jiang}}, \binits{L.}},
\bauthor{\bsnm{{S{\o}rensen}}, \binits{A.S.}},
\bauthor{\bsnm{{Ye}}, \binits{J.}},
\bauthor{\bsnm{{Lukin}}, \binits{M.D.}}:
\batitle{{A quantum network of clocks}}.
\bjtitle{Nature Physics}
\bvolume{10}(\bissue{8}),
\bfpage{582}--\blpage{587}
(\byear{2014})
{\href{https://arxiv.org/abs/1310.6045}{{arXiv:1310.6045}}}
{[quant-ph]}.
\doiurl{10.1038/nphys3000}
\end{barticle}
\endbibitem

\bibitem{network-entangled}
\begin{barticle}
\bauthor{\bsnm{{Nichol}}, \binits{B.C.}},
\bauthor{\bsnm{{Srinivas}}, \binits{R.}},
\bauthor{\bsnm{{Nadlinger}}, \binits{D.P.}},
\bauthor{\bsnm{{Drmota}}, \binits{P.}},
\bauthor{\bsnm{{Main}}, \binits{D.}},
\bauthor{\bsnm{{Araneda}}, \binits{G.}},
\bauthor{\bsnm{{Ballance}}, \binits{C.J.}},
\bauthor{\bsnm{{Lucas}}, \binits{D.M.}}:
\batitle{{An elementary quantum network of entangled optical atomic clocks}}.
\bjtitle{Nature}
\bvolume{609}(\bissue{7928}),
\bfpage{689}--\blpage{694}
(\byear{2022})
{\href{https://arxiv.org/abs/2111.10336}{{arXiv:2111.10336}}}
{[physics.atom-ph]}.
\doiurl{10.1038/s41586-022-05088-z}
\end{barticle}
\endbibitem

\bibitem{nuclear0}
\begin{barticle}
\bauthor{\bsnm{Kazakov}, \binits{G.A.}},
\bauthor{\bsnm{Litvinov}, \binits{A.N.}},
\bauthor{\bsnm{Romanenko}, \binits{V.I.}},
\bauthor{\bsnm{Yatsenko}, \binits{L.P.}},
\bauthor{\bsnm{Romanenko}, \binits{A.V.}},
\bauthor{\bsnm{Schreitl}, \binits{M.}},
\bauthor{\bsnm{Winkler}, \binits{G.}},
\bauthor{\bsnm{Schumm}, \binits{T.}}:
\batitle{Performance of a $^{229}${T}horium solid-state nuclear clock}.
\bjtitle{New Journal of Physics}
\bvolume{14}(\bissue{8}),
\bfpage{083019}
(\byear{2012}).
\doiurl{10.1088/1367-2630/14/8/083019}
\end{barticle}
\endbibitem

\bibitem{nuclear2}
\begin{barticle}
\bauthor{\bsnm{{von der Wense}}, \binits{L.}},
\bauthor{\bsnm{{Seiferle}}, \binits{B.}},
\bauthor{\bsnm{{Laatiaoui}}, \binits{M.}},
\bauthor{\bsnm{{Neumayr}}, \binits{J.B.}},
\bauthor{\bsnm{{Maier}}, \binits{H.-J.}},
\bauthor{\bsnm{{Wirth}}, \binits{H.-F.}},
\bauthor{\bsnm{{Mokry}}, \binits{C.}},
\bauthor{\bsnm{{Runke}}, \binits{J.}},
\bauthor{\bsnm{{Eberhardt}}, \binits{K.}},
\bauthor{\bsnm{{D{\"u}llmann}}, \binits{C.E.}},
\bauthor{\bsnm{{Trautmann}}, \binits{N.G.}},
\bauthor{\bsnm{{Thirolf}}, \binits{P.G.}}:
\batitle{{Direct detection of the $^{229}$Th nuclear clock transition}}.
\bjtitle{Nature}
\bvolume{533}(\bissue{7601}),
\bfpage{47}--\blpage{51}
(\byear{2016})
{\href{https://arxiv.org/abs/1710.11398}{{arXiv:1710.11398}}}
{[nucl-ex]}.
\doiurl{10.1038/nature17669}
\end{barticle}
\endbibitem

\bibitem{nuclear3}
\begin{barticle}
\bauthor{\bsnm{{Masuda}}, \binits{T.}},
\bauthor{\bsnm{{Yoshimi}}, \binits{A.}},
\bauthor{\bsnm{{Fujieda}}, \binits{A.}},
\bauthor{\bsnm{{Fujimoto}}, \binits{H.}},
\bauthor{\bsnm{{Haba}}, \binits{H.}},
\bauthor{\bsnm{{Hara}}, \binits{H.}},
\bauthor{\bsnm{{Hiraki}}, \binits{T.}},
\bauthor{\bsnm{{Kaino}}, \binits{H.}},
\bauthor{\bsnm{{Kasamatsu}}, \binits{Y.}},
\bauthor{\bsnm{{Kitao}}, \binits{S.}},
\bauthor{\bsnm{{Konashi}}, \binits{K.}},
\bauthor{\bsnm{{Miyamoto}}, \binits{Y.}},
\bauthor{\bsnm{{Okai}}, \binits{K.}},
\bauthor{\bsnm{{Okubo}}, \binits{S.}},
\bauthor{\bsnm{{Sasao}}, \binits{N.}},
\bauthor{\bsnm{{Seto}}, \binits{M.}},
\bauthor{\bsnm{{Schumm}}, \binits{T.}},
\bauthor{\bsnm{{Shigekawa}}, \binits{Y.}},
\bauthor{\bsnm{{Suzuki}}, \binits{K.}},
\bauthor{\bsnm{{Stellmer}}, \binits{S.}},
\bauthor{\bsnm{{Tamasaku}}, \binits{K.}},
\bauthor{\bsnm{{Uetake}}, \binits{S.}},
\bauthor{\bsnm{{Watanabe}}, \binits{M.}},
\bauthor{\bsnm{{Watanabe}}, \binits{T.}},
\bauthor{\bsnm{{Yasuda}}, \binits{Y.}},
\bauthor{\bsnm{{Yamaguchi}}, \binits{A.}},
\bauthor{\bsnm{{Yoda}}, \binits{Y.}},
\bauthor{\bsnm{{Yokokita}}, \binits{T.}},
\bauthor{\bsnm{{Yoshimura}}, \binits{M.}},
\bauthor{\bsnm{{Yoshimura}}, \binits{K.}}:
\batitle{{{X}-ray pumping of the $^{229}$Th nuclear clock isomer}}.
\bjtitle{Nature}
\bvolume{573}(\bissue{7773}),
\bfpage{238}--\blpage{242}
(\byear{2019})
{\href{https://arxiv.org/abs/1902.04823}{{arXiv:1902.04823}}}
{[nucl-ex]}.
\doiurl{10.1038/s41586-019-1542-3}
\end{barticle}
\endbibitem

\bibitem{nuclear4}
\begin{barticle}
\bauthor{\bsnm{{Seiferle}}, \binits{B.}},
\bauthor{\bsnm{{von der Wense}}, \binits{L.}},
\bauthor{\bsnm{{Bilous}}, \binits{P.V.}},
\bauthor{\bsnm{{Amersdorffer}}, \binits{I.}},
\bauthor{\bsnm{{Lemell}}, \binits{C.}},
\bauthor{\bsnm{{Libisch}}, \binits{F.}},
\bauthor{\bsnm{{Stellmer}}, \binits{S.}},
\bauthor{\bsnm{{Schumm}}, \binits{T.}},
\bauthor{\bsnm{{D{\"u}llmann}}, \binits{C.E.}},
\bauthor{\bsnm{{P{\'a}lffy}}, \binits{A.}},
\bauthor{\bsnm{{Thirolf}}, \binits{P.G.}}:
\batitle{{Energy of the $^{229}$Th nuclear clock transition}}.
\bjtitle{Nature}
\bvolume{573}(\bissue{7773}),
\bfpage{243}--\blpage{246}
(\byear{2019})
{\href{https://arxiv.org/abs/1905.06308}{{arXiv:1905.06308}}}
{[nucl-ex]}.
\doiurl{10.1038/s41586-019-1533-4}
\end{barticle}
\endbibitem

\bibitem{nuclear5}
\begin{barticle}
\bauthor{\bsnm{{Sikorsky}}, \binits{T.}},
\bauthor{\bsnm{{Geist}}, \binits{J.}},
\bauthor{\bsnm{{Hengstler}}, \binits{D.}},
\bauthor{\bsnm{{Kempf}}, \binits{S.}},
\bauthor{\bsnm{{Gastaldo}}, \binits{L.}},
\bauthor{\bsnm{{Enss}}, \binits{C.}},
\bauthor{\bsnm{{Mokry}}, \binits{C.}},
\bauthor{\bsnm{{Runke}}, \binits{J.}},
\bauthor{\bsnm{{D{\"u}llmann}}, \binits{C.E.}},
\bauthor{\bsnm{{Wobrauschek}}, \binits{P.}},
\bauthor{\bsnm{{Beeks}}, \binits{K.}},
\bauthor{\bsnm{{Rosecker}}, \binits{V.}},
\bauthor{\bsnm{{Sterba}}, \binits{J.H.}},
\bauthor{\bsnm{{Kazakov}}, \binits{G.}},
\bauthor{\bsnm{{Schumm}}, \binits{T.}},
\bauthor{\bsnm{{Fleischmann}}, \binits{A.}}:
\batitle{{Measurement of the $^{229}$Th Isomer Energy with a Magnetic Microcalorimeter}}.
\bjtitle{Physical Review Letters}
\bvolume{125}(\bissue{14}),
\bfpage{142503}
(\byear{2020})
{\href{https://arxiv.org/abs/2005.13340}{{arXiv:2005.13340}}}
{[nucl-ex]}.
\doiurl{10.1103/PhysRevLett.125.142503}
\end{barticle}
\endbibitem

\bibitem{nuclear6}
\begin{barticle}
\bauthor{\bsnm{{Kraemer}}, \binits{S.}},
\bauthor{\bsnm{{Moens}}, \binits{J.}},
\bauthor{\bsnm{{Athanasakis-Kaklamanakis}}, \binits{M.}},
\bauthor{\bsnm{{Bara}}, \binits{S.}},
\bauthor{\bsnm{{Beeks}}, \binits{K.}},
\bauthor{\bsnm{{Chhetri}}, \binits{P.}},
\bauthor{\bsnm{{Chrysalidis}}, \binits{K.}},
\bauthor{\bsnm{{Claessens}}, \binits{A.}},
\bauthor{\bsnm{{Cocolios}}, \binits{T.E.}},
\bauthor{\bsnm{{Correia}}, \binits{J.G.M.}},
\bauthor{\bsnm{{Witte}}, \binits{H.D.}},
\bauthor{\bsnm{{Ferrer}}, \binits{R.}},
\bauthor{\bsnm{{Geldhof}}, \binits{S.}},
\bauthor{\bsnm{{Heinke}}, \binits{R.}},
\bauthor{\bsnm{{Hosseini}}, \binits{N.}},
\bauthor{\bsnm{{Huyse}}, \binits{M.}},
\bauthor{\bsnm{{K{\"o}ster}}, \binits{U.}},
\bauthor{\bsnm{{Kudryavtsev}}, \binits{Y.}},
\bauthor{\bsnm{{Laatiaoui}}, \binits{M.}},
\bauthor{\bsnm{{Lica}}, \binits{R.}},
\bauthor{\bsnm{{Magchiels}}, \binits{G.}},
\bauthor{\bsnm{{Manea}}, \binits{V.}},
\bauthor{\bsnm{{Merckling}}, \binits{C.}},
\bauthor{\bsnm{{Pereira}}, \binits{L.M.C.}},
\bauthor{\bsnm{{Raeder}}, \binits{S.}},
\bauthor{\bsnm{{Schumm}}, \binits{T.}},
\bauthor{\bsnm{{Sels}}, \binits{S.}},
\bauthor{\bsnm{{Thirolf}}, \binits{P.G.}},
\bauthor{\bsnm{{Tunhuma}}, \binits{S.M.}},
\bauthor{\bsnm{{Van Den Bergh}}, \binits{P.}},
\bauthor{\bsnm{{Van Duppen}}, \binits{P.}},
\bauthor{\bsnm{{Vantomme}}, \binits{A.}},
\bauthor{\bsnm{{Verlinde}}, \binits{M.}},
\bauthor{\bsnm{{Villarreal}}, \binits{R.}},
\bauthor{\bsnm{{Wahl}}, \binits{U.}}:
\batitle{{Observation of the radiative decay of the $^{229}$Th nuclear clock isomer}}.
\bjtitle{Nature}
\bvolume{617}(\bissue{7962}),
\bfpage{706}--\blpage{710}
(\byear{2023})
{\href{https://arxiv.org/abs/2209.10276}{{arXiv:2209.10276}}}
{[nucl-ex]}.
\doiurl{10.1038/s41586-023-05894-z}
\end{barticle}
\endbibitem

\bibitem{nuclear7}
\begin{barticle}
\bauthor{\bsnm{Tiedau}, \binits{J.}},
\bauthor{\bsnm{Okhapkin}, \binits{M.V.}},
\bauthor{\bsnm{Zhang}, \binits{K.}},
\bauthor{\bsnm{Thielking}, \binits{J.}},
\bauthor{\bsnm{Zitzer}, \binits{G.}},
\bauthor{\bsnm{Peik}, \binits{E.}},
\bauthor{\bsnm{Schaden}, \binits{F.}},
\bauthor{\bsnm{Pronebner}, \binits{T.}},
\bauthor{\bsnm{Morawetz}, \binits{I.}},
\bauthor{\bsnm{De~Col}, \binits{L.T.}},
\bauthor{\bsnm{Schneider}, \binits{F.}},
\bauthor{\bsnm{Leitner}, \binits{A.}},
\bauthor{\bsnm{Pressler}, \binits{M.}},
\bauthor{\bsnm{Kazakov}, \binits{G.A.}},
\bauthor{\bsnm{Beeks}, \binits{K.}},
\bauthor{\bsnm{Sikorsky}, \binits{T.}},
\bauthor{\bsnm{Schumm}, \binits{T.}}:
\batitle{Laser excitation of the {T}h-229 nucleus}.
\bjtitle{Phys. Rev. Lett.}
\bvolume{132},
\bfpage{182501}
(\byear{2024}).
\doiurl{10.1103/PhysRevLett.132.182501}
\end{barticle}
\endbibitem

\bibitem{nuclear8}
\begin{barticle}
\bauthor{\bsnm{Campbell}, \binits{C.J.}},
\bauthor{\bsnm{Radnaev}, \binits{A.G.}},
\bauthor{\bsnm{Kuzmich}, \binits{A.}},
\bauthor{\bsnm{Dzuba}, \binits{V.A.}},
\bauthor{\bsnm{Flambaum}, \binits{V.V.}},
\bauthor{\bsnm{Derevianko}, \binits{A.}}:
\batitle{Single-ion nuclear clock for metrology at the 19th decimal place}.
\bjtitle{Phys. Rev. Lett.}
\bvolume{108},
\bfpage{120802}
(\byear{2012}).
\doiurl{10.1103/PhysRevLett.108.120802}
\end{barticle}
\endbibitem

\bibitem{th_freq}
\begin{barticle}
\bauthor{\bsnm{{Zhang}}, \binits{C.}},
\bauthor{\bsnm{{Ooi}}, \binits{T.}},
\bauthor{\bsnm{{Higgins}}, \binits{J.S.}},
\bauthor{\bsnm{{Doyle}}, \binits{J.F.}},
\bauthor{\bsnm{{von der Wense}}, \binits{L.}},
\bauthor{\bsnm{{Beeks}}, \binits{K.}},
\bauthor{\bsnm{{Leitner}}, \binits{A.}},
\bauthor{\bsnm{{Kazakov}}, \binits{G.A.}},
\bauthor{\bsnm{{Li}}, \binits{P.}},
\bauthor{\bsnm{{Thirolf}}, \binits{P.G.}},
\bauthor{\bsnm{{Schumm}}, \binits{T.}},
\bauthor{\bsnm{{Ye}}, \binits{J.}}:
\batitle{{Frequency ratio of the $^{229\textrm{m}}$Th nuclear isomeric transition and the $^{87}$Sr atomic clock}}.
\bjtitle{Nature}
\bvolume{633}(\bissue{8028}),
\bfpage{63}--\blpage{70}
(\byear{2024})
{\href{https://arxiv.org/abs/2406.18719}{{arXiv:2406.18719}}}
{[physics.atom-ph]}.
\doiurl{10.1038/s41586-024-07839-6}
\end{barticle}
\endbibitem

\bibitem{storage}
\begin{barticle}
\bauthor{\bsnm{Shvyd'ko}, \binits{Y.V.}},
\bauthor{\bsnm{Hertrich}, \binits{T.}},
\bauthor{\bparticle{van} \bsnm{B\"urck}, \binits{U.}},
\bauthor{\bsnm{Gerdau}, \binits{E.}},
\bauthor{\bsnm{Leupold}, \binits{O.}},
\bauthor{\bsnm{Metge}, \binits{J.}},
\bauthor{\bsnm{R\"uter}, \binits{H.D.}},
\bauthor{\bsnm{Schwendy}, \binits{S.}},
\bauthor{\bsnm{Smirnov}, \binits{G.V.}},
\bauthor{\bsnm{Potzel}, \binits{W.}},
\bauthor{\bsnm{Schindelmann}, \binits{P.}}:
\batitle{Storage of nuclear excitation energy through magnetic switching}.
\bjtitle{Phys. Rev. Lett.}
\bvolume{77},
\bfpage{3232}--\blpage{3235}
(\byear{1996}).
\doiurl{10.1103/PhysRevLett.77.3232}
\end{barticle}
\endbibitem

\bibitem{45Sc}
\begin{barticle}
\bauthor{\bsnm{Shvyd'ko}, \binits{Y.}},
\bauthor{\bsnm{Röhlsberger}, \binits{R.}},
\bauthor{\bsnm{Kocharovskaya}, \binits{O.}},
\bauthor{\bsnm{Evers}, \binits{J.}},
\bauthor{\bsnm{Geloni}, \binits{G.A.}},
\bauthor{\bsnm{Liu}, \binits{P.}},
\bauthor{\bsnm{Shu}, \binits{D.}},
\bauthor{\bsnm{Miceli}, \binits{A.}},
\bauthor{\bsnm{Stone}, \binits{B.}},
\bauthor{\bsnm{Hippler}, \binits{W.}},
\bauthor{\bsnm{Marx-Glowna}, \binits{B.}},
\bauthor{\bsnm{Uschmann}, \binits{I.}},
\bauthor{\bsnm{Loetzsch}, \binits{R.}},
\bauthor{\bsnm{Leupold}, \binits{O.}},
\bauthor{\bsnm{Wille}, \binits{H.-C.}},
\bauthor{\bsnm{Sergeev}, \binits{I.}},
\bauthor{\bsnm{Gerharz}, \binits{M.}},
\bauthor{\bsnm{Zhang}, \binits{X.}},
\bauthor{\bsnm{Grech}, \binits{C.}},
\bauthor{\bsnm{Guetg}, \binits{M.}},
\bauthor{\bsnm{Kocharyan}, \binits{V.}},
\bauthor{\bsnm{Kujala}, \binits{N.}},
\bauthor{\bsnm{Liu}, \binits{S.}},
\bauthor{\bsnm{Qin}, \binits{W.}},
\bauthor{\bsnm{Zozulya}, \binits{A.}},
\bauthor{\bsnm{Hallmann}, \binits{J.}},
\bauthor{\bsnm{Boesenberg}, \binits{U.}},
\bauthor{\bsnm{Jo}, \binits{W.}},
\bauthor{\bsnm{Möller}, \binits{J.}},
\bauthor{\bsnm{Rodriguez-Fernandez}, \binits{A.}},
\bauthor{\bsnm{Youssef}, \binits{M.}},
\bauthor{\bsnm{Madsen}, \binits{A.}},
\bauthor{\bsnm{Kolodziej}, \binits{T.}}:
\batitle{Resonant {X}-ray excitation of the nuclear clock isomer $^{45}${S}c}.
\bjtitle{Nature}
\bvolume{622}(\bissue{7983}),
\bfpage{471}--\blpage{475}
(\byear{2023}).
\doiurl{10.1038/s41586-023-06491-w}
\end{barticle}
\endbibitem

\bibitem{xfelo}
\begin{botherref}
\oauthor{\bsnm{{Adams}}, \binits{B.}},
\oauthor{\bsnm{{Aeppli}}, \binits{G.}},
\oauthor{\bsnm{{Allison}}, \binits{T.}},
\oauthor{\bsnm{{Baron}}, \binits{A.Q.R.}},
\oauthor{\bsnm{{Bucksbaum}}, \binits{P.}},
\oauthor{\bsnm{{Chumakov}}, \binits{A.I.}},
\oauthor{\bsnm{{Corder}}, \binits{C.}},
\oauthor{\bsnm{{Cramer}}, \binits{S.P.}},
\oauthor{\bsnm{{DeBeer}}, \binits{S.}},
\oauthor{\bsnm{{Ding}}, \binits{Y.}},
\oauthor{\bsnm{{Evers}}, \binits{J.}},
\oauthor{\bsnm{{Frisch}}, \binits{J.}},
\oauthor{\bsnm{{Fuchs}}, \binits{M.}},
\oauthor{\bsnm{{Gr{\"u}bel}}, \binits{G.}},
\oauthor{\bsnm{{Hastings}}, \binits{J.B.}},
\oauthor{\bsnm{{Heyl}}, \binits{C.M.}},
\oauthor{\bsnm{{Holberg}}, \binits{L.}},
\oauthor{\bsnm{{Huang}}, \binits{Z.}},
\oauthor{\bsnm{{Ishikawa}}, \binits{T.}},
\oauthor{\bsnm{{Kaldun}}, \binits{A.}},
\oauthor{\bsnm{{Kim}}, \binits{K.-J.}},
\oauthor{\bsnm{{Kolodziej}}, \binits{T.}},
\oauthor{\bsnm{{Krzywinski}}, \binits{J.}},
\oauthor{\bsnm{{Li}}, \binits{Z.}},
\oauthor{\bsnm{{Liao}}, \binits{W.-T.}},
\oauthor{\bsnm{{Lindberg}}, \binits{R.}},
\oauthor{\bsnm{{Madsen}}, \binits{A.}},
\oauthor{\bsnm{{Maxwell}}, \binits{T.}},
\oauthor{\bsnm{{Monaco}}, \binits{G.}},
\oauthor{\bsnm{{Nelson}}, \binits{K.}},
\oauthor{\bsnm{{Palffy}}, \binits{A.}},
\oauthor{\bsnm{{Porat}}, \binits{G.}},
\oauthor{\bsnm{{Qin}}, \binits{W.}},
\oauthor{\bsnm{{Raubenheimer}}, \binits{T.}},
\oauthor{\bsnm{{Reis}}, \binits{D.A.}},
\oauthor{\bsnm{{R{\"o}hlsberger}}, \binits{R.}},
\oauthor{\bsnm{{Santra}}, \binits{R.}},
\oauthor{\bsnm{{Schoenlein}}, \binits{R.}},
\oauthor{\bsnm{{Sch{\"u}nemann}}, \binits{V.}},
\oauthor{\bsnm{{Shpyrko}}, \binits{O.}},
\oauthor{\bsnm{{Shvyd'ko}}, \binits{Y.}},
\oauthor{\bsnm{{Shwartz}}, \binits{S.}},
\oauthor{\bsnm{{Singer}}, \binits{A.}},
\oauthor{\bsnm{{Sinha}}, \binits{S.K.}},
\oauthor{\bsnm{{Sutton}}, \binits{M.}},
\oauthor{\bsnm{{Tamasaku}}, \binits{K.}},
\oauthor{\bsnm{{Wille}}, \binits{H.-C.}},
\oauthor{\bsnm{{Yabashi}}, \binits{M.}},
\oauthor{\bsnm{{Ye}}, \binits{J.}},
\oauthor{\bsnm{{Zhu}}, \binits{D.}}:
{Scientific Opportunities with an X-ray Free-Electron Laser Oscillator}.
arXiv e-prints,
1903--09317
(2019)
{\href{https://arxiv.org/abs/1903.09317}{{arXiv:1903.09317}}}
{[physics.ins-det]}.
\doiurl{10.48550/arXiv.1903.09317}
\end{botherref}
\endbibitem

\bibitem{hhg}
\begin{bbook}
\bauthor{\bsnm{Jin}, \binits{C.}}:
\bbtitle{Introduction to High-Order Harmonic Generation},
pp. \bfpage{1}--\blpage{23}.
\bpublisher{Springer},
\blocation{Cham}
(\byear{2013}).
\doiurl{10.1007/978-3-319-01625-2_1}.
\burl{https://doi.org/10.1007/978-3-319-01625-2_1}
\end{bbook}
\endbibitem

\bibitem{progX}
\begin{bchapter}
\bauthor{\bsnm{Marcus}, \binits{G.}}, \betal:
\bctitle{{Cavity-Based Free-Electron Laser Research and Development: A Joint Argonne National Laboratory and {SLAC} {N}ational {L}aboratory {C}ollaboration}}.
In: \bbtitle{39th International Free Electron Laser Conference},
p. \bfpage{04}
(\byear{2019}).
\doiurl{10.18429/JACoW-FEL2019-TUD04}
\end{bchapter}
\endbibitem

\bibitem{progX2}
\begin{barticle}
\bauthor{\bsnm{Rauer}, \binits{P.}},
\bauthor{\bsnm{Decking}, \binits{W.}},
\bauthor{\bsnm{Lipka}, \binits{D.}},
\bauthor{\bsnm{Thoden}, \binits{D.}},
\bauthor{\bsnm{Wohlenberg}, \binits{T.}},
\bauthor{\bsnm{Bahns}, \binits{I.}},
\bauthor{\bsnm{Brueggmann}, \binits{U.}},
\bauthor{\bsnm{Casalbuoni}, \binits{S.}},
\bauthor{\bsnm{Di~Felice}, \binits{M.}},
\bauthor{\bsnm{Dommach}, \binits{M.}},
\bauthor{\bsnm{Gr\"unert}, \binits{J.}},
\bauthor{\bsnm{Karabekyan}, \binits{S.}},
\bauthor{\bsnm{Koch}, \binits{A.}},
\bauthor{\bsnm{La~Civita}, \binits{D.}},
\bauthor{\bsnm{Rio}, \binits{B.}},
\bauthor{\bsnm{Samoylova}, \binits{L.}},
\bauthor{\bsnm{Sinn}, \binits{H.}},
\bauthor{\bsnm{Vannoni}, \binits{M.}},
\bauthor{\bsnm{Youngman}, \binits{C.}},
\bauthor{\bsnm{Hillert}, \binits{W.}},
\bauthor{\bsnm{Rossbach}, \binits{J.}}:
\batitle{Cavity based {X}-ray free electron laser demonstrator at the european {X}-ray free electron laser facility}.
\bjtitle{Phys. Rev. Accel. Beams}
\bvolume{26},
\bfpage{020701}
(\byear{2023}).
\doiurl{10.1103/PhysRevAccelBeams.26.020701}
\end{barticle}
\endbibitem

\bibitem{molecular}
\begin{barticle}
\bauthor{\bsnm{Tannor}, \binits{D.J.}},
\bauthor{\bsnm{Kosloff}, \binits{R.}},
\bauthor{\bsnm{Rice}, \binits{S.A.}}:
\batitle{Coherent pulse sequence induced control of selectivity of reactions: Exact quantum mechanical calculations}.
\bjtitle{The Journal of Chemical Physics}
\bvolume{85}(\bissue{10}),
\bfpage{5805}--\blpage{5820}
(\byear{1986})
{\href{https://arxiv.org/abs/https://pubs.aip.org/aip/jcp/article-pdf/85/10/5805/18961430/5805\_1\_online.pdf}{{https://pubs.aip.org/aip/jcp/article-pdf/85/10/5805/18961430/5805\_1\_online.pdf}}}.
\doiurl{10.1063/1.451542}
\end{barticle}
\endbibitem

\bibitem{electrondynamics}
\begin{barticle}
\bauthor{\bsnm{{Bouchene, M. A.}}},
\bauthor{\bsnm{{Blanchet, V.}}},
\bauthor{\bsnm{{Nicole, C.}}},
\bauthor{\bsnm{{Melikechi, N.}}},
\bauthor{\bsnm{{Girard, B.}}},
\bauthor{\bsnm{{Ruppe, H.}}},
\bauthor{\bsnm{{Rutz, S.}}},
\bauthor{\bsnm{{Schreiber, E.}}},
\bauthor{\bsnm{{Wöste, L.}}}:
\batitle{Temporal coherent control induced by wave packet interferences in one and two photon atomic transitions}.
\bjtitle{Eur. Phys. J. D}
\bvolume{2}(\bissue{2}),
\bfpage{131}--\blpage{141}
(\byear{1998}).
\doiurl{10.1007/s100530050122}
\end{barticle}
\endbibitem

\bibitem{electrondynamics2}
\begin{barticle}
\bauthor{\bsnm{Pr\"akelt}, \binits{A.}},
\bauthor{\bsnm{Wollenhaupt}, \binits{M.}},
\bauthor{\bsnm{Sarpe-Tudoran}, \binits{C.}},
\bauthor{\bsnm{Baumert}, \binits{T.}}:
\batitle{Phase control of a two-photon transition with shaped femtosecond laser-pulse sequences}.
\bjtitle{Phys. Rev. A}
\bvolume{70},
\bfpage{063407}
(\byear{2004}).
\doiurl{10.1103/PhysRevA.70.063407}
\end{barticle}
\endbibitem

\bibitem{packets}
\begin{barticle}
\bauthor{\bsnm{Wollenhaupt}, \binits{M.}},
\bauthor{\bsnm{Assion}, \binits{A.}},
\bauthor{\bsnm{Liese}, \binits{D.}},
\bauthor{\bsnm{Sarpe-Tudoran}, \binits{C.}},
\bauthor{\bsnm{Baumert}, \binits{T.}},
\bauthor{\bsnm{Zamith}, \binits{S.}},
\bauthor{\bsnm{Bouchene}, \binits{M.A.}},
\bauthor{\bsnm{Girard}, \binits{B.}},
\bauthor{\bsnm{Flettner}, \binits{A.}},
\bauthor{\bsnm{Weichmann}, \binits{U.}},
\bauthor{\bsnm{Gerber}, \binits{G.}}:
\batitle{Interferences of ultrashort free electron wave packets}.
\bjtitle{Phys. Rev. Lett.}
\bvolume{89},
\bfpage{173001}
(\byear{2002}).
\doiurl{10.1103/PhysRevLett.89.173001}
\end{barticle}
\endbibitem

\bibitem{compact}
\begin{barticle}
\bauthor{\bsnm{{Pr{\"a}kelt}}, \binits{A.}},
\bauthor{\bsnm{{Wollenhaupt}}, \binits{M.}},
\bauthor{\bsnm{{Assion}}, \binits{A.}},
\bauthor{\bsnm{{Horn}}, \binits{C.}},
\bauthor{\bsnm{{Sarpe-Tudoran}}, \binits{C.}},
\bauthor{\bsnm{{Winter}}, \binits{M.}},
\bauthor{\bsnm{{Baumert}}, \binits{T.}}:
\batitle{{Compact, robust, and flexible setup for femtosecond pulse shaping}}.
\bjtitle{Review of Scientific Instruments}
\bvolume{74}(\bissue{11}),
\bfpage{4950}--\blpage{4953}
(\byear{2003}).
\doiurl{10.1063/1.1611998}
\end{barticle}
\endbibitem

\bibitem{zepto680}
\begin{barticle}
\bauthor{\bsnm{Tross}, \binits{J.}},
\bauthor{\bsnm{Kolliopoulos}, \binits{G.}},
\bauthor{\bsnm{Trallero-Herrero}, \binits{C.A.}}:
\batitle{Self referencing attosecond interferometer with zeptosecond precision}.
\bjtitle{Opt. Express}
\bvolume{27}(\bissue{16}),
\bfpage{22960}--\blpage{22969}
(\byear{2019}).
\doiurl{10.1364/OE.27.022960}
\end{barticle}
\endbibitem

\bibitem{atto_photo}
\begin{barticle}
\bauthor{\bsnm{Driver}, \binits{T.}},
\bauthor{\bsnm{Mountney}, \binits{M.}},
\bauthor{\bsnm{Wang}, \binits{J.}},
\bauthor{\bsnm{Ortmann}, \binits{L.}},
\bauthor{\bsnm{Al-Haddad}, \binits{A.}},
\bauthor{\bsnm{Berrah}, \binits{N.}},
\bauthor{\bsnm{Bostedt}, \binits{C.}},
\bauthor{\bsnm{Champenois}, \binits{E.G.}},
\bauthor{\bsnm{DiMauro}, \binits{L.F.}},
\bauthor{\bsnm{Duris}, \binits{J.}},
\bauthor{\bsnm{Garratt}, \binits{D.}},
\bauthor{\bsnm{Glownia}, \binits{J.M.}},
\bauthor{\bsnm{Guo}, \binits{Z.}},
\bauthor{\bsnm{Haxton}, \binits{D.}},
\bauthor{\bsnm{Isele}, \binits{E.}},
\bauthor{\bsnm{Ivanov}, \binits{I.}},
\bauthor{\bsnm{Ji}, \binits{J.}},
\bauthor{\bsnm{Kamalov}, \binits{A.}},
\bauthor{\bsnm{Li}, \binits{S.}},
\bauthor{\bsnm{Lin}, \binits{M.-F.}},
\bauthor{\bsnm{Marangos}, \binits{J.P.}},
\bauthor{\bsnm{Obaid}, \binits{R.}},
\bauthor{\bsnm{O’Neal}, \binits{J.T.}},
\bauthor{\bsnm{Rosenberger}, \binits{P.}},
\bauthor{\bsnm{Shivaram}, \binits{N.H.}},
\bauthor{\bsnm{Wang}, \binits{A.L.}},
\bauthor{\bsnm{Walter}, \binits{P.}},
\bauthor{\bsnm{Wolf}, \binits{T.J.A.}},
\bauthor{\bsnm{W\"{o}rner}, \binits{H.J.}},
\bauthor{\bsnm{Zhang}, \binits{Z.}},
\bauthor{\bsnm{Bucksbaum}, \binits{P.H.}},
\bauthor{\bsnm{Kling}, \binits{M.F.}},
\bauthor{\bsnm{Landsman}, \binits{A.S.}},
\bauthor{\bsnm{Lucchese}, \binits{R.R.}},
\bauthor{\bsnm{Emmanouilidou}, \binits{A.}},
\bauthor{\bsnm{Marinelli}, \binits{A.}},
\bauthor{\bsnm{Cryan}, \binits{J.P.}}:
\batitle{Attosecond delays in {X}-ray molecular ionization}.
\bjtitle{Nature}
\bvolume{632}(\bissue{8026}),
\bfpage{762}--\blpage{767}
(\byear{2024}).
\doiurl{10.1038/s41586-024-07771-9}
\end{barticle}
\endbibitem

\bibitem{rabbitt2}
\begin{barticle}
\bauthor{\bsnm{Mairesse}, \binits{Y.}},
\bauthor{\bparticle{de} \bsnm{Bohan}, \binits{A.}},
\bauthor{\bsnm{Frasinski}, \binits{L.J.}},
\bauthor{\bsnm{Merdji}, \binits{H.}},
\bauthor{\bsnm{Dinu}, \binits{L.C.}},
\bauthor{\bsnm{Monchicourt}, \binits{P.}},
\bauthor{\bsnm{Breger}, \binits{P.}},
\bauthor{\bsnm{Kovačev}, \binits{M.}},
\bauthor{\bsnm{Taïeb}, \binits{R.}},
\bauthor{\bsnm{Carré}, \binits{B.}},
\bauthor{\bsnm{Muller}, \binits{H.G.}},
\bauthor{\bsnm{Agostini}, \binits{P.}},
\bauthor{\bsnm{Salières}, \binits{P.}}:
\batitle{Attosecond synchronization of high-harmonic soft {X}-rays}.
\bjtitle{Science}
\bvolume{302}(\bissue{5650}),
\bfpage{1540}--\blpage{1543}
(\byear{2003})
{\href{https://arxiv.org/abs/https://www.science.org/doi/pdf/10.1126/science.1090277}{{https://www.science.org/doi/pdf/10.1126/science.1090277}}}.
\doiurl{10.1126/science.1090277}
\end{barticle}
\endbibitem

\bibitem{atto_beat}
\begin{barticle}
\bauthor{\bsnm{Jiang}, \binits{W.}},
\bauthor{\bsnm{Armstrong}, \binits{G.S.J.}},
\bauthor{\bsnm{Han}, \binits{L.}},
\bauthor{\bsnm{Xu}, \binits{Y.}},
\bauthor{\bsnm{Zuo}, \binits{Z.}},
\bauthor{\bsnm{Tong}, \binits{J.}},
\bauthor{\bsnm{Lu}, \binits{P.}},
\bauthor{\bsnm{Dahlstr\"om}, \binits{J.M.}},
\bauthor{\bsnm{Ueda}, \binits{K.}},
\bauthor{\bsnm{Brown}, \binits{A.C.}},
\bauthor{\bparticle{van~der} \bsnm{Hart}, \binits{H.W.}},
\bauthor{\bsnm{Gong}, \binits{X.}},
\bauthor{\bsnm{Wu}, \binits{J.}}:
\batitle{Resolving quantum interference black box through attosecond photoionization spectroscopy}.
\bjtitle{Phys. Rev. Lett.}
\bvolume{131},
\bfpage{203201}
(\byear{2023}).
\doiurl{10.1103/PhysRevLett.131.203201}
\end{barticle}
\endbibitem

\bibitem{frog}
\begin{barticle}
\bauthor{\bsnm{Mairesse}, \binits{Y.}},
\bauthor{\bsnm{Qu\'er\'e}, \binits{F.}}:
\batitle{Frequency-resolved optical gating for complete reconstruction of attosecond bursts}.
\bjtitle{Phys. Rev. A}
\bvolume{71},
\bfpage{011401}
(\byear{2005}).
\doiurl{10.1103/PhysRevA.71.011401}
\end{barticle}
\endbibitem

\bibitem{zepto_camera}
\begin{botherref}
\oauthor{\bsnm{{Ni}}, \binits{H.}},
\oauthor{\bsnm{{Donsa}}, \binits{S.}},
\oauthor{\bsnm{{Gong}}, \binits{X.}},
\oauthor{\bsnm{{Ueda}}, \binits{K.}},
\oauthor{\bsnm{{Wu}}, \binits{J.}},
\oauthor{\bsnm{{Burgd{\"o}rfer}}, \binits{J.}}:
{Zeptosecond Angular Streak Camera}.
arXiv e-prints,
2206--00581
(2022)
{\href{https://arxiv.org/abs/2206.00581}{{arXiv:2206.00581}}}
{[physics.atom-ph]}.
\doiurl{10.48550/arXiv.2206.00581}
\end{botherref}
\endbibitem

\bibitem{eit_first}
\begin{barticle}
\bauthor{\bsnm{Boller}, \binits{K.-J.}},
\bauthor{\bparticle{Imamo\ifmmode~\breve{g}\else} \bsnm{\u{g}\fi{}lu}, \binits{A.}},
\bauthor{\bsnm{Harris}, \binits{S.E.}}:
\batitle{Observation of electromagnetically induced transparency}.
\bjtitle{Phys. Rev. Lett.}
\bvolume{66},
\bfpage{2593}--\blpage{2596}
(\byear{1991}).
\doiurl{10.1103/PhysRevLett.66.2593}
\end{barticle}
\endbibitem

\bibitem{eit}
\begin{barticle}
\bauthor{\bsnm{Fleischhauer}, \binits{M.}},
\bauthor{\bsnm{Imamoglu}, \binits{A.}},
\bauthor{\bsnm{Marangos}, \binits{J.P.}}:
\batitle{Electromagnetically induced transparency: Optics in coherent media}.
\bjtitle{Rev. Mod. Phys.}
\bvolume{77},
\bfpage{633}--\blpage{673}
(\byear{2005}).
\doiurl{10.1103/RevModPhys.77.633}
\end{barticle}
\endbibitem

\bibitem{transparency}
\begin{barticle}
\bauthor{\bsnm{{Zhao}}, \binits{X.}},
\bauthor{\bsnm{{Huang}}, \binits{R.}},
\bauthor{\bsnm{{Du}}, \binits{X.}},
\bauthor{\bsnm{{Zhang}}, \binits{Z.}},
\bauthor{\bsnm{{Li}}, \binits{G.}}:
\batitle{{Ultrahigh-Q Metasurface Transparency Band Induced by Collective{\textendash}Collective Coupling}}.
\bjtitle{Nano Letters}
\bvolume{24}(\bissue{4}),
\bfpage{1238}--\blpage{1245}
(\byear{2024}).
\doiurl{10.1021/acs.nanolett.3c04174}
\end{barticle}
\endbibitem

\bibitem{gas17}
\begin{barticle}
\bauthor{\bsnm{Hau}, \binits{L.V.}},
\bauthor{\bsnm{Harris}, \binits{S.E.}},
\bauthor{\bsnm{Dutton}, \binits{Z.}},
\bauthor{\bsnm{Behroozi}, \binits{C.H.}}:
\batitle{Light speed reduction to 17 metres per second in an ultracold atomic gas}.
\bjtitle{Nature}
\bvolume{397},
\bfpage{594}--\blpage{598}
(\byear{1999})
\end{barticle}
\endbibitem

\bibitem{vapor}
\begin{barticle}
\bauthor{\bsnm{Budker}, \binits{D.}},
\bauthor{\bsnm{Kimball}, \binits{D.F.}},
\bauthor{\bsnm{Rochester}, \binits{S.M.}},
\bauthor{\bsnm{Yashchuk}, \binits{V.V.}}:
\batitle{Nonlinear magneto-optics and reduced group velocity of light in atomic vapor with slow ground state relaxation}.
\bjtitle{Phys. Rev. Lett.}
\bvolume{83},
\bfpage{1767}--\blpage{1770}
(\byear{1999}).
\doiurl{10.1103/PhysRevLett.83.1767}
\end{barticle}
\endbibitem

\bibitem{solid}
\begin{bchapter}
\bauthor{\bsnm{Turukhin}, \binits{A.}},
\bauthor{\bsnm{Sudarshanam}, \binits{V.S.}},
\bauthor{\bsnm{Shahriar}, \binits{M.S.}},
\bauthor{\bsnm{Musser}, \binits{J.A.}},
\bauthor{\bsnm{Hemmer}, \binits{P.R.}}:
\bctitle{First observation of ultraslow group velocity of light in a solid}.
In: \bbtitle{Technical Digest. Summaries of Papers Presented at the Quantum Electronics and Laser Science Conference. Postconference Technical Digest (IEEE Cat. No.01CH37172)},
pp. \bfpage{6}--\blpage{7}
(\byear{2001}).
\doiurl{10.1109/QELS.2001.961775}
\end{bchapter}
\endbibitem

\bibitem{opto}
\begin{barticle}
\bauthor{\bsnm{Yan}, \binits{X.-B.}}:
\batitle{Optomechanically induced ultraslow and ultrafast light}.
\bjtitle{Physica E: Low-dimensional Systems and Nanostructures}
\bvolume{131},
\bfpage{114759}
(\byear{2021}).
\doiurl{10.1016/j.physe.2021.114759}
\end{barticle}
\endbibitem

\bibitem{mikaeili22}
\begin{botherref}
\oauthor{\bsnm{Mikaeili}, \binits{H.}},
\oauthor{\bsnm{Dalafi}, \binits{A.}},
\oauthor{\bsnm{Ghanaatshoar}, \binits{M.}},
\oauthor{\bsnm{Askari}, \binits{B.}}:
Ultraslow light realization using an interacting {B}ose–{E}instein condensate trapped in a shallow optical lattice.
Scientific Reports
\textbf{12}(1)
(2022).
\doiurl{10.1038/s41598-022-08250-9}
\end{botherref}
\endbibitem

\bibitem{100mus}
\begin{barticle}
\bauthor{\bsnm{Phillips}, \binits{D.F.}},
\bauthor{\bsnm{Fleischhauer}, \binits{A.}},
\bauthor{\bsnm{Mair}, \binits{A.}},
\bauthor{\bsnm{Walsworth}, \binits{R.L.}},
\bauthor{\bsnm{Lukin}, \binits{M.D.}}:
\batitle{Storage of light in atomic vapor}.
\bjtitle{Phys. Rev. Lett.}
\bvolume{86},
\bfpage{783}--\blpage{786}
(\byear{2001}).
\doiurl{10.1103/PhysRevLett.86.783}
\end{barticle}
\endbibitem

\bibitem{halted}
\begin{barticle}
\bauthor{\bsnm{{Liu}}, \binits{C.}},
\bauthor{\bsnm{{Dutton}}, \binits{Z.}},
\bauthor{\bsnm{{Behroozi}}, \binits{C.H.}},
\bauthor{\bsnm{{Hau}}, \binits{L.V.}}:
\batitle{{Observation of coherent optical information storage in an atomic medium using halted light pulses}}.
\bjtitle{Nature}
\bvolume{409}(\bissue{6819}),
\bfpage{490}--\blpage{493}
(\byear{2001}).
\doiurl{10.1038/35054017}
\end{barticle}
\endbibitem

\bibitem{1sec}
\begin{barticle}
\bauthor{\bsnm{Zhang}, \binits{R.}},
\bauthor{\bsnm{Garner}, \binits{S.R.}},
\bauthor{\bsnm{Hau}, \binits{L.V.}}:
\batitle{Creation of long-term coherent optical memory via controlled nonlinear interactions in bose-einstein condensates}.
\bjtitle{Phys. Rev. Lett.}
\bvolume{103},
\bfpage{233602}
(\byear{2009}).
\doiurl{10.1103/PhysRevLett.103.233602}
\end{barticle}
\endbibitem

\bibitem{1sec_b}
\begin{barticle}
\bauthor{\bsnm{Schnorrberger}, \binits{U.}},
\bauthor{\bsnm{Thompson}, \binits{J.D.}},
\bauthor{\bsnm{Trotzky}, \binits{S.}},
\bauthor{\bsnm{Pugatch}, \binits{R.}},
\bauthor{\bsnm{Davidson}, \binits{N.}},
\bauthor{\bsnm{Kuhr}, \binits{S.}},
\bauthor{\bsnm{Bloch}, \binits{I.}}:
\batitle{Electromagnetically induced transparency and light storage in an atomic mott insulator}.
\bjtitle{Phys. Rev. Lett.}
\bvolume{103},
\bfpage{033003}
(\byear{2009}).
\doiurl{10.1103/PhysRevLett.103.033003}
\end{barticle}
\endbibitem

\bibitem{2sec}
\begin{barticle}
\bauthor{\bsnm{Longdell}, \binits{J.J.}},
\bauthor{\bsnm{Fraval}, \binits{E.}},
\bauthor{\bsnm{Sellars}, \binits{M.J.}},
\bauthor{\bsnm{Manson}, \binits{N.B.}}:
\batitle{Stopped light with storage times greater than one second using electromagnetically induced transparency in a solid}.
\bjtitle{Phys. Rev. Lett.}
\bvolume{95},
\bfpage{063601}
(\byear{2005}).
\doiurl{10.1103/PhysRevLett.95.063601}
\end{barticle}
\endbibitem

\bibitem{1min}
\begin{barticle}
\bauthor{\bsnm{Dudin}, \binits{Y.O.}},
\bauthor{\bsnm{Li}, \binits{L.}},
\bauthor{\bsnm{Kuzmich}, \binits{A.}}:
\batitle{Light storage on the time scale of a minute}.
\bjtitle{Phys. Rev. A}
\bvolume{87},
\bfpage{031801}
(\byear{2013}).
\doiurl{10.1103/PhysRevA.87.031801}
\end{barticle}
\endbibitem

\bibitem{1min_b}
\begin{barticle}
\bauthor{\bsnm{Heinze}, \binits{G.}},
\bauthor{\bsnm{Hubrich}, \binits{C.}},
\bauthor{\bsnm{Halfmann}, \binits{T.}}:
\batitle{Stopped light and image storage by electromagnetically induced transparency up to the regime of one minute}.
\bjtitle{Phys. Rev. Lett.}
\bvolume{111},
\bfpage{033601}
(\byear{2013}).
\doiurl{10.1103/PhysRevLett.111.033601}
\end{barticle}
\endbibitem

\bibitem{hollow}
\begin{barticle}
\bauthor{\bsnm{Russell}, \binits{P.}},
\bauthor{\bsnm{H{\"o}lzer}, \binits{P.}},
\bauthor{\bsnm{Chang}, \binits{W.}},
\bauthor{\bsnm{Abdolvand}, \binits{A.}},
\bauthor{\bsnm{Travers}, \binits{J.}}:
\batitle{Hollow-core photonic crystal fibres for gas-based nonlinear optics}.
\bjtitle{Nature Photonics}
\bvolume{8}(\bissue{4}),
\bfpage{278}--\blpage{286}
(\byear{2014}).
\doiurl{10.1038/nphoton.2013.312}
\end{barticle}
\endbibitem

\bibitem{hollow2}
\begin{barticle}
\bauthor{\bsnm{Travers}, \binits{J.C.}},
\bauthor{\bsnm{Chang}, \binits{W.}},
\bauthor{\bsnm{Nold}, \binits{J.}},
\bauthor{\bsnm{Joly}, \binits{N.Y.}},
\bauthor{\bsnm{Russell}, \binits{P.S.J.}}:
\batitle{Ultrafast nonlinear optics in gas-filled hollow-core photonic crystal fibers}.
\bjtitle{J. Opt. Soc. Am. B}
\bvolume{28}(\bissue{12}),
\bfpage{11}--\blpage{26}
(\byear{2011}).
\doiurl{10.1364/JOSAB.28.000A11}
\end{barticle}
\endbibitem

\bibitem{silica}
\begin{bchapter}
\bauthor{\bsnm{Tamura}, \binits{Y.}}:
\bctitle{Ultra-low loss silica core fiber}.
In: \bbtitle{Optical Fiber Communication Conference}
(\byear{2018}).
\doiurl{10.1364/OFC.2018.M4B.1}.
\burl{https://opg.optica.org/abstract.cfm?URI=OFC-2018-M4B.1}
\end{bchapter}
\endbibitem

\bibitem{silica2}
\begin{bchapter}
\bauthor{\bsnm{Yokota}, \binits{H.}},
\bauthor{\bsnm{Kanamori}, \binits{H.}},
\bauthor{\bsnm{Ishiguro}, \binits{Y.}},
\bauthor{\bsnm{Tanaka}, \binits{G.}},
\bauthor{\bsnm{Tanaka}, \binits{S.}},
\bauthor{\bsnm{Takada}, \binits{H.}},
\bauthor{\bsnm{Watanabe}, \binits{M.}},
\bauthor{\bsnm{Suzuki}, \binits{S.}},
\bauthor{\bsnm{Yano}, \binits{K.}},
\bauthor{\bsnm{Hoshikawa}, \binits{M.}},
\bauthor{\bsnm{Shimba}, \binits{H.}}:
\bctitle{Ultra-low-loss pure-silica-core single-mode fiber and transmission experiment}.
In: \bbtitle{Optical Fiber Communication}
(\byear{1986}).
\doiurl{10.1364/OFC.1986.PD3}.
\burl{https://opg.optica.org/abstract.cfm?URI=OFC-1986-PD3}
\end{bchapter}
\endbibitem

\bibitem{wmg}
\begin{barticle}
\bauthor{\bsnm{Dumeige}, \binits{Y.}},
\bauthor{\bsnm{F\'eron}, \binits{P.}}:
\batitle{Whispering-gallery-mode analysis of phase-matched doubly resonant second-harmonic generation}.
\bjtitle{Phys. Rev. A}
\bvolume{74},
\bfpage{063804}
(\byear{2006}).
\doiurl{10.1103/PhysRevA.74.063804}
\end{barticle}
\endbibitem

\bibitem{wmg2}
\begin{barticle}
\bauthor{\bsnm{Lin}, \binits{G.}},
\bauthor{\bsnm{Coillet}, \binits{A.}},
\bauthor{\bsnm{Chembo}, \binits{Y.K.}}:
\batitle{Nonlinear photonics with high-q whispering-gallery-mode resonators}.
\bjtitle{Adv. Opt. Photon.}
\bvolume{9}(\bissue{4}),
\bfpage{828}--\blpage{890}
(\byear{2017}).
\doiurl{10.1364/AOP.9.000828}
\end{barticle}
\endbibitem

\bibitem{wmg3}
\begin{barticle}
\bauthor{\bsnm{Huet}, \binits{V.}},
\bauthor{\bsnm{Rasoloniaina}, \binits{A.}},
\bauthor{\bsnm{Guillem\'e}, \binits{P.}},
\bauthor{\bsnm{Rochard}, \binits{P.}},
\bauthor{\bsnm{F\'eron}, \binits{P.}},
\bauthor{\bsnm{Mortier}, \binits{M.}},
\bauthor{\bsnm{Levenson}, \binits{A.}},
\bauthor{\bsnm{Bencheikh}, \binits{K.}},
\bauthor{\bsnm{Yacomotti}, \binits{A.}},
\bauthor{\bsnm{Dumeige}, \binits{Y.}}:
\batitle{Millisecond photon lifetime in a slow-light microcavity}.
\bjtitle{Phys. Rev. Lett.}
\bvolume{116},
\bfpage{133902}
(\byear{2016}).
\doiurl{10.1103/PhysRevLett.116.133902}
\end{barticle}
\endbibitem

\bibitem{crr}
\begin{barticle}
\bauthor{\bsnm{{Guo}}, \binits{Z.}},
\bauthor{\bsnm{{Huang}}, \binits{Q.}}:
\batitle{{Photon Storage in a Dynamic Two-Ring-Two-Bus System}}.
\bjtitle{Optics and Photonics Journal}
\bvolume{9}(\bissue{8}),
\bfpage{20}--\blpage{25}
(\byear{2019}).
\doiurl{10.4236/opj.2019.98B003}
\end{barticle}
\endbibitem

\bibitem{crr2}
\begin{bchapter}
\bauthor{\bsnm{Ciminelli}, \binits{C.}},
\bauthor{\bsnm{Dell'Olio}, \binits{F.}},
\bauthor{\bsnm{Campanella}, \binits{C.E.}},
\bauthor{\bsnm{Armenise}, \binits{M.N.}}:
\bctitle{Coupled ring resonators: Physical effects and potential applications}.
In: \bbtitle{IEEE Photonics Conference 2012},
pp. \bfpage{135}--\blpage{136}
(\byear{2012}).
\doiurl{10.1109/IPCon.2012.6358526}
\end{bchapter}
\endbibitem

\bibitem{fabry1}
\begin{botherref}
\oauthor{\bsnm{Gallego}, \binits{J.}},
\oauthor{\bsnm{Ghosh}, \binits{S.}},
\oauthor{\bsnm{Alavi}, \binits{S.K.}},
\oauthor{\bsnm{Alt}, \binits{W.}},
\oauthor{\bsnm{Martinez-Dorantes}, \binits{M.}},
\oauthor{\bsnm{Meschede}, \binits{D.}},
\oauthor{\bsnm{Ratschbacher}, \binits{L.}}:
High-finesse fiber {Fabry--Perot} cavities: stabilization and mode matching analysis.
Appl. Phys. B
\textbf{122}(3)
(2016)
\end{botherref}
\endbibitem

\bibitem{fabry2}
\begin{barticle}
\bauthor{\bsnm{Hunger}, \binits{D.}},
\bauthor{\bsnm{Steinmetz}, \binits{T.}},
\bauthor{\bsnm{Colombe}, \binits{Y.}},
\bauthor{\bsnm{Deutsch}, \binits{C.}},
\bauthor{\bsnm{Hänsch}, \binits{T.W.}},
\bauthor{\bsnm{Reichel}, \binits{J.}}:
\batitle{A fiber {F}abry–{P}erot cavity with high finesse}.
\bjtitle{New Journal of Physics}
\bvolume{12}(\bissue{6}),
\bfpage{065038}
(\byear{2010}).
\doiurl{10.1088/1367-2630/12/6/065038}
\end{barticle}
\endbibitem

\bibitem{fabry3}
\begin{barticle}
\bauthor{\bsnm{Chen}, \binits{X.}},
\bauthor{\bsnm{Chardin}, \binits{C.}},
\bauthor{\bsnm{Makles}, \binits{K.}},
\bauthor{\bsnm{Ca{\"e}r}, \binits{C.}},
\bauthor{\bsnm{Chua}, \binits{S.}},
\bauthor{\bsnm{Braive}, \binits{R.}},
\bauthor{\bsnm{Robert-Philip}, \binits{I.}},
\bauthor{\bsnm{Briant}, \binits{T.}},
\bauthor{\bsnm{Cohadon}, \binits{P.-F.}},
\bauthor{\bsnm{Heidmann}, \binits{A.}},
\bauthor{\bsnm{Jacqmin}, \binits{T.}},
\bauthor{\bsnm{Del{\'e}glise}, \binits{S.}}:
\batitle{High-finesse {Fabry--Perot} cavities with bidimensional {Si$_3$N$_4$} photonic-crystal slabs}.
\bjtitle{Light Sci. Appl.}
\bvolume{6}(\bissue{1}),
\bfpage{16190}--\blpage{16190}
(\byear{2016})
\end{barticle}
\endbibitem

\bibitem{fabry_micro}
\begin{botherref}
\oauthor{\bsnm{Bitarafan}, \binits{M.H.}},
\oauthor{\bsnm{DeCorby}, \binits{R.G.}}:
On-chip high-finesse {F}abry-{P}erot microcavities for optical sensing and quantum information.
Sensors
\textbf{17}(8)
(2017).
\doiurl{10.3390/s17081748}
\end{botherref}
\endbibitem

\bibitem{fabry_long}
\begin{barticle}
\bauthor{\bsnm{Valle}, \binits{F.D.}},
\bauthor{\bsnm{Milotti}, \binits{E.}},
\bauthor{\bsnm{Ejlli}, \binits{A.}},
\bauthor{\bsnm{Gastaldi}, \binits{U.}},
\bauthor{\bsnm{Messineo}, \binits{G.}},
\bauthor{\bsnm{Piemontese}, \binits{L.}},
\bauthor{\bsnm{Zavattini}, \binits{G.}},
\bauthor{\bsnm{Pengo}, \binits{R.}},
\bauthor{\bsnm{Ruoso}, \binits{G.}}:
\batitle{Extremely long decay time optical cavity}.
\bjtitle{Opt. Express}
\bvolume{22}(\bissue{10}),
\bfpage{11570}--\blpage{11577}
(\byear{2014}).
\doiurl{10.1364/OE.22.011570}
\end{barticle}
\endbibitem

\bibitem{fabry_nl1}
\begin{barticle}
\bauthor{\bsnm{{Heuck}}, \binits{M.}},
\bauthor{\bsnm{{Jacobs}}, \binits{K.}},
\bauthor{\bsnm{{Englund}}, \binits{D.R.}}:
\batitle{{Photon-photon interactions in dynamically coupled cavities}}.
\bjtitle{Phys. Rev. A}
\bvolume{101}(\bissue{4}),
\bfpage{042322}
(\byear{2020})
{\href{https://arxiv.org/abs/1905.02134}{{arXiv:1905.02134}}}
{[quant-ph]}.
\doiurl{10.1103/PhysRevA.101.042322}
\end{barticle}
\endbibitem

\bibitem{fabry_nl2}
\begin{barticle}
\bauthor{\bsnm{Jabri}, \binits{H.}},
\bauthor{\bsnm{Eleuch}, \binits{H.}}:
\batitle{Light squeezing enhancement by coupling nonlinear optical cavities}.
\bjtitle{Sci. Rep.}
\bvolume{14}(\bissue{1}),
\bfpage{7753}
(\byear{2024})
\end{barticle}
\endbibitem

\bibitem{ee}
\begin{botherref}
\oauthor{\bsnm{Hsu}, \binits{C.W.}},
\oauthor{\bsnm{Zhen}, \binits{B.}},
\oauthor{\bsnm{Stone}, \binits{A.D.}},
\oauthor{\bsnm{Joannopoulos}, \binits{J.D.}},
\oauthor{\bsnm{Solja{\v c}i{\'c}}, \binits{M.}}:
Bound states in the continuum.
Nat. Rev. Mater.
\textbf{1}(9)
(2016)
\end{botherref}
\endbibitem

\bibitem{ee2}
\begin{barticle}
\bauthor{\bsnm{Friedrich}, \binits{H.}},
\bauthor{\bsnm{Wintgen}, \binits{D.}}:
\batitle{Interfering resonances and bound states in the continuum}.
\bjtitle{Phys. Rev. A}
\bvolume{32},
\bfpage{3231}--\blpage{3242}
(\byear{1985}).
\doiurl{10.1103/PhysRevA.32.3231}
\end{barticle}
\endbibitem

\bibitem{ee3}
\begin{barticle}
\bauthor{\bsnm{Monticone}, \binits{F.}},
\bauthor{\bsnm{Al\`u}, \binits{A.}}:
\batitle{Embedded photonic eigenvalues in 3d nanostructures}.
\bjtitle{Phys. Rev. Lett.}
\bvolume{112},
\bfpage{213903}
(\byear{2014}).
\doiurl{10.1103/PhysRevLett.112.213903}
\end{barticle}
\endbibitem

\bibitem{infopres}
\begin{barticle}
\bauthor{\bsnm{Cotrufo}, \binits{M.}},
\bauthor{\bsnm{Al\`{u}}, \binits{A.}}:
\batitle{Excitation of single-photon embedded eigenstates in coupled cavity atom systems}.
\bjtitle{Optica}
\bvolume{6}(\bissue{6}),
\bfpage{799}--\blpage{804}
(\byear{2019}).
\doiurl{10.1364/OPTICA.6.000799}
\end{barticle}
\endbibitem

\bibitem{1hour}
\begin{barticle}
\bauthor{\bsnm{{Ma}}, \binits{Y.}},
\bauthor{\bsnm{{Ma}}, \binits{Y.-Z.}},
\bauthor{\bsnm{{Zhou}}, \binits{Z.-Q.}},
\bauthor{\bsnm{{Li}}, \binits{C.-F.}},
\bauthor{\bsnm{{Guo}}, \binits{G.-C.}}:
\batitle{{One-hour coherent optical storage in an atomic frequency comb memory}}.
\bjtitle{Nature Communications}
\bvolume{12},
\bfpage{2381}
(\byear{2021})
{\href{https://arxiv.org/abs/2012.14605}{{arXiv:2012.14605}}}
{[quant-ph]}.
\doiurl{10.1038/s41467-021-22706-y}
\end{barticle}
\endbibitem

\bibitem{muares}
\begin{barticle}
\bauthor{\bsnm{{Sesana et al.}}, \binits{A.}}:
\batitle{{Unveiling the gravitational universe at {\ensuremath{\mu}}{H}z frequencies}}.
\bjtitle{Experimental Astronomy}
\bvolume{51}(\bissue{3}),
\bfpage{1333}--\blpage{1383}
(\byear{2021})
{\href{https://arxiv.org/abs/1908.11391}{{arXiv:1908.11391}}}
{[astro-ph.IM]}.
\doiurl{10.1007/s10686-021-09709-9}
\end{barticle}
\endbibitem

\bibitem{pta0}
\begin{bchapter}
\bauthor{\bsnm{{Sesana}}, \binits{A.}}:
\bctitle{{Pulsar Timing Arrays and the Challenge of Massive Black Hole Binary Astrophysics}}.
In: \bbtitle{Gravitational Wave Astrophysics}.
\bsertitle{Astrophysics and Space Science Proceedings},
vol. \bseriesno{40},
p. \bfpage{147}
(\byear{2015}).
\doiurl{10.1007/978-3-319-10488-1\_13}
\end{bchapter}
\endbibitem

\bibitem{pta1}
\begin{barticle}
\bauthor{\bsnm{{Verbiest et al.}}, \binits{J.P.W.}}:
\batitle{{The International Pulsar Timing Array: First data release}}.
\bjtitle{Monthly Notices of the Royal Astronomical Society}
\bvolume{458}(\bissue{2}),
\bfpage{1267}--\blpage{1288}
(\byear{2016})
{\href{https://arxiv.org/abs/1602.03640}{{arXiv:1602.03640}}}
{[astro-ph.IM]}.
\doiurl{10.1093/mnras/stw347}
\end{barticle}
\endbibitem

\bibitem{pta2}
\begin{bchapter}
\bauthor{\bsnm{Verbiest}, \binits{J.P.W.}},
\bauthor{\bsnm{Os{\l}owski}, \binits{S.}},
\bauthor{\bsnm{Burke-Spolaor}, \binits{S.}}:
\bctitle{Pulsar timing array experiments}.
In: \beditor{\bsnm{Bambi}, \binits{C.}},
\beditor{\bsnm{Katsanevas}, \binits{S.}},
\beditor{\bsnm{Kokkotas}, \binits{K.D.}} (eds.)
\bbtitle{Handbook of Gravitational Wave Astronomy},
pp. \bfpage{1}--\blpage{42}.
\bpublisher{Springer},
\blocation{Singapore}
(\byear{2021}).
\doiurl{10.1007/978-981-15-4702-7_4-1}.
\burl{https://doi.org/10.1007/978-981-15-4702-7_4-1}
\end{bchapter}
\endbibitem

\bibitem{pta3}
\begin{bchapter}
\bauthor{\bsnm{{Manchester}}, \binits{R.N.}}:
\bctitle{{Pulsar Timing Arrays and their Applications}}.
In: \beditor{\bsnm{{Burgay}}, \binits{M.}},
\beditor{\bsnm{{D'Amico}}, \binits{N.}},
\beditor{\bsnm{{Esposito}}, \binits{P.}},
\beditor{\bsnm{{Pellizzoni}}, \binits{A.}},
\beditor{\bsnm{{Possenti}}, \binits{A.}} (eds.)
\bbtitle{Radio Pulsars: An Astrophysical Key to Unlock the Secrets of the Universe}.
\bsertitle{American Institute of Physics Conference Series},
vol. \bseriesno{1357},
pp. \bfpage{65}--\blpage{72}
(\byear{2011}).
\doiurl{10.1063/1.3615080}
\end{bchapter}
\endbibitem

\bibitem{airy0}
\begin{barticle}
\bauthor{\bsnm{Berry}, \binits{M.V.}},
\bauthor{\bsnm{Balazs}, \binits{N.L.}}:
\batitle{Nonspreading wave packets}.
\bjtitle{American Journal of Physics}
\bvolume{47}(\bissue{3}),
\bfpage{264}--\blpage{267}
(\byear{1979})
{\href{https://arxiv.org/abs/https://pubs.aip.org/aapt/ajp/article-pdf/47/3/264/11943535/264\_1\_online.pdf}{{https://pubs.aip.org/aapt/ajp/article-pdf/47/3/264/11943535/264\_1\_online.pdf}}}.
\doiurl{10.1119/1.11855}
\end{barticle}
\endbibitem

\bibitem{airy2}
\begin{barticle}
\bauthor{\bsnm{Maruca}, \binits{S.}},
\bauthor{\bsnm{Kumar}, \binits{S.}},
\bauthor{\bsnm{Sua}, \binits{Y.M.}},
\bauthor{\bsnm{Chen}, \binits{J.-Y.}},
\bauthor{\bsnm{Shahverdi}, \binits{A.}},
\bauthor{\bsnm{Huang}, \binits{Y.-P.}}:
\batitle{Quantum {A}iry photons}.
\bjtitle{J. Phys. B At. Mol. Opt. Phys.}
\bvolume{51}(\bissue{17}),
\bfpage{175501}
(\byear{2018})
\end{barticle}
\endbibitem

\bibitem{bessel_1}
\begin{barticle}
\bauthor{\bsnm{Fahrbach}, \binits{F.O.}},
\bauthor{\bsnm{Simon}, \binits{P.}},
\bauthor{\bsnm{Rohrbach}, \binits{A.}}:
\batitle{Microscopy with self-reconstructing beams}.
\bjtitle{Nature Photonics}
\bvolume{4}(\bissue{11}),
\bfpage{780}--\blpage{785}
(\byear{2010}).
\doiurl{10.1038/nphoton.2010.204}
\end{barticle}
\endbibitem

\bibitem{bessel_2}
\begin{barticle}
\bauthor{\bsnm{Mitri}, \binits{F.G.}}:
\batitle{Arbitrary scattering of an electromagnetic zero-order {B}essel beam by a dielectric sphere}.
\bjtitle{Opt. Lett.}
\bvolume{36}(\bissue{5}),
\bfpage{766}--\blpage{768}
(\byear{2011}).
\doiurl{10.1364/OL.36.000766}
\end{barticle}
\endbibitem

\bibitem{airy_beams}
\begin{barticle}
\bauthor{\bsnm{Siviloglou}, \binits{G.A.}},
\bauthor{\bsnm{Broky}, \binits{J.}},
\bauthor{\bsnm{Dogariu}, \binits{A.}},
\bauthor{\bsnm{Christodoulides}, \binits{D.N.}}:
\batitle{Observation of accelerating {A}iry beams}.
\bjtitle{Phys. Rev. Lett.}
\bvolume{99},
\bfpage{213901}
(\byear{2007}).
\doiurl{10.1103/PhysRevLett.99.213901}
\end{barticle}
\endbibitem

\bibitem{airy_pre}
\begin{barticle}
\bauthor{\bsnm{{Berry}}, \binits{M.V.}},
\bauthor{\bsnm{{Balazs}}, \binits{N.L.}}:
\batitle{{Nonspreading wave packets}}.
\bjtitle{American Journal of Physics}
\bvolume{47}(\bissue{3}),
\bfpage{264}--\blpage{267}
(\byear{1979}).
\doiurl{10.1119/1.11855}
\end{barticle}
\endbibitem

\bibitem{parabolic}
\begin{barticle}
\bauthor{\bsnm{Bandres}, \binits{M.}}:
\batitle{Accelerating parabolic beams}.
\bjtitle{Optics Letters}
\bvolume{33},
\bfpage{1678}--\blpage{1680}
(\byear{2008}).
\doiurl{10.1364/OL.33.001678}
\end{barticle}
\endbibitem

\bibitem{airy_e}
\begin{barticle}
\bauthor{\bsnm{Voloch-Bloch}, \binits{N.}},
\bauthor{\bsnm{Lereah}, \binits{Y.}},
\bauthor{\bsnm{Lilach}, \binits{Y.}},
\bauthor{\bsnm{Gover}, \binits{A.}},
\bauthor{\bsnm{Arie}, \binits{A.}}:
\batitle{Generation of electron airy beams}.
\bjtitle{Nature}
\bvolume{494}(\bissue{7437}),
\bfpage{331}--\blpage{335}
(\byear{2013}).
\doiurl{10.1038/nature11840}
\end{barticle}
\endbibitem

\bibitem{airy_n}
\begin{barticle}
\bauthor{\bsnm{Sarenac}, \binits{D.}},
\bauthor{\bsnm{Lailey}, \binits{O.}},
\bauthor{\bsnm{Henderson}, \binits{M.E.}},
\bauthor{\bsnm{Ekinci}, \binits{H.}},
\bauthor{\bsnm{Clark}, \binits{C.W.}},
\bauthor{\bsnm{Cory}, \binits{D.G.}},
\bauthor{\bsnm{DeBeer-Schmitt}, \binits{L.}},
\bauthor{\bsnm{Huber}, \binits{M.G.}},
\bauthor{\bsnm{White}, \binits{J.S.}},
\bauthor{\bsnm{Zhernenkov}, \binits{K.}},
\bauthor{\bsnm{Pushin}, \binits{D.A.}}:
\batitle{Generation of neutron {A}iry beams}.
\bjtitle{Phys. Rev. Lett.}
\bvolume{134},
\bfpage{153401}
(\byear{2025}).
\doiurl{10.1103/PhysRevLett.134.153401}
\end{barticle}
\endbibitem

\bibitem{airy_period1}
\begin{barticle}
\bauthor{\bsnm{Chremmos}, \binits{I.D.}},
\bauthor{\bsnm{Efremidis}, \binits{N.K.}}:
\batitle{Band-specific phase engineering for curving and focusing light in waveguide arrays}.
\bjtitle{Phys. Rev. A}
\bvolume{85},
\bfpage{063830}
(\byear{2012}).
\doiurl{10.1103/PhysRevA.85.063830}
\end{barticle}
\endbibitem

\bibitem{airy_period2}
\begin{barticle}
\bauthor{\bsnm{Efremidis}, \binits{N.K.}},
\bauthor{\bsnm{Chremmos}, \binits{I.D.}}:
\batitle{Caustic design in periodic lattices}.
\bjtitle{Opt. Lett.}
\bvolume{37}(\bissue{7}),
\bfpage{1277}--\blpage{1279}
(\byear{2012}).
\doiurl{10.1364/OL.37.001277}
\end{barticle}
\endbibitem

\bibitem{airy_propag2}
\begin{barticle}
\bauthor{\bsnm{Aadhi}, \binits{A.}},
\bauthor{\bsnm{Chaitanya}, \binits{N.A.}},
\bauthor{\bsnm{Jabir}, \binits{M.V.}},
\bauthor{\bsnm{Vaity}, \binits{P.}},
\bauthor{\bsnm{Singh}, \binits{R.P.}},
\bauthor{\bsnm{Samanta}, \binits{G.K.}}:
\batitle{Airy beam optical parametric oscillator}.
\bjtitle{Sci. Rep.}
\bvolume{6}(\bissue{1}),
\bfpage{25245}
(\byear{2016})
\end{barticle}
\endbibitem

\bibitem{airy_comp1}
\begin{barticle}
\bauthor{\bsnm{Preciado}, \binits{M.A.}},
\bauthor{\bsnm{Dholakia}, \binits{K.}},
\bauthor{\bsnm{Mazilu}, \binits{M.}}:
\batitle{Generation of attenuation-compensating {A}iry beams}.
\bjtitle{Opt. Lett.}
\bvolume{39}(\bissue{16}),
\bfpage{4950}--\blpage{4953}
(\byear{2014}).
\doiurl{10.1364/OL.39.004950}
\end{barticle}
\endbibitem

\bibitem{airy_comp}
\begin{botherref}
\oauthor{\bsnm{Schley}, \binits{R.}},
\oauthor{\bsnm{Kaminer}, \binits{I.}},
\oauthor{\bsnm{Greenfield}, \binits{E.}},
\oauthor{\bsnm{Bekenstein}, \binits{R.}},
\oauthor{\bsnm{Lumer}, \binits{Y.}},
\oauthor{\bsnm{Segev}, \binits{M.}}:
Loss-proof self-accelerating beams and their use in non-paraxial manipulation of particles’ trajectories.
Nature Communications
\textbf{5}(1)
(2014).
\doiurl{10.1038/ncomms6189}
\end{botherref}
\endbibitem

\bibitem{airy1}
\begin{barticle}
\bauthor{\bsnm{Wang}, \binits{J.}},
\bauthor{\bsnm{Zuo}, \binits{Y.}},
\bauthor{\bsnm{Wang}, \binits{X.}},
\bauthor{\bsnm{Christodoulides}, \binits{D.N.}},
\bauthor{\bsnm{Siviloglou}, \binits{G.A.}},
\bauthor{\bsnm{Chen}, \binits{J.F.}}:
\batitle{Spatiotemporal single-photon {A}iry bullets}.
\bjtitle{Phys. Rev. Lett.}
\bvolume{132},
\bfpage{143601}
(\byear{2024}).
\doiurl{10.1103/PhysRevLett.132.143601}
\end{barticle}
\endbibitem

\bibitem{airy_propag}
\begin{barticle}
\bauthor{\bsnm{Ji}, \binits{X.}},
\bauthor{\bsnm{Eyyubo\u{g}lu}, \binits{H.T.}},
\bauthor{\bsnm{Ji}, \binits{G.}},
\bauthor{\bsnm{Jia}, \binits{X.}}:
\batitle{Propagation of an {A}iry beam through the atmosphere}.
\bjtitle{Opt. Express}
\bvolume{21}(\bissue{2}),
\bfpage{2154}--\blpage{2164}
(\byear{2013}).
\doiurl{10.1364/OE.21.002154}
\end{barticle}
\endbibitem

\bibitem{bessel_propag3}
\begin{botherref}
\oauthor{\bsnm{Qi}, \binits{H.}},
\oauthor{\bsnm{Li}, \binits{Z.}},
\oauthor{\bsnm{Wang}, \binits{Y.}},
\oauthor{\bsnm{Chen}, \binits{X.}},
\oauthor{\bsnm{Pan}, \binits{H.}},
\oauthor{\bsnm{Wu}, \binits{E.}},
\oauthor{\bsnm{Wu}, \binits{G.}}:
Bessel-beam single-photon high-resolution imaging in time and space.
Photonics
\textbf{11}(8)
(2024).
\doiurl{10.3390/photonics11080704}
\end{botherref}
\endbibitem

\bibitem{bessel_propag1}
\begin{barticle}
\bauthor{\bsnm{{Negri}}, \binits{E.}},
\bauthor{\bsnm{{Giusti}}, \binits{F.}},
\bauthor{\bsnm{{Fuscaldo}}, \binits{W.}},
\bauthor{\bsnm{{Burghignoli}}, \binits{P.}},
\bauthor{\bsnm{{Martini}}, \binits{E.}},
\bauthor{\bsnm{{Galli}}, \binits{A.}}:
\batitle{{Ultra-long-range {B}essel beams via leaky waves with mitigated open stopband}}.
\bjtitle{Applied Physics Letters}
\bvolume{126}(\bissue{12}),
\bfpage{121703}
(\byear{2025})
{\href{https://arxiv.org/abs/2412.13338}{{arXiv:2412.13338}}}
{[physics.optics]}.
\doiurl{10.1063/5.0253371}
\end{barticle}
\endbibitem

\bibitem{bessel_propag2}
\begin{barticle}
\bauthor{\bsnm{Birch}, \binits{P.}},
\bauthor{\bsnm{Ituen}, \binits{I.}},
\bauthor{\bsnm{Young}, \binits{R.}},
\bauthor{\bsnm{Chatwin}, \binits{C.}}:
\batitle{Long-distance {B}essel beam propagation through kolmogorov turbulence}.
\bjtitle{J. Opt. Soc. Am. A}
\bvolume{32}(\bissue{11}),
\bfpage{2066}--\blpage{2073}
(\byear{2015}).
\doiurl{10.1364/JOSAA.32.002066}
\end{barticle}
\endbibitem

\bibitem{counting}
\begin{bbook}
\bauthor{\bsnm{Becker}, \binits{W.}}:
\bbtitle{Advanced Time-correlated Single Photon Counting Techniques},
\bedition{2005} edn.
\bsertitle{Springer series in chemical physics}.
\bpublisher{Springer},
\blocation{Berlin, Germany}
(\byear{2005})
\end{bbook}
\endbibitem

\bibitem{diode}
\begin{barticle}
\bauthor{\bsnm{Spinelli}, \binits{A.}},
\bauthor{\bsnm{Davis}, \binits{L.}},
\bauthor{\bsnm{Dautet}, \binits{H.}}:
\batitle{Actively quenched single-photon avalanche diode for high repetition rate time-gated photon counting}.
\bjtitle{Review of Scientific Instruments}
\bvolume{67},
\bfpage{55}--\blpage{61}
(\byear{1996}).
\doiurl{10.1063/1.1146551}
\end{barticle}
\endbibitem

\bibitem{germanium-diode}
\begin{barticle}
\bauthor{\bsnm{{Na}}, \binits{N.}},
\bauthor{\bsnm{{Lu}}, \binits{Y.-C.}},
\bauthor{\bsnm{{Liu}}, \binits{Y.-H.}},
\bauthor{\bsnm{{Chen}}, \binits{P.-W.}},
\bauthor{\bsnm{{Lai}}, \binits{Y.-C.}},
\bauthor{\bsnm{{Lin}}, \binits{Y.-R.}},
\bauthor{\bsnm{{Lin}}, \binits{C.-C.}},
\bauthor{\bsnm{{Shia}}, \binits{T.}},
\bauthor{\bsnm{{Cheng}}, \binits{C.-H.}},
\bauthor{\bsnm{{Chen}}, \binits{S.-L.}}:
\batitle{{Room temperature operation of germanium-silicon single-photon avalanche diode}}.
\bjtitle{Nature}
\bvolume{627}(\bissue{8003}),
\bfpage{295}--\blpage{300}
(\byear{2024})
{\href{https://arxiv.org/abs/2407.10379}{{arXiv:2407.10379}}}
{[physics.ins-det]}.
\doiurl{10.1038/s41586-024-07076-x}
\end{barticle}
\endbibitem

\bibitem{spapd}
\begin{barticle}
\bauthor{\bsnm{Cova}, \binits{S.}},
\bauthor{\bsnm{Ghioni}, \binits{M.}},
\bauthor{\bsnm{Lacaita}, \binits{A.}},
\bauthor{\bsnm{Samori}, \binits{C.}},
\bauthor{\bsnm{Zappa}, \binits{F.}}:
\batitle{Avalanche photodiodes and quenching circuits for single-photon detection}.
\bjtitle{Appl. Opt.}
\bvolume{35}(\bissue{12}),
\bfpage{1956}--\blpage{1976}
(\byear{1996}).
\doiurl{10.1364/AO.35.001956}
\end{barticle}
\endbibitem

\bibitem{spnano_rev0}
\begin{bchapter}
\bauthor{\bsnm{Nam}, \binits{S.W.}},
\bauthor{\bsnm{Verma}, \binits{V.B.}},
\bauthor{\bsnm{Allman}, \binits{M.S.}},
\bauthor{\bsnm{Horansky}, \binits{R.}},
\bauthor{\bsnm{Lita}, \binits{A.}},
\bauthor{\bsnm{Marsili}, \binits{F.}},
\bauthor{\bsnm{Stern}, \binits{J.A.}},
\bauthor{\bsnm{Shaw}, \binits{M.D.}},
\bauthor{\bsnm{Beyer}, \binits{A.D.}},
\bauthor{\bsnm{Mirin}, \binits{R.P.}}:
\bctitle{Nanowire superconducting single photon detectors progress and promise}.
In: \bbtitle{{CLEO}: 2014}.
\bpublisher{OSA},
\blocation{Washington, D.C.}
(\byear{2014}).
\doiurl{10.1364/CLEO_AT.2014.AW3P.1}.
\burl{https://opg.optica.org/abstract.cfm?URI=CLEO_AT-2014-AW3P.1}
\end{bchapter}
\endbibitem

\bibitem{spnano_rev1}
\begin{barticle}
\bauthor{\bsnm{Dauler}, \binits{E.A.}},
\bauthor{\bsnm{Grein}, \binits{M.E.}},
\bauthor{\bsnm{Kerman}, \binits{A.J.}},
\bauthor{\bsnm{Marsili}, \binits{F.}},
\bauthor{\bsnm{Miki}, \binits{S.}},
\bauthor{\bsnm{Nam}, \binits{S.W.}},
\bauthor{\bsnm{Shaw}, \binits{M.D.}},
\bauthor{\bsnm{Terai}, \binits{H.}},
\bauthor{\bsnm{Verma}, \binits{V.B.}},
\bauthor{\bsnm{Yamashita}, \binits{T.}}:
\batitle{{Review of superconducting nanowire single-photon detector system design options and demonstrated performance}}.
\bjtitle{Optical Engineering}
\bvolume{53}(\bissue{8}),
\bfpage{081907}
(\byear{2014}).
\doiurl{10.1117/1.OE.53.8.081907}
\end{barticle}
\endbibitem

\bibitem{spnano_rev2}
\begin{barticle}
\bauthor{\bsnm{Esmaeil~Zadeh}, \binits{I.}},
\bauthor{\bsnm{Chang}, \binits{J.}},
\bauthor{\bsnm{Los}, \binits{J.W.N.}},
\bauthor{\bsnm{Gyger}, \binits{S.}},
\bauthor{\bsnm{Elshaari}, \binits{A.W.}},
\bauthor{\bsnm{Steinhauer}, \binits{S.}},
\bauthor{\bsnm{Dorenbos}, \binits{S.N.}},
\bauthor{\bsnm{Zwiller}, \binits{V.}}:
\batitle{Superconducting nanowire single-photon detectors: A perspective on evolution, state-of-the-art, future developments, and applications}.
\bjtitle{Applied Physics Letters}
\bvolume{118}(\bissue{19}),
\bfpage{190502}
(\byear{2021})
{\href{https://arxiv.org/abs/https://pubs.aip.org/aip/apl/article-pdf/doi/10.1063/5.0045990/20021815/190502\_1\_5.0045990.pdf}{{https://pubs.aip.org/aip/apl/article-pdf/doi/10.1063/5.0045990/20021815/190502\_1\_5.0045990.pdf}}}.
\doiurl{10.1063/5.0045990}
\end{barticle}
\endbibitem

\bibitem{spnano}
\begin{bchapter}
\bauthor{\bsnm{Nam}, \binits{S.}},
\bauthor{\bsnm{Calkins}, \binits{B.}},
\bauthor{\bsnm{Gerritts}, \binits{T.}},
\bauthor{\bsnm{Harrington}, \binits{S.}},
\bauthor{\bsnm{Lita}, \binits{A.E.}},
\bauthor{\bsnm{Marsili}, \binits{F.}},
\bauthor{\bsnm{Verma}, \binits{V.B.}},
\bauthor{\bsnm{Vayshenker}, \binits{I.}},
\bauthor{\bsnm{Mirin}, \binits{R.P.}},
\bauthor{\bsnm{Shaw}, \binits{M.}},
\bauthor{\bsnm{Farr}, \binits{W.}},
\bauthor{\bsnm{Stern}, \binits{J.A.}}:
\bctitle{Superconducting nanowire avalanche photodetectors}.
In: \bbtitle{2013 IEEE Photonics Conference},
pp. \bfpage{366}--\blpage{367}
(\byear{2013}).
\doiurl{10.1109/IPCon.2013.6656589}
\end{bchapter}
\endbibitem

\bibitem{spd}
\begin{barticle}
\bauthor{\bsnm{Hadfield}, \binits{R.H.}}:
\batitle{Single-photon detectors for optical quantum information applications}.
\bjtitle{Nat. Photonics}
\bvolume{3}(\bissue{12}),
\bfpage{696}--\blpage{705}
(\byear{2009})
\end{barticle}
\endbibitem

\bibitem{passabanda}
\begin{barticle}
\bauthor{\bsnm{Fan}, \binits{Y.}},
\bauthor{\bsnm{Zhao}, \binits{L.}},
\bauthor{\bsnm{Qin}, \binits{J.}},
\bauthor{\bsnm{Li}, \binits{J.}},
\bauthor{\bsnm{Huang}, \binits{S.}},
\bauthor{\bsnm{Cao}, \binits{Z.}}:
\batitle{{High-precision time measurement electronics using the bandpass sampling method}}.
\bjtitle{Review of Scientific Instruments}
\bvolume{95}(\bissue{1}),
\bfpage{014705}
(\byear{2024}).
\doiurl{10.1063/5.0178312}
\end{barticle}
\endbibitem

\bibitem{fft}
\begin{barticle}
\bauthor{\bsnm{Goda}, \binits{K.}},
\bauthor{\bsnm{Jalali}, \binits{B.}}:
\batitle{Dispersive fourier transformation for fast continuous single-shot measurements}.
\bjtitle{Nature Photonics}
\bvolume{7}(\bissue{2}),
\bfpage{102}--\blpage{112}
(\byear{2013}).
\doiurl{10.1038/nphoton.2012.359}
\end{barticle}
\endbibitem

\bibitem{fft2}
\begin{barticle}
\bauthor{\bsnm{Mahjoubfar}, \binits{A.}},
\bauthor{\bsnm{Churkin}, \binits{D.V.}},
\bauthor{\bsnm{Barland}, \binits{S.}},
\bauthor{\bsnm{Broderick}, \binits{N.}},
\bauthor{\bsnm{Turitsyn}, \binits{S.K.}},
\bauthor{\bsnm{Jalali}, \binits{B.}}:
\batitle{Time stretch and its applications}.
\bjtitle{Nat. Photonics}
\bvolume{11}(\bissue{6}),
\bfpage{341}--\blpage{351}
(\byear{2017})
\end{barticle}
\endbibitem

\bibitem{Jung2015}
\begin{botherref}
\oauthor{\bsnm{Jung}, \binits{K.}},
\oauthor{\bsnm{Kim}, \binits{J.}}:
All-fibre photonic signal generator for attosecond timing and ultralow-noise microwave.
Scientific Reports
\textbf{5}(1)
(2015).
\doiurl{10.1038/srep16250}
\end{botherref}
\endbibitem

\bibitem{att_jitter}
\begin{barticle}
\bauthor{\bsnm{Song}, \binits{Y.}},
\bauthor{\bsnm{Zhou}, \binits{F.}},
\bauthor{\bsnm{Tian}, \binits{H.}},
\bauthor{\bsnm{Hu}, \binits{M.}}:
\batitle{Attosecond timing jitter within a temporal soliton molecule}.
\bjtitle{Optica}
\bvolume{7}(\bissue{11}),
\bfpage{1531}--\blpage{1534}
(\byear{2020}).
\doiurl{10.1364/OPTICA.397897}
\end{barticle}
\endbibitem

\bibitem{zs_jitter}
\begin{barticle}
\bauthor{\bsnm{Hou}, \binits{D.}},
\bauthor{\bsnm{Lee}, \binits{C.-C.}},
\bauthor{\bsnm{Yang}, \binits{Z.}},
\bauthor{\bsnm{Schibli}, \binits{T.R.}}:
\batitle{Timing jitter characterization of mode-locked lasers with $<1$ zs/$\surd{}${H}z resolution using a simple optical heterodyne technique}.
\bjtitle{Opt. Lett.}
\bvolume{40}(\bissue{13}),
\bfpage{2985}--\blpage{2988}
(\byear{2015}).
\doiurl{10.1364/OL.40.002985}
\end{barticle}
\endbibitem

\bibitem{ys_jitter}
\begin{botherref}
\oauthor{\bsnm{Xu}, \binits{H.}},
\oauthor{\bsnm{Wu}, \binits{H.}},
\oauthor{\bsnm{Hou}, \binits{D.}},
\oauthor{\bsnm{Lu}, \binits{H.}},
\oauthor{\bsnm{Li}, \binits{Z.}},
\oauthor{\bsnm{Zhao}, \binits{J.}}:
Yoctosecond timing jitter sensitivity in tightly synchronized mode-locked {T}i:sapphire lasers.
Photonics
\textbf{9}(8)
(2022).
\doiurl{10.3390/photonics9080569}
\end{botherref}
\endbibitem

\bibitem{Kim:16}
\begin{barticle}
\bauthor{\bsnm{Kim}, \binits{J.}},
\bauthor{\bsnm{Song}, \binits{Y.}}:
\batitle{Ultralow-noise mode-locked fiber lasers and frequency combs: principles, status, and applications}.
\bjtitle{Adv. Opt. Photon.}
\bvolume{8}(\bissue{3}),
\bfpage{465}--\blpage{540}
(\byear{2016}).
\doiurl{10.1364/AOP.8.000465}
\end{barticle}
\endbibitem

\bibitem{hom}
\begin{barticle}
\bauthor{\bsnm{Lyons}, \binits{A.}},
\bauthor{\bsnm{Knee}, \binits{G.C.}},
\bauthor{\bsnm{Bolduc}, \binits{E.}},
\bauthor{\bsnm{Roger}, \binits{T.}},
\bauthor{\bsnm{Leach}, \binits{J.}},
\bauthor{\bsnm{Gauger}, \binits{E.M.}},
\bauthor{\bsnm{Faccio}, \binits{D.}}:
\batitle{Attosecond-resolution {H}ong-{O}u-{M}andel interferometry}.
\bjtitle{Science Advances}
\bvolume{4}(\bissue{5}),
\bfpage{9416}
(\byear{2018})
{\href{https://arxiv.org/abs/https://www.science.org/doi/pdf/10.1126/sciadv.aap9416}{{https://www.science.org/doi/pdf/10.1126/sciadv.aap9416}}}.
\doiurl{10.1126/sciadv.aap9416}
\end{barticle}
\endbibitem

\bibitem{GR-chronometry}
\begin{botherref}
\oauthor{\bsnm{{Philipp}}, \binits{D.}},
\oauthor{\bsnm{{Hackmann}}, \binits{E.}},
\oauthor{\bsnm{{Hackstein}}, \binits{J.P.}},
\oauthor{\bsnm{{L{\"a}mmerzahl}}, \binits{C.}}:
{General Relativistic Chronometry with Clocks on Ground and in Space}.
arXiv e-prints,
2310--11576
(2023)
{\href{https://arxiv.org/abs/2310.11576}{{arXiv:2310.11576}}}
{[gr-qc]}.
\doiurl{10.48550/arXiv.2310.11576}
\end{botherref}
\endbibitem

\bibitem{Takamoto2020}
\begin{barticle}
\bauthor{\bsnm{Takamoto}, \binits{M.}},
\bauthor{\bsnm{Ushijima}, \binits{I.}},
\bauthor{\bsnm{Ohmae}, \binits{N.}},
\bauthor{\bsnm{Yahagi}, \binits{T.}},
\bauthor{\bsnm{Kokado}, \binits{K.}},
\bauthor{\bsnm{Shinkai}, \binits{H.}},
\bauthor{\bsnm{Katori}, \binits{H.}}:
\batitle{Test of general relativity by a pair of transportable optical lattice clocks}.
\bjtitle{Nature Photonics}
\bvolume{14}(\bissue{7}),
\bfpage{411}--\blpage{415}
(\byear{2020}).
\doiurl{10.1038/s41566-020-0619-8}
\end{barticle}
\endbibitem

\bibitem{Schiller2008}
\begin{barticle}
\bauthor{\bsnm{Schiller}, \binits{S.}},
\bauthor{\bsnm{Tino}, \binits{G.M.}},
\bauthor{\bsnm{Gill}, \binits{P.}},
\bauthor{\bsnm{Salomon}, \binits{C.}},
\bauthor{\bsnm{Sterr}, \binits{U.}},
\bauthor{\bsnm{Peik}, \binits{E.}},
\bauthor{\bsnm{Nevsky}, \binits{A.}},
\bauthor{\bsnm{G\"{o}rlitz}, \binits{A.}},
\bauthor{\bsnm{Svehla}, \binits{D.}},
\bauthor{\bsnm{Ferrari}, \binits{G.}},
\bauthor{\bsnm{Poli}, \binits{N.}},
\bauthor{\bsnm{Lusanna}, \binits{L.}},
\bauthor{\bsnm{Klein}, \binits{H.}},
\bauthor{\bsnm{Margolis}, \binits{H.}},
\bauthor{\bsnm{Lemonde}, \binits{P.}},
\bauthor{\bsnm{Laurent}, \binits{P.}},
\bauthor{\bsnm{Santarelli}, \binits{G.}},
\bauthor{\bsnm{Clairon}, \binits{A.}},
\bauthor{\bsnm{Ertmer}, \binits{W.}},
\bauthor{\bsnm{Rasel}, \binits{E.}},
\bauthor{\bsnm{M\"{u}ller}, \binits{J.}},
\bauthor{\bsnm{Iorio}, \binits{L.}},
\bauthor{\bsnm{L\"{a}mmerzahl}, \binits{C.}},
\bauthor{\bsnm{Dittus}, \binits{H.}},
\bauthor{\bsnm{Gill}, \binits{E.}},
\bauthor{\bsnm{Rothacher}, \binits{M.}},
\bauthor{\bsnm{Flechner}, \binits{F.}},
\bauthor{\bsnm{Schreiber}, \binits{U.}},
\bauthor{\bsnm{Flambaum}, \binits{V.}},
\bauthor{\bsnm{Ni}, \binits{W.-T.}},
\bauthor{\bsnm{Liu}, \binits{L.}},
\bauthor{\bsnm{Chen}, \binits{X.}},
\bauthor{\bsnm{Chen}, \binits{J.}},
\bauthor{\bsnm{Gao}, \binits{K.}},
\bauthor{\bsnm{Cacciapuoti}, \binits{L.}},
\bauthor{\bsnm{Holzwarth}, \binits{R.}},
\bauthor{\bsnm{Heß}, \binits{M.P.}},
\bauthor{\bsnm{Sch\"{a}fer}, \binits{W.}}:
\batitle{Einstein gravity explorer–a medium-class fundamental physics mission}.
\bjtitle{Experimental Astronomy}
\bvolume{23}(\bissue{2}),
\bfpage{573}--\blpage{610}
(\byear{2008}).
\doiurl{10.1007/s10686-008-9126-5}
\end{barticle}
\endbibitem

\bibitem{deloc}
\begin{barticle}
\bauthor{\bsnm{{Hu}}, \binits{Y.}},
\bauthor{\bsnm{{Lock}}, \binits{M.P.E.}},
\bauthor{\bsnm{{Woods}}, \binits{M.P.}}:
\batitle{{On the feasibility of detecting quantum delocalization effects on relativistic time dilation in optical clocks}}.
\bjtitle{Quantum Science and Technology}
\bvolume{9}(\bissue{4}),
\bfpage{045052}
(\byear{2024})
{\href{https://arxiv.org/abs/2307.08938}{{arXiv:2307.08938}}}
{[quant-ph]}.
\doiurl{10.1088/2058-9565/ad752c}
\end{barticle}
\endbibitem

\end{thebibliography}

\end{document}